\documentclass[
 reprint,preprintnumbers,
 amsmath,amssymb,
 aps, prd, superscriptaddress,
]{revtex4-1}

\usepackage{graphicx}
\usepackage{dcolumn}
\usepackage{bm}
\usepackage{hyperref}
\usepackage[mathlines]{lineno}
\usepackage{soul}
\usepackage{placeins}
\usepackage[export]{adjustbox}
\usepackage{ amssymb }
\usepackage{rotating}
\usepackage{siunitx}

\usepackage{lineno}
\usepackage{tabularx}
\usepackage[utf8]{inputenc}
\usepackage[T1]{fontenc}

\hypersetup{
    pdfnewwindow=true,    
    colorlinks=true,      
    linkcolor=blue,       
    citecolor=blue,       
    filecolor=blue,       
    urlcolor=blue         
}

\usepackage[normalem]{ulem}

\usepackage[usenames, dvipsnames]{color}

\begin{document}

\preprint{MIT-CTP/5037}

\title{First HAWC Observations of the Sun Constrain Steady TeV Gamma-Ray Emission}

\author{A.~Albert}\affiliation{Physics Division, Los Alamos National Laboratory, Los Alamos, NM, USA }
\author{R.~Alfaro}\affiliation{Instituto de F\'{i}sica, Universidad Nacional Autónoma de México, Ciudad de Mexico, Mexico }
\author{C.~Alvarez}\affiliation{Universidad Autónoma de Chiapas, Tuxtla Gutiérrez, Chiapas, México}
\author{R.~Arceo}\affiliation{Universidad Autónoma de Chiapas, Tuxtla Gutiérrez, Chiapas, México}
\author{J.C.~Arteaga-Velázquez}\affiliation{Universidad Michoacana de San Nicolás de Hidalgo, Morelia, Mexico }
\author{D.~Avila Rojas}\affiliation{Instituto de F\'{i}sica, Universidad Nacional Autónoma de México, Ciudad de Mexico, Mexico }
\author{H.A.~Ayala Solares}\affiliation{Department of Physics, Pennsylvania State University, University Park, PA, USA }
\author{E.~Belmont-Moreno}\affiliation{Instituto de F\'{i}sica, Universidad Nacional Autónoma de México, Ciudad de Mexico, Mexico }
\author{S.Y.~BenZvi}\affiliation{Department of Physics \& Astronomy, University of Rochester, Rochester, NY , USA }
\author{C.~Brisbois}\affiliation{Department of Physics, Michigan Technological University, Houghton, MI, USA }
\author{K.S.~Caballero-Mora}\affiliation{Universidad Autónoma de Chiapas, Tuxtla Gutiérrez, Chiapas, México}
\author{T.~Capistrán}\affiliation{Instituto Nacional de Astrof\'{i}sica, Óptica y Electrónica, Puebla, Mexico }
\author{A.~Carramiñana}\affiliation{Instituto Nacional de Astrof\'{i}sica, Óptica y Electrónica, Puebla, Mexico }
\author{S.~Casanova}\affiliation{Institute of Nuclear Physics Polish Academy of Sciences, PL-31342 IFJ-PAN, Krakow, Poland }
\author{M.~Castillo}\affiliation{Universidad Michoacana de San Nicolás de Hidalgo, Morelia, Mexico }
\author{J.~Cotzomi}\affiliation{Facultad de Ciencias F\'{i}sico Matemáticas, Benemérita Universidad Autónoma de Puebla, Puebla, Mexico }
\author{S.~Coutiño de León}\affiliation{Instituto Nacional de Astrof\'{i}sica, Óptica y Electrónica, Puebla, Mexico }
\author{C.~De León}\affiliation{Facultad de Ciencias F\'{i}sico Matemáticas, Benemérita Universidad Autónoma de Puebla, Puebla, Mexico }
\author{E.~De la Fuente}\affiliation{Departamento de F\'{i}sica, Centro Universitario de Ciencias Exactase Ingenierias, Universidad de Guadalajara, Guadalajara, Mexico }
\author{S.~Dichiara}\affiliation{Instituto de Astronom\'{i}a, Universidad Nacional Autónoma de México, Ciudad de Mexico, Mexico }
\author{B.L.~Dingus}\affiliation{Physics Division, Los Alamos National Laboratory, Los Alamos, NM, USA }
\author{M.A.~DuVernois}\affiliation{Department of Physics, University of Wisconsin-Madison, Madison, WI, USA }
\author{J.C.~Díaz-Vélez}\affiliation{Departamento de F\'{i}sica, Centro Universitario de Ciencias Exactase Ingenierias, Universidad de Guadalajara, Guadalajara, Mexico }
\author{K.~Engel}\affiliation{Department of Physics, University of Maryland, College Park, MD, USA }
\author{O.~Enríquez-Rivera}\affiliation{Instituto de Geof\'{i}sica, Universidad Nacional Autónoma de México, Ciudad de Mexico, Mexico }
\author{C.~Espinoza}\affiliation{Instituto de F\'{i}sica, Universidad Nacional Autónoma de México, Ciudad de Mexico, Mexico }
\author{H.~Fleischhack}\affiliation{Department of Physics, Michigan Technological University, Houghton, MI, USA }
\author{N.~Fraija}\affiliation{Instituto de Astronom\'{i}a, Universidad Nacional Autónoma de México, Ciudad de Mexico, Mexico }
\author{J.A.~García-González}\affiliation{Instituto de F\'{i}sica, Universidad Nacional Autónoma de México, Ciudad de Mexico, Mexico }
\author{F.~Garfias}\affiliation{Instituto de Astronom\'{i}a, Universidad Nacional Autónoma de México, Ciudad de Mexico, Mexico }
\author{M.M.~González}\affiliation{Instituto de Astronom\'{i}a, Universidad Nacional Autónoma de México, Ciudad de Mexico, Mexico }
\author{J.A.~Goodman}\affiliation{Department of Physics, University of Maryland, College Park, MD, USA }
\author{Z.~Hampel-Arias}\affiliation{Department of Physics, University of Wisconsin-Madison, Madison, WI, USA }
\author{J.P.~Harding}\affiliation{Physics Division, Los Alamos National Laboratory, Los Alamos, NM, USA }
\author{S.~Hernandez}\affiliation{Instituto de F\'{i}sica, Universidad Nacional Autónoma de México, Ciudad de Mexico, Mexico }
\author{B.~Hona}\affiliation{Department of Physics, Michigan Technological University, Houghton, MI, USA }
\author{F.~Hueyotl-Zahuantitla}\affiliation{Universidad Autónoma de Chiapas, Tuxtla Gutiérrez, Chiapas, México}
\author{P.~Hüntemeyer}\affiliation{Department of Physics, Michigan Technological University, Houghton, MI, USA }
\author{A.~Iriarte}\affiliation{Instituto de Astronom\'{i}a, Universidad Nacional Autónoma de México, Ciudad de Mexico, Mexico }
\author{A.~Jardin-Blicq}\affiliation{Max-Planck Institute for Nuclear Physics, 69117 Heidelberg, Germany}
\author{V.~Joshi}\affiliation{Max-Planck Institute for Nuclear Physics, 69117 Heidelberg, Germany}
\author{S.~Kaufmann}\affiliation{Universidad Autónoma de Chiapas, Tuxtla Gutiérrez, Chiapas, México}
\author{H.~León Vargas}\affiliation{Instituto de F\'{i}sica, Universidad Nacional Autónoma de México, Ciudad de Mexico, Mexico }
\author{G.~Luis-Raya}\affiliation{Universidad Politecnica de Pachuca, Pachuca, Hgo, Mexico }
\author{J.~Lundeen}\affiliation{Department of Physics and Astronomy, Michigan State University, East Lansing, MI, USA }
\author{R.~López-Coto}\affiliation{INFN and Universita di Padova, via Marzolo 8, I-35131,Padova,Italy}
\author{K.~Malone}\affiliation{Department of Physics, Pennsylvania State University, University Park, PA, USA }
\author{S.S.~Marinelli}\affiliation{Department of Physics and Astronomy, Michigan State University, East Lansing, MI, USA }
\author{O.~Martinez}\affiliation{Facultad de Ciencias F\'{i}sico Matemáticas, Benemérita Universidad Autónoma de Puebla, Puebla, Mexico }
\author{I.~Martinez-Castellanos}\affiliation{Department of Physics, University of Maryland, College Park, MD, USA }
\author{J.~Martínez-Castro}\affiliation{Centro de Investigaci\'on en Computaci\'on, Instituto Polit\'ecnico Nacional, M\'exico City, M\'exico.}
\author{P.~Miranda-Romagnoli}\affiliation{Universidad Autónoma del Estado de Hidalgo, Pachuca, Mexico }
\author{E.~Moreno}\affiliation{Facultad de Ciencias F\'{i}sico Matemáticas, Benemérita Universidad Autónoma de Puebla, Puebla, Mexico }
\author{M.~Mostafá}\affiliation{Department of Physics, Pennsylvania State University, University Park, PA, USA }
\author{A.~Nayerhoda}\affiliation{Institute of Nuclear Physics Polish Academy of Sciences, PL-31342 IFJ-PAN, Krakow, Poland }
\author{L.~Nellen}\affiliation{Instituto de Ciencias Nucleares, Universidad Nacional Autónoma de Mexico, Ciudad de Mexico, Mexico }
\author{M.~Newbold}\affiliation{Department of Physics and Astronomy, University of Utah, Salt Lake City, UT, USA }
\author{M.U.~Nisa}\email{Corresponding author \\Email: mnisa@ur.rochester.edu}\affiliation{Department of Physics \& Astronomy, University of Rochester, Rochester, NY , USA }
\author{R.~Noriega-Papaqui}\affiliation{Universidad Autónoma del Estado de Hidalgo, Pachuca, Mexico }
\author{J.~Pretz}\affiliation{Department of Physics, Pennsylvania State University, University Park, PA, USA }
\author{E.G.~Pérez-Pérez}\affiliation{Universidad Politecnica de Pachuca, Pachuca, Hgo, Mexico }
\author{Z.~Ren}\affiliation{Dept of Physics and Astronomy, University of New Mexico, Albuquerque, NM, USA }
\author{C.D.~Rho}\affiliation{Department of Physics \& Astronomy, University of Rochester, Rochester, NY , USA }
\author{C.~Rivière}\affiliation{Department of Physics, University of Maryland, College Park, MD, USA }
\author{D.~Rosa-González}\affiliation{Instituto Nacional de Astrof\'{i}sica, Óptica y Electrónica, Puebla, Mexico }
\author{M.~Rosenberg}\affiliation{Department of Physics, Pennsylvania State University, University Park, PA, USA }
\author{E.~Ruiz-Velasco}\affiliation{Max-Planck Institute for Nuclear Physics, 69117 Heidelberg, Germany}
\author{H.~Salazar}\affiliation{Facultad de Ciencias F\'{i}sico Matemáticas, Benemérita Universidad Autónoma de Puebla, Puebla, Mexico }
\author{F.~Salesa Greus}\affiliation{Institute of Nuclear Physics Polish Academy of Sciences, PL-31342 IFJ-PAN, Krakow, Poland }
\author{A.~Sandoval}\affiliation{Instituto de F\'{i}sica, Universidad Nacional Autónoma de México, Ciudad de Mexico, Mexico }
\author{M.~Schneider}\affiliation{Department of Physics, University of Maryland, College Park, MD, USA }
\author{H.~Schoorlemmer}\affiliation{Max-Planck Institute for Nuclear Physics, 69117 Heidelberg, Germany}
\author{M.~Seglar Arroyo}\affiliation{Department of Physics, Pennsylvania State University, University Park, PA, USA }
\author{G.~Sinnis}\affiliation{Physics Division, Los Alamos National Laboratory, Los Alamos, NM, USA }
\author{A.J.~Smith}\affiliation{Department of Physics, University of Maryland, College Park, MD, USA }
\author{R.W.~Springer}\affiliation{Department of Physics and Astronomy, University of Utah, Salt Lake City, UT, USA }
\author{P.~Surajbali}\affiliation{Max-Planck Institute for Nuclear Physics, 69117 Heidelberg, Germany}
\author{I.~Taboada}\affiliation{School of Physics and Center for Relativistic Astrophysics - Georgia Institute of Technology, Atlanta, GA, USA 30332 }
\author{O.~Tibolla}\affiliation{Universidad Autónoma de Chiapas, Tuxtla Gutiérrez, Chiapas, México}
\author{K.~Tollefson}\affiliation{Department of Physics and Astronomy, Michigan State University, East Lansing, MI, USA }
\author{I.~Torres}\affiliation{Instituto Nacional de Astrof\'{i}sica, Óptica y Electrónica, Puebla, Mexico }
\author{L.~Villaseñor}\affiliation{Facultad de Ciencias F\'{i}sico Matemáticas, Benemérita Universidad Autónoma de Puebla, Puebla, Mexico }
\author{T.~Weisgarber}\affiliation{Department of Physics, University of Wisconsin-Madison, Madison, WI, USA }
\author{S.~Westerhoff}\affiliation{Department of Physics, University of Wisconsin-Madison, Madison, WI, USA }
\author{I.G.~Wisher}\affiliation{Department of Physics, University of Wisconsin-Madison, Madison, WI, USA }
\author{J.~Wood}\affiliation{Department of Physics, University of Wisconsin-Madison, Madison, WI, USA }
\author{T.~Yapici}\affiliation{Department of Physics \& Astronomy, University of Rochester, Rochester, NY , USA }
\author{A.~Zepeda}\affiliation{Physics Department, Centro de Investigacion y de Estudios Avanzados del IPN, Mexico City, DF, Mexico }
\author{H.~Zhou}\affiliation{Physics Division, Los Alamos National Laboratory, Los Alamos, NM, USA }
\author{J.D.~Álvarez}\affiliation{Universidad Michoacana de San Nicolás de Hidalgo, Morelia, Mexico }

\collaboration{HAWC Collaboration\vspace{-0.3cm}}
\author{J. F. Beacom}
\affiliation{Center for Cosmology and AstroParticle Physics (CCAPP), Ohio State University, Columbus, Ohio 43210, USA}
\affiliation{Department of Physics, Ohio State University, Columbus, Ohio 43210, USA}
\affiliation{Department of Astronomy, Ohio State University, Columbus, Ohio 43210, USA}

\author{R. K.\ Leane}
\affiliation{Center for Theoretical Physics, Massachusetts Institute of Technology, Cambridge, MA 02139, USA}

\author{T. Linden}
\affiliation{Center for Cosmology and AstroParticle Physics (CCAPP), Ohio State University, Columbus, Ohio 43210, USA}

\author{K. C. Y. Ng}
\affiliation{Department of Particle Physics and Astrophysics, Weizmann Institute of Science, Rehovot 76100, Israel}

\author{A. H. G. Peter}
\affiliation{Center for Cosmology and AstroParticle Physics (CCAPP), Ohio State University, Columbus, Ohio 43210, USA}
\affiliation{Department of Physics, Ohio State University, Columbus, Ohio 43210, USA}
\affiliation{Department of Astronomy, Ohio State University, Columbus, Ohio 43210, USA}

\author{B. Zhou \vspace{0.5cm}}
\affiliation{Center for Cosmology and AstroParticle Physics (CCAPP), Ohio State University, Columbus, Ohio 43210, USA}
\affiliation{Department of Physics, Ohio State University, Columbus, Ohio 43210, USA}

\date{\today}

\begin{abstract}
Steady gamma-ray emission up to at least 200 GeV has been detected from the solar disk in the Fermi-LAT data, with the brightest, hardest emission occurring during solar minimum.  The likely cause is hadronic cosmic rays undergoing collisions in the Sun's atmosphere after being redirected from ingoing to outgoing in magnetic fields, though the exact mechanism is not understood.  An important new test of the gamma-ray production mechanism will follow from observations at higher energies.  Only the High Altitude Water Cherenkov (HAWC) Observatory has the required sensitivity to effectively probe the Sun in the TeV range.  Using three years of HAWC data from November 2014 to December 2017, just prior to the solar minimum, we search for 1--100 TeV gamma rays from the solar disk.  No evidence of a signal is observed, and we set strong upper limits on the flux at a few $10^{-12}$ TeV$^{-1}$ cm$^{-2}$ s$^{-1}$ at 1 TeV. Our limit, which is the most constraining result on TeV gamma rays from the Sun, is $\sim 10\%$ of the theoretical maximum flux (based on a model where all incoming cosmic rays produce outgoing photons), which in turn is comparable to the Fermi-LAT data near 100 GeV. The prospects for a first TeV detection of the Sun by HAWC are especially high during solar minimum, which began in early 2018.

\end{abstract}
\maketitle


\section{\label{sec:intro}Introduction}
The Sun is an established source of MeV-GeV gamma rays, containing both transient and steady components. Solar flares, accelerating particles in explosive bursts, produce gamma rays up to 4 GeV via bremsstrahlung and pion decay. This gamma-ray emission has been observed since the 1980s \cite{2014ApJ...787...15A, Kafexhiu:2018wmh,Strong327,Lin2002,2014ApJ...787...15A, Pesce-Rollins:2015hpa,Share:2017tgw}. On the other hand, the observational study of steady-state gamma-ray emission from the Sun --- occurring during both the quiescent and active phases --- has only become possible in the last decade with space-based missions. The definitive evidence of GeV gamma rays from the Sun, first hinted at in archival EGRET data \cite{2008A&A...480..847O}, was found in the initial eighteen months of the Fermi-LAT data \cite{0004-637X-734-2-116}. The gamma rays come in two distinct spatial components: a halo extending up to $20^\circ$ produced by inverse-Compton scattering of low-energy solar photons by cosmic-ray (CR) electrons, and the solar-disk emission, expected to arise from cosmic rays interacting with the solar atmosphere. While the extended emission agrees well with models of inverse-Compton gamma rays \cite{2008ICRC....2..505O,Orlando:2006zs,Moskalenko:2006ta,Orlando:2013pza,Orlando:2017iyc}, there are no good theoretical explanations for the GeV observations of the disk emission.

Hadronic interactions between Galactic cosmic rays and the solar atmosphere have long been theorized as the main source of steady emission from the solar disk \cite{1991ApJ...382..652S,1989gros.work.....J,doi:10.1029/JZ071i023p05778}. In the model by Seckel, Stanev and Gaisser, cosmic rays interact with the Sun's atmosphere, undergo reflection in magnetic flux tubes and produce particle cascades (including gamma rays) on their way out \cite{1991ApJ...382..652S}. The theoretical upper bound on the flux from this process, which we denote as CR upper bound, is derived by assuming the maximal production of gamma rays from interactions between the incoming cosmic rays and the solar surface \cite{2018arXiv180305436L}. Surprisingly, the observed flux above 1 GeV is higher than the nominal predictions in Ref. \cite{1991ApJ...382..652S} by almost a factor of 7.

\begin{figure}[t!]     
\centering
\includegraphics[width=0.52\textwidth]{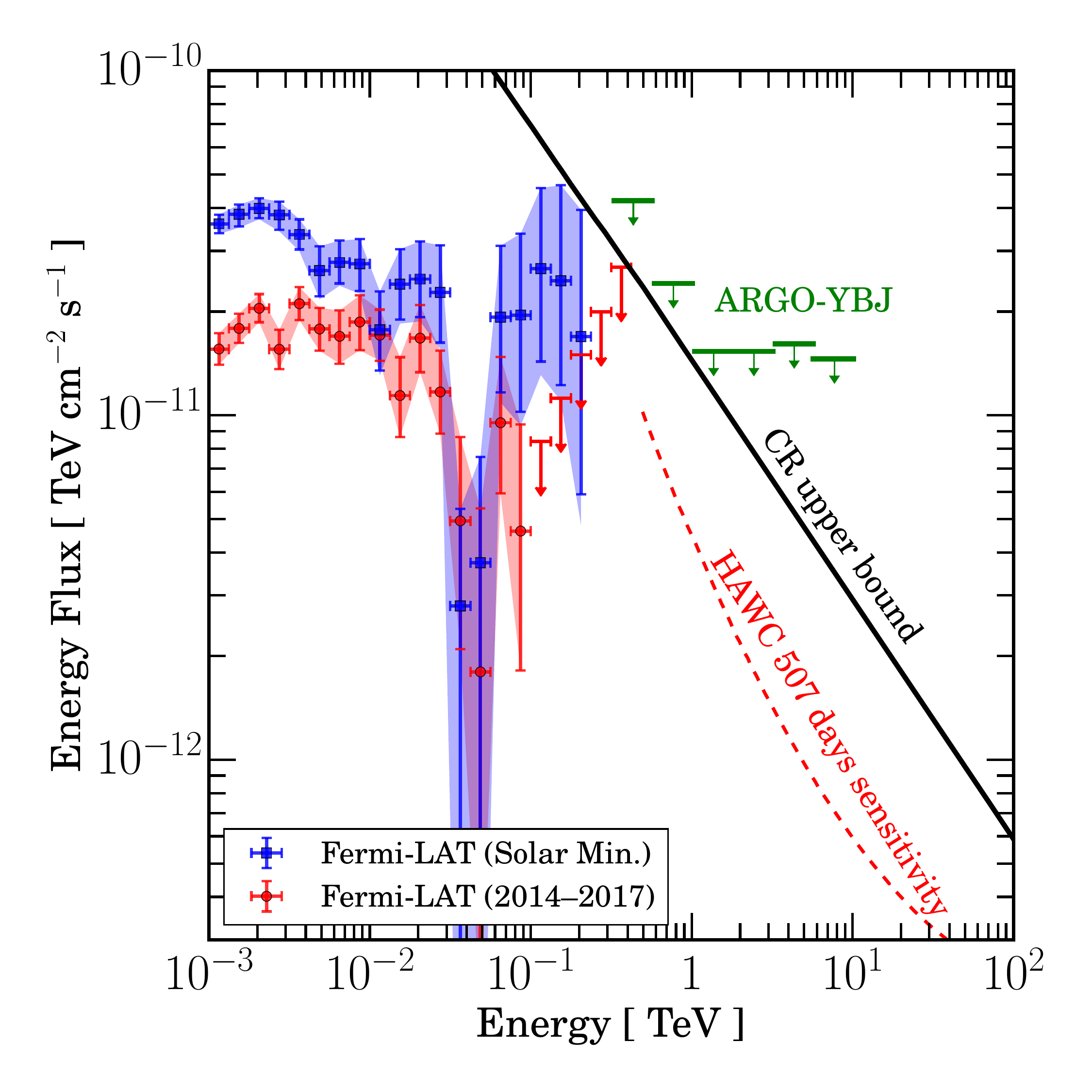}
\caption{The solar atmospheric gamma-ray flux measured in the GeV range~\cite{2018arXiv180305436L,2018arXiv180406846T} and its observational limits \cite{2016ARGOGAMMA} and prospects in the TeV range with HAWC \cite{2017ApJ...843...39A}.  We show the 1 year sensitivity of HAWC for a $E^{-2.63}$ source for scale, and compute the actual sensitivity to Sun in this work. We focus on the disk emission, for which Fermi-LAT data is approaching the theoretical maximum. The inverse-Compton emission from the halo is expected to be small in the TeV range \cite{2016arXiv161202420Z}.}
\label{fig:intro}
\end{figure}

The disk emission has been confirmed in follow-up studies utilizing 6 years and 9 years of the Fermi-LAT data with the highest-energy observations extending above 200 GeV~\cite{Ng:2015gya,2018arXiv180305436L,2018arXiv180406846T}. In addition, the gamma-ray flux between 1--100 GeV has been observed to be anti-correlated with solar activity, varying by a factor $\gtrsim3$ between solar minimum and maximum (see Fig. \ref{fig:intro}, which is explained below).  An unexplained dip near 40 GeV in the spectrum was also found, and resolved disk images shows polar and equatorial components whose strength varies through the solar cycle \cite{2018arXiv180305436L,2018arXiv180406846T}. Interestingly, the observed spectrum during the last solar minimum (cycle 24) is much harder~($\sim E^{-2.2}$) than that predicted in Ref.~\cite{1991ApJ...382..652S}~($\sim E^{-2.7}$), and reached almost $\sim 10\%$ of the CR upper bound~\cite{2018arXiv180406846T}. This flux, if continued into the TeV range would represent a flux as high as 10$\%$ of the Crab nebula \cite{2017ApJ...843...39A}; this strongly motivates extending the measurements into the TeV range.


A further motivation comes from the search for new physics in the TeV range. The Sun may capture and accumulate dark matter at its core, which then annihilates into Standard Model particles to produce observable neutrinos~\cite{Gould:1991hx,PhysRevLett.55.257,1995NuPhS..43..265E,2004PhRvD..69l3505L,2009arXiv0908.0899F,2009PhRvD..79j3532P,2011JCAP...09..029R,Danninger:2014xza,Choi:2015ara,Aartsen:2016zhm,2017JCAP...05..046W,Garani:2017jcj,Baum:2016oow}, and in some dark matter models~\cite{Meade:2009mu, 2010PhRvD..81g5004B,2010PhRvD..81a6002S, Bell:2011sn, Feng:2016ijc, Adrian-Martinez:2016gti, 2017PhRvD..95l3016L, Arina:2017sng, Smolinsky:2017fvb}, other observables including gamma rays. 
Observing the Sun at the highest accessible energies would not only help in fully understanding the hadronic emission of gamma-rays and the accompanying high-energy neutrinos~\cite{0954-3899-19-9-019,Ingelman:1996mj, Andersen:2011dz, 2017PhRvD..96j3006N,2017JCAP...07..024A, Edsjo:2017kjk}, but also in searching for dark matter. 

Figure \ref{fig:intro} summarizes the status of solar disk gamma-ray measurements above 1 GeV and their potential extension into the TeV range. It shows the Fermi-LAT observation during solar minimum~\cite{2018arXiv180406846T} and the 2014--2017 spectrum and upper limit that covers the same time period as the HAWC dataset in this work~(see Sec. \ref{sec:data} below). Also shown is the 1 year HAWC sensitivity: the energy flux required to obtain a 5$\sigma$ detection 50$\%$ of the time, for a point source at the Crab declination~\cite{2017ApJ...843...39A}. This comparison highlights the potential power of HAWC for observing gamma rays from the Sun, especially during solar minimum, which could help identify the expected rigidity cutoff when the gyroradius of the primary cosmic rays reaches the extent of the Sun~ \cite{Ng:2015gya}, as well as understanding the modulation of the gamma-ray flux \cite{2018arXiv180406846T,2014A&ARv..22...78W}

Because the maximum gamma-ray energies accessible to satellite experiments like Fermi-LAT are limited to a few hundred GeV \cite{2009ApJ...697.1071A,2017APh....95....6C,PhysRevLett.119.181101}, the Sun can only be studied in TeV gamma rays by ground-based observatories. Most TeV gamma-ray experiments rely on the Imaging Air Cherenkov technique and only take data at night \cite{2013JInst...8P6008A,LESSARD1999243,2014APh....54...67B,2017APh....94...29A}. The High Altitude Water Cherenkov (HAWC) Observatory, offering continuous daytime observations, is one of the few running experiments capable of observing the Sun at TeV energies. HAWC has been collecting data from the Sun since beginning full operations in November 2014 and will continue to monitor the Sun throughout the upcoming solar minimum. The long-term analysis will allow us to study the time variation of the flux at TeV energies. 

In this paper, which serves as a prelude to the upcoming solar minimum analysis, we describe our first three years of observations of the solar disk conducted in a relatively active portion of the solar cycle. Section \ref{HAWC} briefly introduces HAWC and the procedure of data collection. Section \ref{sec:obs} describes the analysis and the computation of upper limits on the gamma-ray flux. The sensitivity of the measurement with simulations and a discussion of systematic uncertainties are presented in Section \ref{sec:sim}. Section \ref{conc} discusses the results and concludes. The HAWC results have important implications for dark matter searches from the Sun. We explore that aspect of the study in detail in a companion paper \cite{DMPaper}.

\section{\label{HAWC}The HAWC Gamma-ray Observatory}
HAWC is a wide field-of-view ground-based array of 300 detectors that uses the water Cherenkov technique to survey the TeV sky in gamma rays and cosmic rays \cite{Abeysekara:2013qka,2017ApJ...843...39A}. The observatory is located at an altitude of 4100 m above sea level at Sierra Negra in Mexico and covers an area of 22,000 m$^2$. The individual detectors consist of four upward facing photomultiplier tubes (PMTs) anchored to the bottom of a cylindrical water tank 4.5 m high and 7.3 m wide. Gamma rays and cosmic rays from astrophysical sources produce extensive air showers of particles, which cascade through the Earth's atmosphere. The PMTs are triggered by the Cherenkov light produced by the muons and electrons in the air shower as they pass through the water in the tanks. HAWC is triggered at a rate of 25 kHz and has a duty cycle of $>95\%$. 

To select candidate air-shower events for analysis, we use a multiplicity condition of 28 PMTs (channels) within a 150 ns window. Further cuts ensure that at least $6\%$ of the operational channels are triggered, setting an initial angular resolution of $\sim 1.2^\circ$, which is then improved using additional data-quality cuts \cite{2014ApJ...796..108A}. The spatial and temporal distribution of the charge measured by the PMTs over the array are used to determine shower properties including the arrival direction of the primary particle, and the position of the core (the projection of the shower axis onto the array) of each air-shower event. The topology of each shower is also used to determine whether the primary particle is a cosmic ray or a gamma ray. In contrast to hadronic showers, the electromagnetic showers have a very compact distribution of charge as a function of distance from the core on the array. These differences are parameterized in the gamma-hadron cuts described in Ref. \cite{2017ApJ...843...39A}. For a complete description of the hardware calibration and reconstruction of the data, see Refs. \cite{Abeysekara:2013qka,2017ApJ...843...39A}.

HAWC can efficiently detect gamma rays and cosmic rays with energies between $\sim1$ TeV and several hundred TeV. With high background rejection and an angular resolution approaching $0.2^\circ$ at the highest energies, HAWC has been measuring very-high-energy gamma-ray emission from point and extended sources since commencing full operations in 2014 \cite{Abeysekara:2017hyn}. At the same time HAWC has also analyzed TeV cosmic rays through studies of the Moon and Sun Shadows \cite{2017arXiv171000890H, Abeysekara:2018syp,Enriquez:2015nva}. In the following section, we review the analysis of the Sun in cosmic rays and extend it further to search for TeV gamma rays.   


\section{\label{sec:obs}Observation of the Solar Disk}
\subsection{\label{sec:data}Data Selection}
This work uses data collected at HAWC between November 2014 and December 2017 corresponding to three years of observation in solar cycle 24. The solar activity in this cycle peaked in April 2014. This implies that the HAWC data samples a period of steadily decreasing solar activity just prior to the imminent solar minimum. The total live time of the data is 1017 days. To avoid signal contamination from other sources, we exclude the days when the Sun's position is within ten degrees of the galactic plane or other bright sources in the HAWC field-of-view, such as the Crab nebula. This cut reduces the data set to 829 days.

The data are divided into nine analysis bins based on the fraction of the total PMTs hit by an air shower. The size of the shower, which is quantified by the fraction of the triggered array is a measure of the energy of an event. Higher-energy showers trigger a greater fraction of the total available number of PMT channels. Table \ref{tab:bins} lists the maximum fractional number of PMTs hit and the estimated median energy in each bin assuming an $E^{-2.7}$ spectrum. The distribution of energies within the bins are correlated and have wide dispersions \cite{2017ApJ...843...39A}. Below, we show that our results have only a mild dependence on the assumed spectrum within a bin (Sec. \ref{sub:limits}). 

For comparison with GeV measurements, we also obtain data from Fermi-LAT covering the same period as the HAWC dataset in this work. The Fermi-LAT 2014--2017 spectrum and upper limit (Fig.\ref{fig:intro}) is obtained with the same procedure as in Ref \cite{2018arXiv180406846T}.  

\begin{table*}[ht!]
\centering
\begin{ruledtabular}
\begin{tabular}{cccccccc}
\textbf{Bin} & \textbf{Fractional Hits} &$\textbf{E}_{\textbf{median}}$\textbf{/TeV} & \textbf{RoI Radius [$\mathbf{^\circ}$]} &\textbf{$\epsilon_{\text{CR}}$ [\%]} & \textbf{$\epsilon_{\gamma}$ [\%]} & \textbf{CR Events ($\times10^9$)} &\textbf{Post-cuts Events ($\times10^9$)}  \\
\hline
1            &  0.067 -- 0.105      &   0.88    & $1.50\pm0.22$    	&  15    &  75    &   231 &     30.5                    \\
2            & 0.105 -- 0.162     &     1.36     &  $1.37\pm0.19 $	&  10  & 80      &   103   &   9.8                   \\
3            & 0.162 -- 0.247     &     2.24      &  $1.10\pm0.09$ 	&  5    &  90      &  52.2        &  2.7                   \\
4            & 0.247 -- 0.356      &    4.23      &   $0.93\pm0.06 $	& 1.50  & 70     &  25.2    &  0.34                   \\
5            & 0.356 -- 0.485      &    6.56       &  $ 0.85\pm0.05 $	&  0.55 & 65    & 12.6     &  0.07                   \\
6            & 0.485 -- 0.618       &   14.7       &   $0.59\pm0.03 $	& 0.2    & 53      & 6.16 & 0.013                    \\
7            & 0.618 -- 0.740    &      17.2        &   $0.74\pm0.05 $    & 0.25 &  70       &2.93  & 0.008            \\
8            & 0.740 -- 0.840 &         24.9       &    $0.41\pm0.03$	&  0.14 & 72       &1.42 & 0.002               \\
9            & 0.840 -- 1.000     &     60.1        &    $0.36\pm0.02$	&  0.2  &    70      &1.45 & 0.003                \\

\end{tabular}
\end{ruledtabular}
\caption{Fractional PMT hits and corresponding median energy for each analysis Bin assuming an $E^{-2.7}$. The RoI radius is the 68$\%$ containment width of the Sun shadow obtained by fitting a 2d Gaussian to the shape of the deficit at the position of the Sun. $\epsilon_{\text{CR}}$ and $\epsilon_{\gamma}$ are the respective fractions of cosmic rays and gamma rays retained after applying gamma-hadron cuts for each bin. The uncertainty on $\epsilon_{\rm CR}$ is between 0.0001--0.001\% for all bins. The gamma-hadron separation efficiency improves with energy, e.g., Bin 1 retains $15\%$ of the hadrons whereas Bin 9 retains only $0.2\%$, while the gamma-ray efficiency stays approximately constant. Also shown are the total number of events before (CR map) and after (post-cuts map) applying the gamma-hadron separation cuts on three years of data.}
\label{tab:bins}
\end{table*}

\subsection{\label{sec:maps} Analysis}
\subsubsection{Sky Maps and Background Estimation}
We search for gamma rays by projecting the air-shower directions onto a sky map of equal area pixels. The HEALPix library \cite{2005ApJ...622..759G} is used to divide the sky into pixels with an angular width of $0.1^\circ$ in equatorial coordinates. The coordinates of each pixel $i$ are given in terms of the right ascension and declination angles, $\alpha$, and $\delta$ respectively. 

The sky map consists of the counts of the arrival directions of events binned in pixels. To look for gamma-ray sources, small-scale anisotropies or other interesting features, we analyze the data in every pixel relative to an expected background.  The expected background, dominantly cosmic rays, is computed using the technique of Direct Integration, as discussed in Refs. \cite{2003ApJ...595..803A,2012ApJ...750...63A}. 
The observed cosmic-ray flux is known to vary slowly over both space and time. It is temporally stable on week-long timescales, and isotropic over the full sky at the level of 10$^{-3}$ \cite{2014ApJ...796..108A,0004-637X-698-2-2121}. We compute the expected cosmic-ray background using an integration time $\Delta t$ of 2 hours, which effectively averages the arrival distribution in right ascension at angular scales of roughly $30^\circ$. The estimated background counts $\langle N(\alpha_i,\delta_i)\rangle$ is the all-sky event rate $R(h-\alpha)$ convolved with the normalized local arrival distribution $A(h, \delta)$ for each $\Delta t$ interval, which is then summed over for the entire duration of the dataset.
\begin{equation}
\langle N(\alpha_i,\delta_i)\rangle = \sum_{\Delta t} \int dh\, R_{\Delta t}(h-\alpha) A_{\Delta t}(h, \delta),
\end{equation}
where $h$ is the hour angle.
For any pixel $i$ of the map, we can define the relative counts $\Delta \mathcal{N}_i$ as the excess or deficit with respect to the average background,
\begin{equation}
\label{eq:rel}
\Delta \mathcal{N}_i = N(\alpha_i,\delta_i) - \langle N(\alpha_i,\delta_i)\rangle,
\end{equation}
where $N(\alpha_i,\delta_i)$ is the observed number of events in the data map. Note that in general, the observed number of events includes both cosmic rays $N_{\text{CR}}$ and gamma rays $N_{\gamma}$. The expected background distribution in terms of its gamma-ray and cosmic-ray components is given by
\begin{equation}
\langle N(\alpha_i,\delta_i)\rangle =  \langle N_{\gamma}(\alpha_i,\delta_i)\rangle + \langle N_{\text{CR}}(\alpha_i,\delta_i)\rangle .
\end{equation}
The isotropic gamma-ray flux, $\langle N_{\gamma}(\alpha_i,\delta_i)\rangle$ is negligible, so the expected background is mostly the isotropic component of cosmic rays, $\langle N(\alpha_i,\delta_i)\rangle \simeq \langle N_{\text{CR}}(\alpha_i,\delta_i)\rangle$. The number of gamma rays in pixel $i$ is $N_{\gamma}^{i}$. Similarly, the cosmic-ray component of the data is given by $N_{\text{CR}}^{i}$. 

We convert the map into Sun-centered coordinates ($\alpha',\delta')$ by subtracting the equatorial coordinates of the Sun from the equatorial coordinates of each event, 
\begin{equation}
\alpha' = \alpha - \alpha_{\text{sun}}, \,
\delta' = \delta - \delta_{\text{sun}}.
\end{equation}
In these coordinates the Sun is centered at (0,0) with an angular diameter of $\sim0.5^\circ$. The analysis uses events only within 3$^\circ$ of the Sun, to good approximation, preserving the angular distances under the transformation. Any variations in the angular distances are mimicked exactly in both data and background, so the net distortion in the map is negligible \cite{2011PhDT........70C}. 

We define a region of interest~(RoI), which is centered at the solar disk (0,0) and has a width equal to the 68$\%$ containment of the Sun shadow~(see below) in the respective bin. The total counts in the RoI are computed by summing all pixels in the region. Thus, Eq. (\ref{eq:rel}) summed over all pixels $i$ in the RoI can be written as
\begin{equation}
\label{eq:RI}
\Delta \mathcal{N}_{\text{RoI}} = \sum_{i \in \text{RoI}} \left(N_{\gamma}^{i} +  N_{\text{CR}}^{i} - \langle N(\alpha'_i,\delta'_i)\rangle \right).
\end{equation}

\subsubsection{The Sun Shadow and the Net Gamma-ray Excess}
\begin{figure*}
\centering
\makebox[0.4\width][c]{
\begin{tabular}{@{}ccc@{}}
\includegraphics[width=0.37\textwidth]{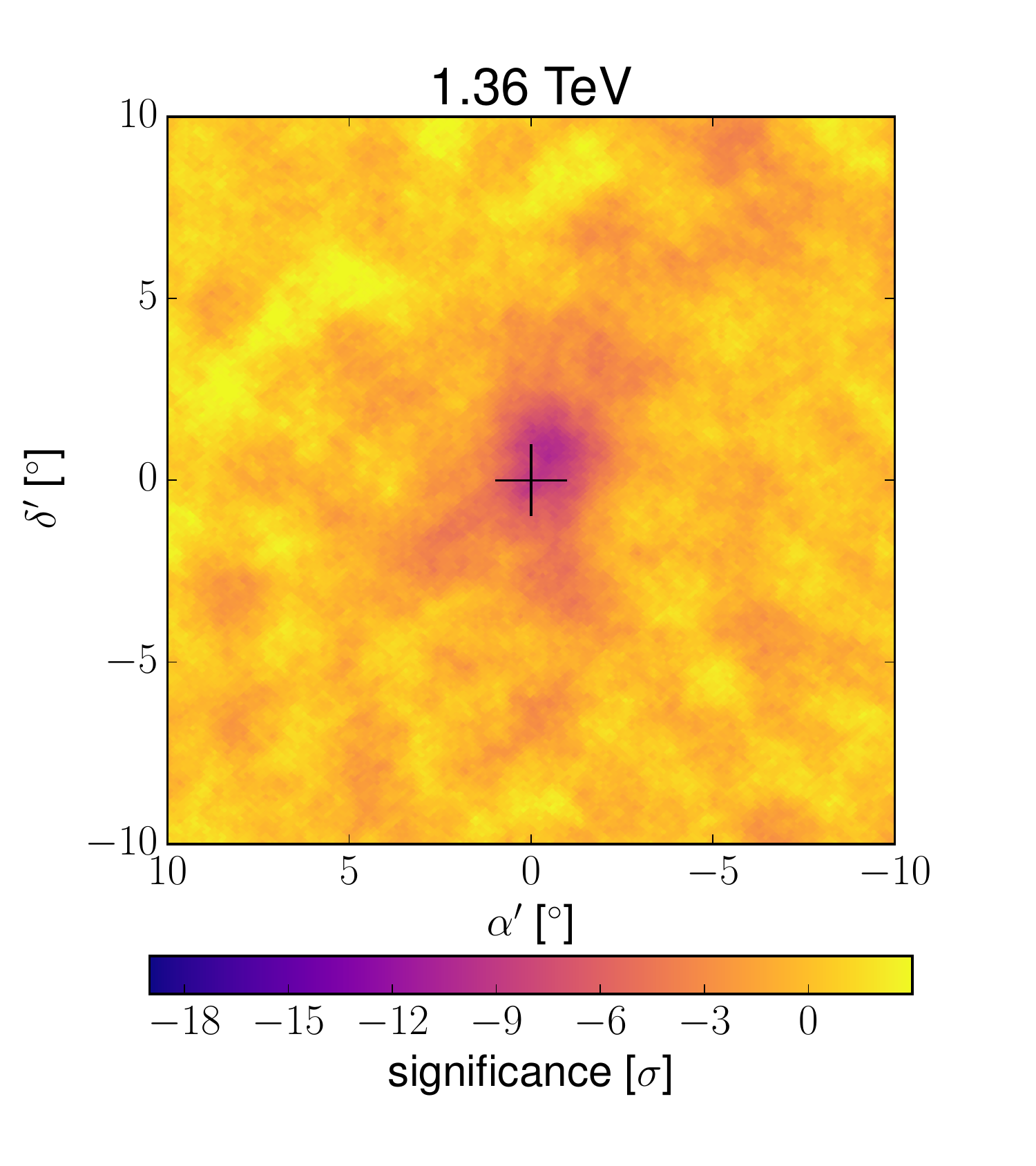} &
\includegraphics[width=0.37\textwidth]{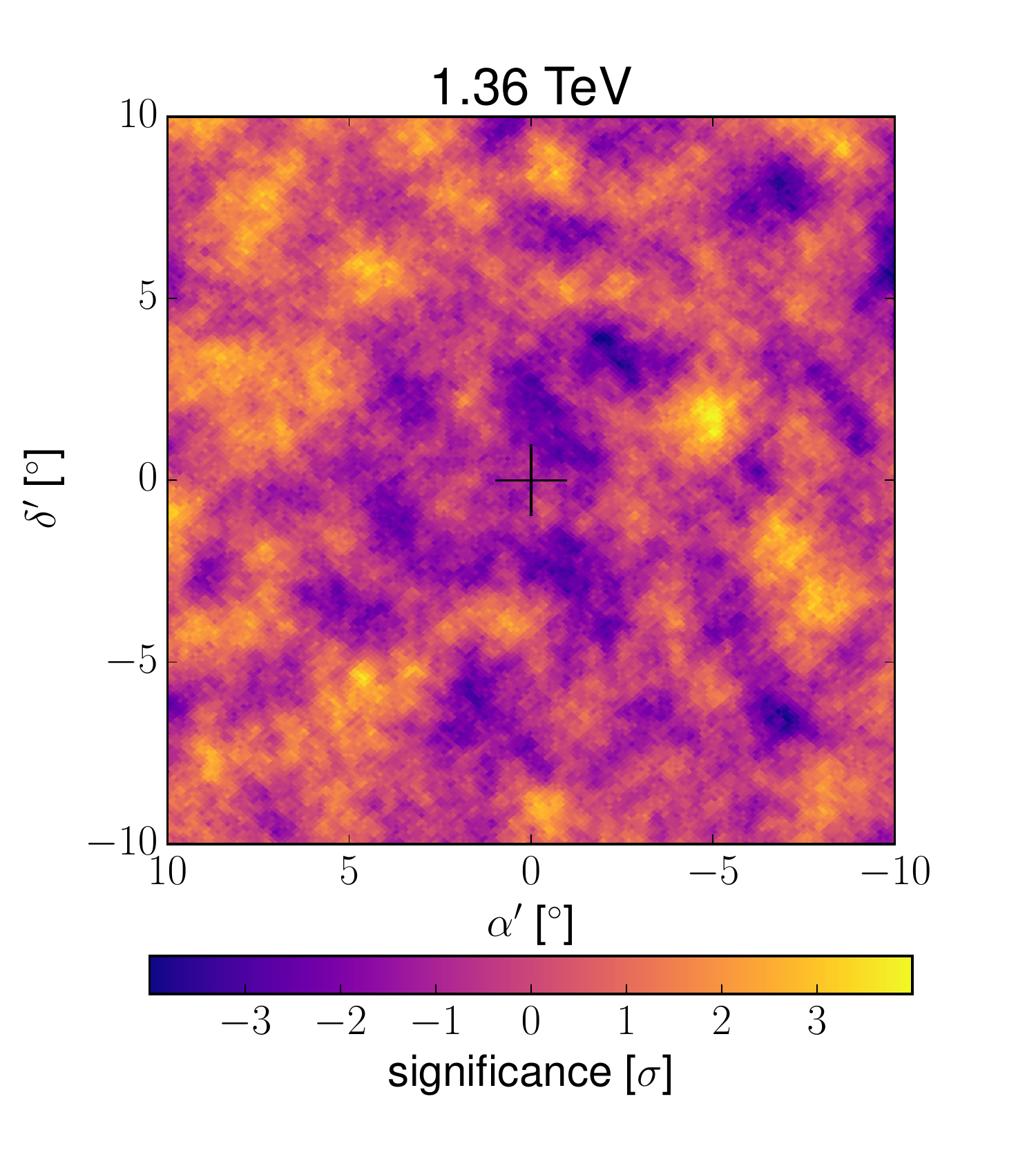} &
\includegraphics[width=0.37\textwidth]{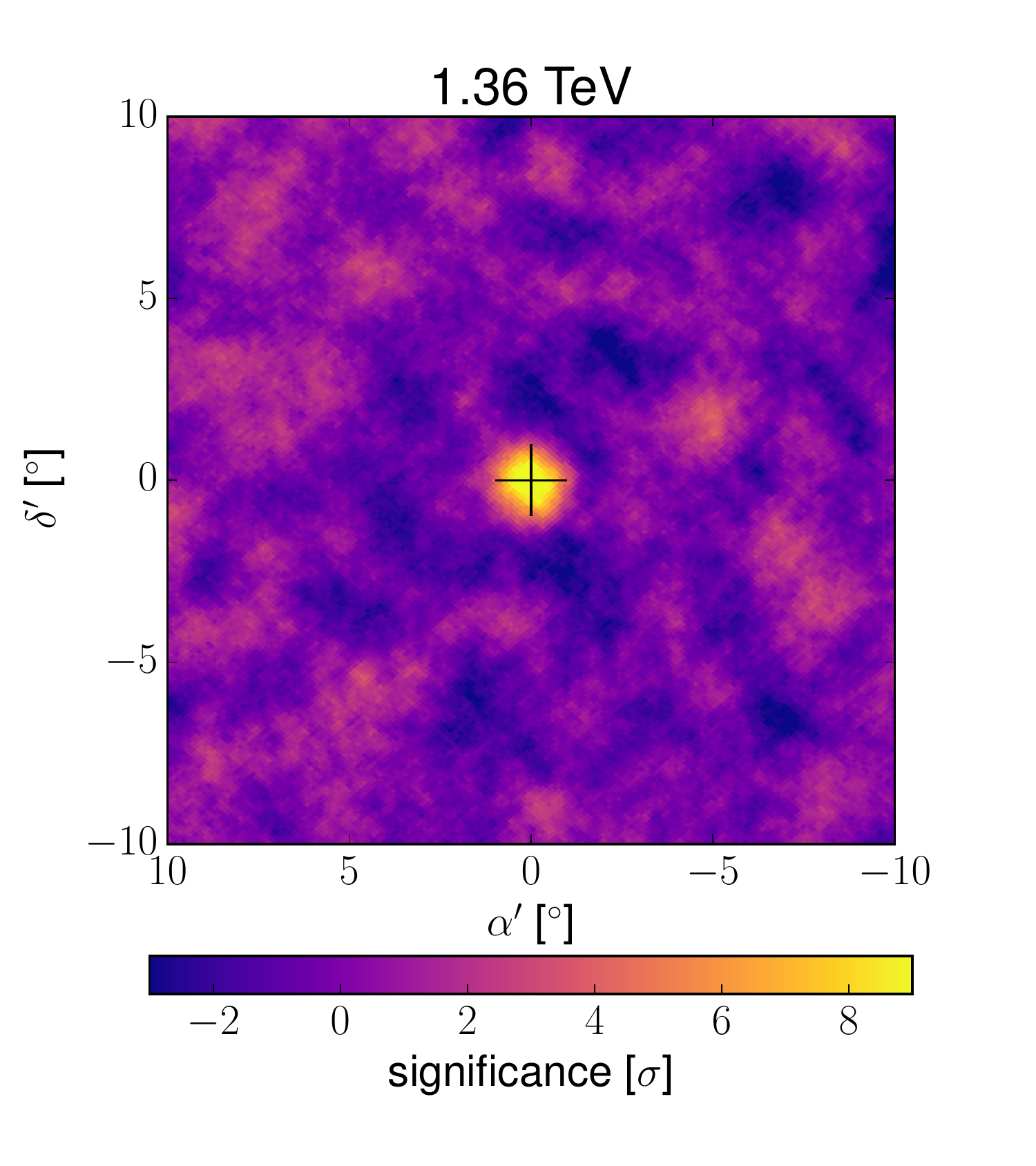} \\
\includegraphics[width=0.37\textwidth]{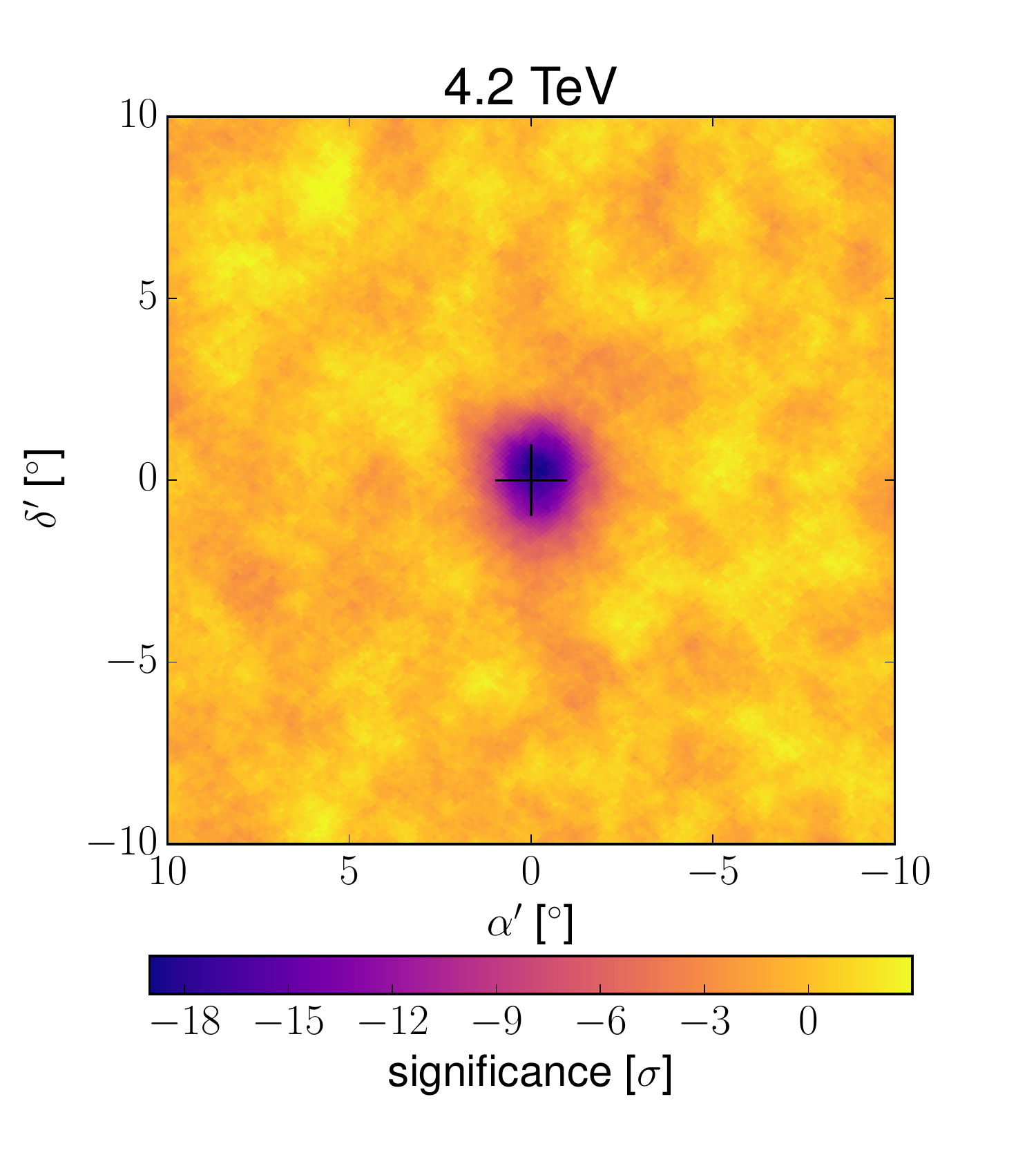} &
\includegraphics[width=0.37\textwidth]{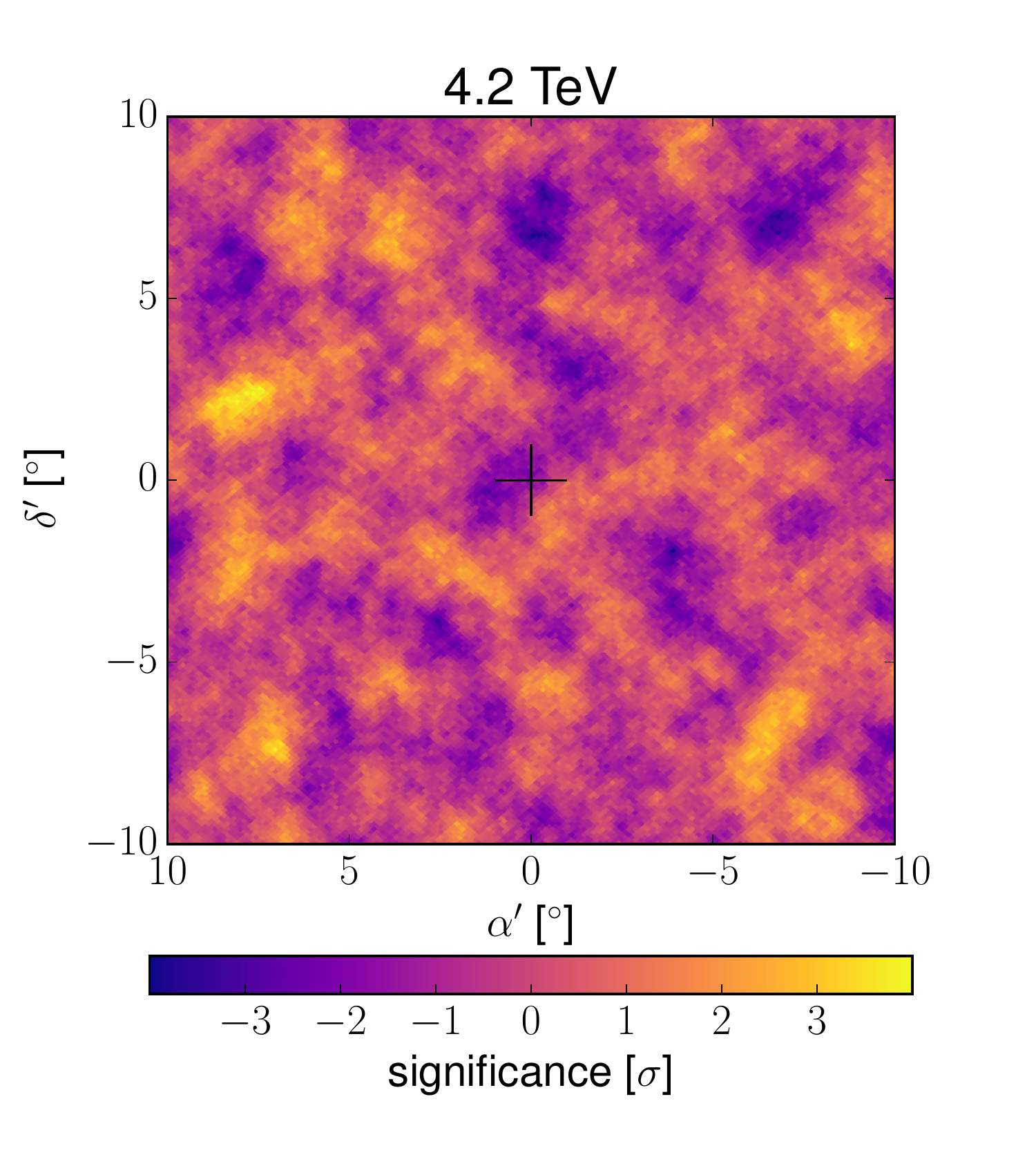}&
 \includegraphics[width=0.37\textwidth]{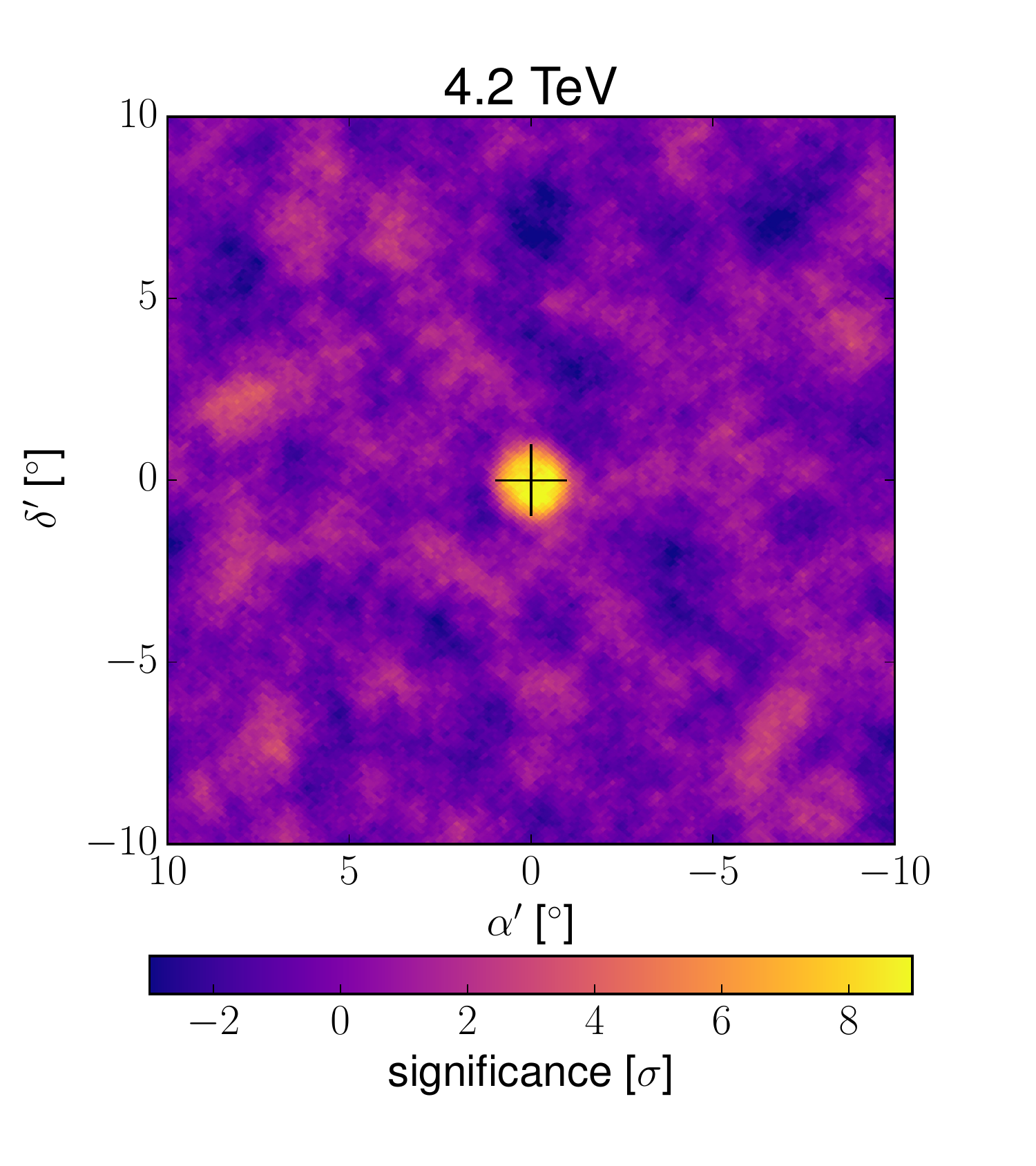}\\
\includegraphics[width=0.37\textwidth]{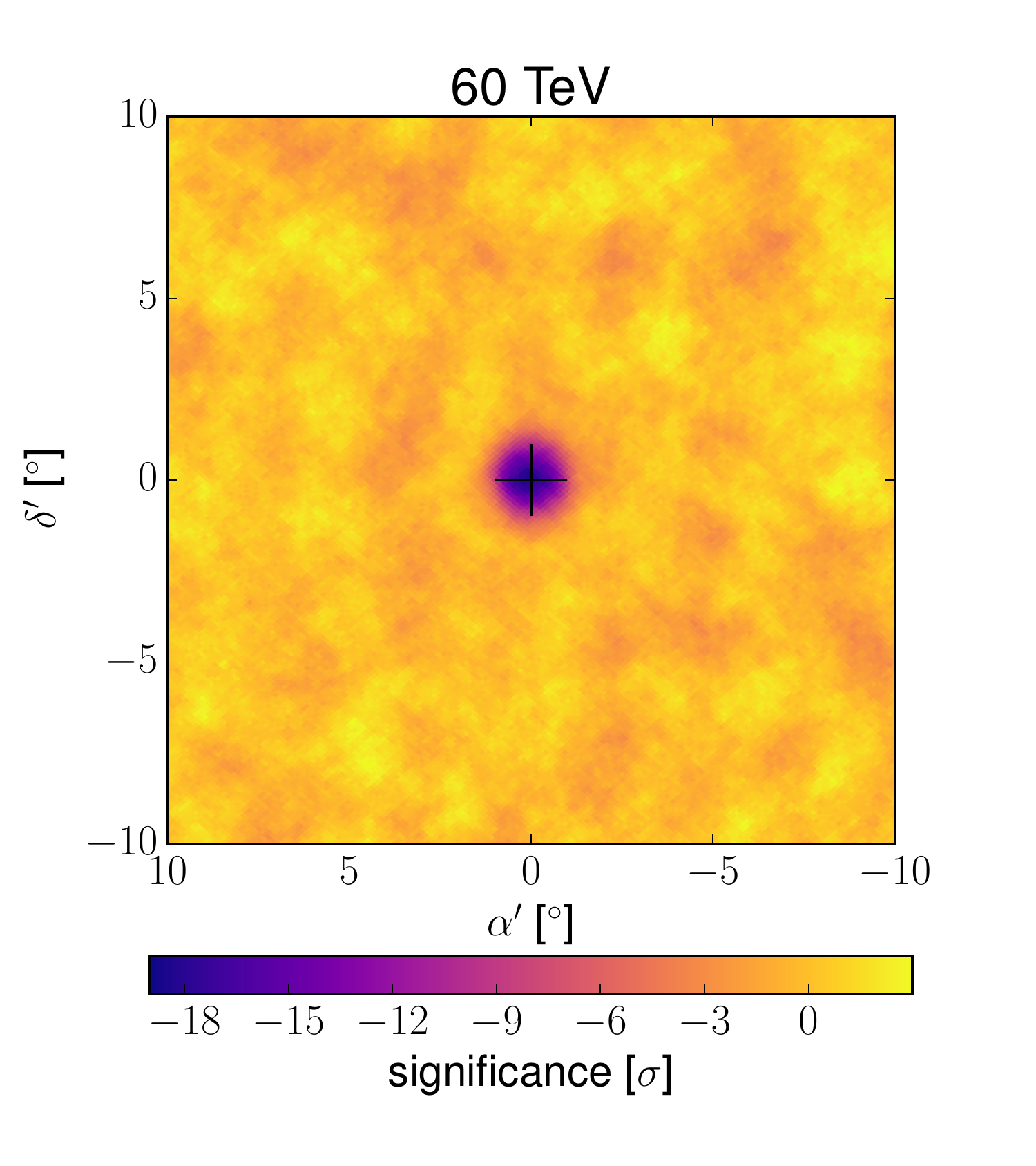} &
\includegraphics[width=0.37\textwidth]{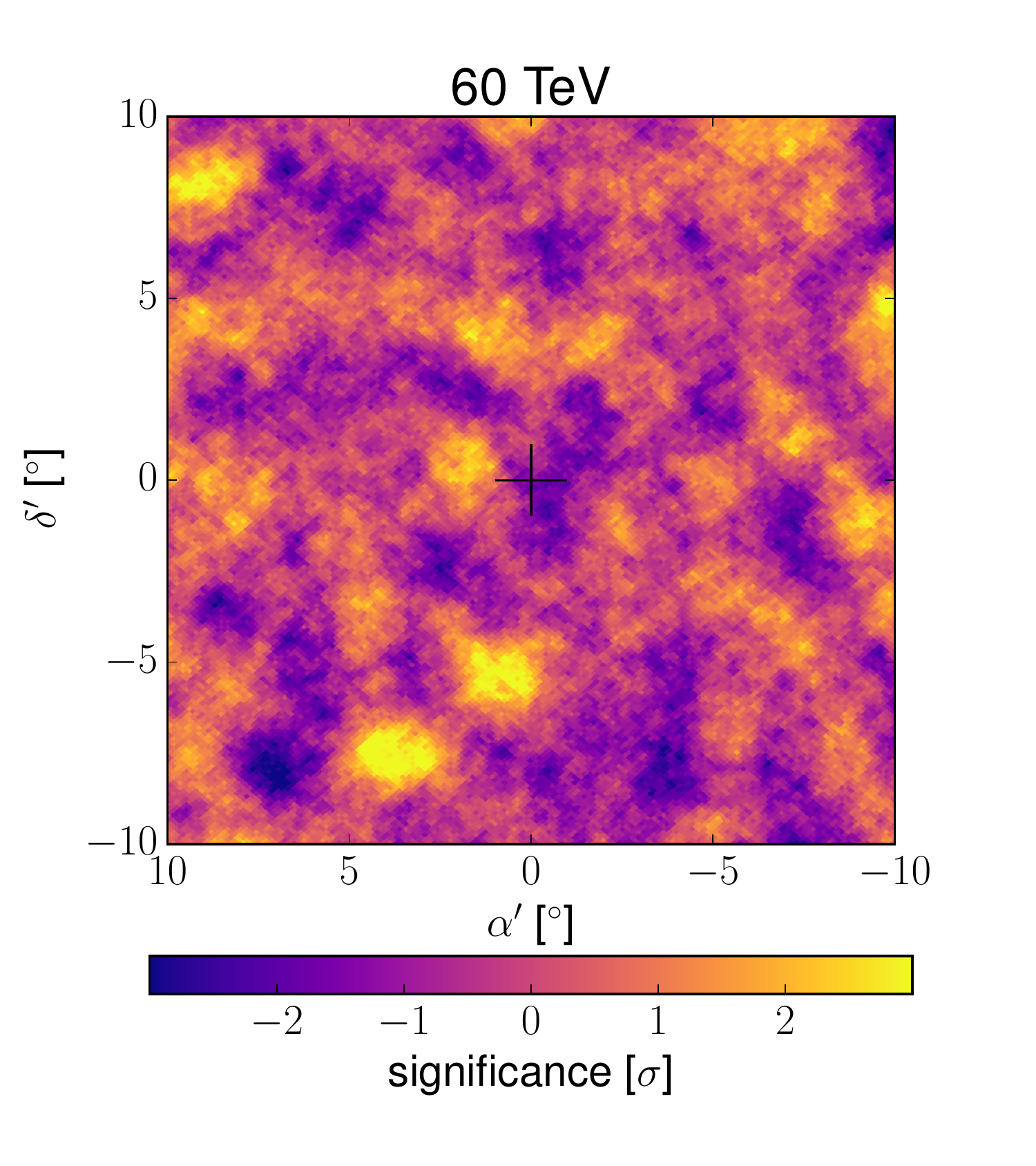}&
\includegraphics[width=0.37\textwidth]{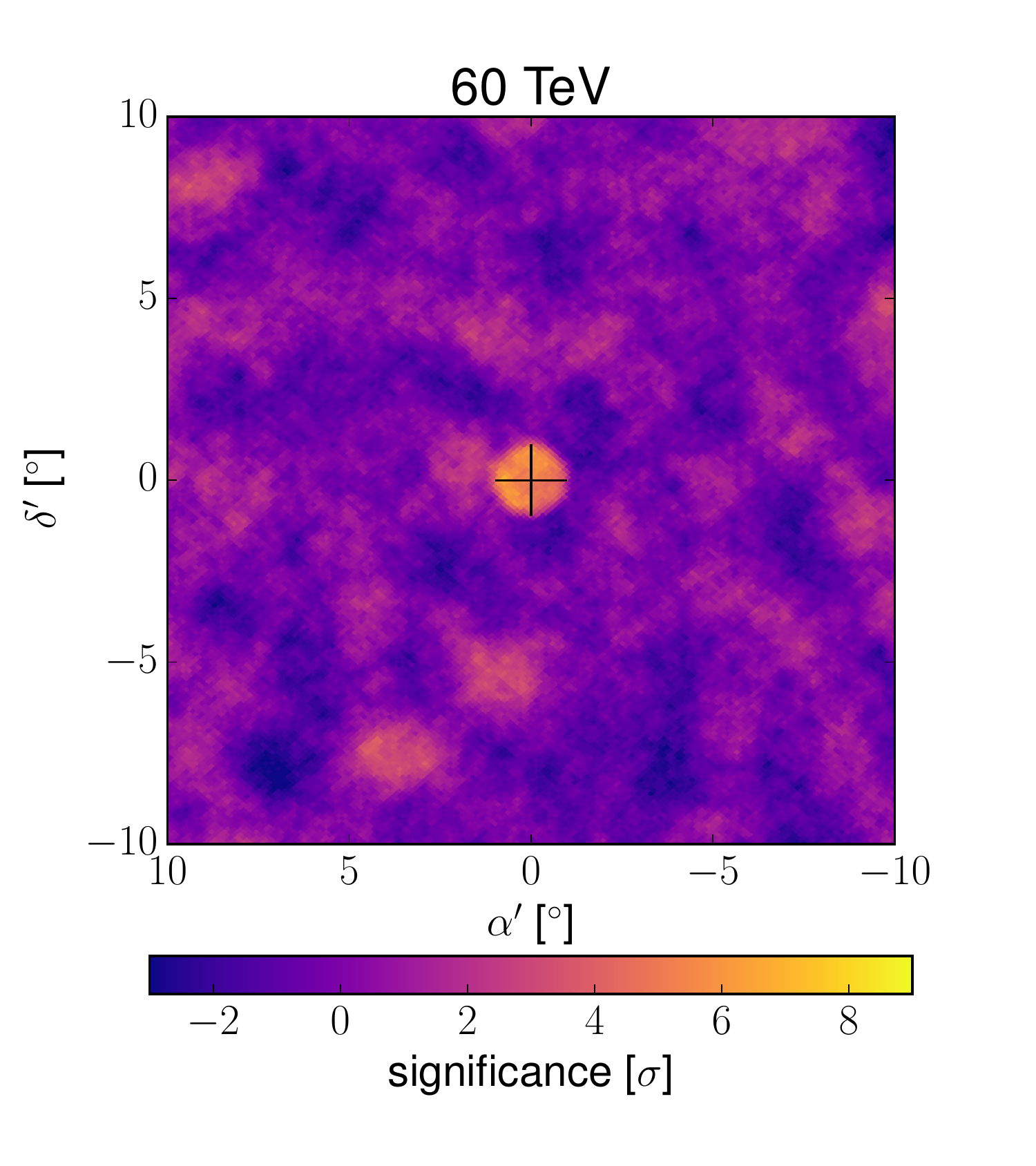} \\
\end{tabular}
}
\caption{\textbf{Left:} Observed Sun shadow, described by Eq.\;(\ref{RI}), at median energies of 1.36, 4.2 and 60 TeV. The 1$\sigma$ width of the shadow is 1.3$^\circ$, 0.9$^\circ$ and 0.3$^\circ$ at the respective energies. \textbf{Center:} Same maps with gamma-hadron cuts applied: Eq.\;(\ref{gh}). \textbf{Right:} The simulated Sun maps for the maximum expected flux from cosmic-ray interactions in the solar atmosphere. The black cross marks the position of the Sun.}
\label{fig:shadow}
\end{figure*}

While we are interested in detecting an excess gamma-ray signal from the Sun, the measurement is complicated by the fact that the Sun blocks incoming cosmic rays, producing a deficit in the observed signal at the position of the solar disk. This feature is also known as the ``Sun shadow'' (Fig. \ref{fig:shadow}). Compared to the Moon shadow observed by HAWC, the Sun shadow at low energies ($\sim$1 TeV) is less significant due to the effect of the solar magnetic fields \cite{Abeysekara:2018syp,Enriquez:2015nva}. The Sun shadow is more pronounced at higher energies, at which the cosmic rays are less deflected by the coronal and interplanetary magnetic fields~\cite{Amenomori:2013own, 2018PhRvL.120c1101A, Amenomori:2018xip}. The evolution of the shadow size with energy is also an illustration of the angular resolution of the detector to cosmic rays, which has been modeled and verified in Refs. \cite{2017ApJ...843...39A, 2014ApJ...796..108A,2017arXiv171000890H}. The presence of the shadow can bias the search for gamma rays in two ways:

1. The deficit can result in an underestimation of the overall cosmic-ray expected background.

2. The deficit may conceal the gamma-ray excess within the shadow.

The first problem is overcome by using RoI masking when computing the expected background distribution \cite{2017ApJ...842...85A}. In this way the process of direct integration does not use any pixels within a $3^\circ$ radius of the Sun and consequently the deficit of cosmic rays does not lower the overall expected background. The second problem poses a challenge for extracting the net gamma-ray excess from the solar-disk observations. We tackle this by analyzing the maps with and without gamma-hadron separation as described below.


The cosmic-ray sky map includes all the data following the standard quality cuts \cite{2017arXiv171000890H}. The Sun shadow can be observed in the cosmic-ray data where the Sun blocks out part of the incoming flux. We observe the Sun shadow at different energies in the combined 3 year data. We calculate the deviation from the expected background in each pixel and assign it a significance based on the method in Ref. \cite{1983ApJ...272..317L}. The significance of the shadow increases from $7 \sigma$ in Bin 1 ($\sim 880$ GeV) to $19 \sigma$ in Bin 9 ($\sim 60$ TeV) and peaks at $21\sigma$ in Bin 6 ($\sim15$ TeV). 

The post-cuts sky maps are produced after applying gamma-hadron cuts to the data, which remove a significant fraction of cosmic rays \cite{2017ApJ...843...39A}. Figure \ref{fig:shadow} shows the cosmic-ray sky maps and the maps after gamma-hadron cuts, which are all smoothed by a 1$^\circ$ top-hat function. The entire cosmic-ray sky map at 1.36 TeV has $103$ billion events while the post-cuts sky map (with no discernible shadow) contains $9.8$ billion events~(See Table~\ref{tab:bins}), the bulk of which are still hadronic cosmic rays. The suppression of cosmic rays by the gamma-hadron cuts brings down the significance of the deficit at the Sun's position to $0.7 \sigma$ in Bin 1, $0.12 \sigma$ in Bin 6 and  $1\sigma$ in Bin 9. If there were a significant gamma-ray signal from the Sun, we would expect it to show up in the map as a bump with a roughly Gaussian profile centered at the solar disk. The cosmic-ray and post-cuts sky maps, along with a knowledge of the efficiency of gamma-hadron separation, can be used to obtain the number of gamma rays above the expected background in the RoI. 

\begin{figure*}[htb!]
\centering
\makebox[0.01\width][c]{\begin{tabular}{c}
\includegraphics[width = 1.12\textwidth]{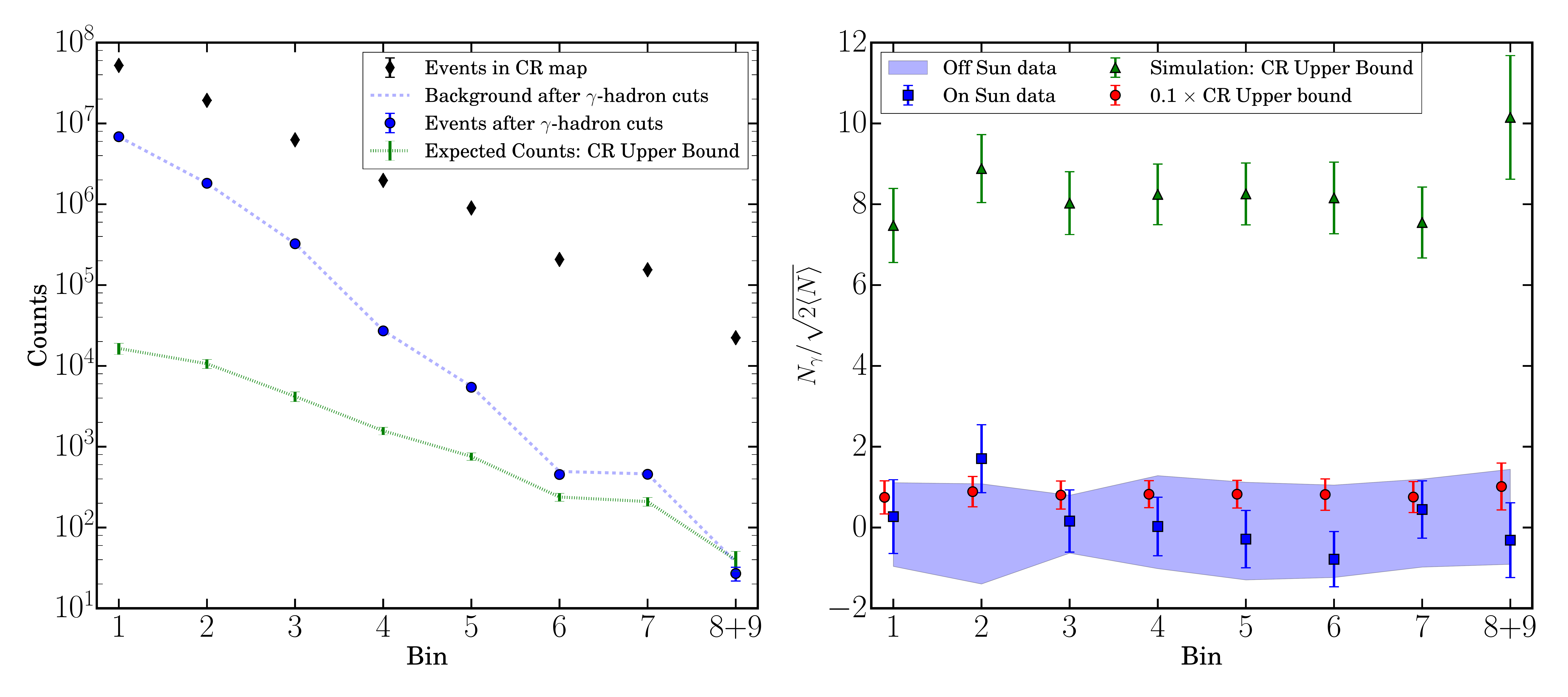}
\end{tabular}}
\caption{\textbf{Left:} Total observed events from the RoI, before and after gamma-hadron separation for eight analysis bins from Table \ref{tab:bins} (Bins 8 and 9 have been combined due to their low counts). The total number of off-source background events passing the gamma-hadron cuts is shown as the dashed line, which closely matches the on-source data. The CR upper bound line shows the expected excess gamma rays from the model. \textbf{Right:} Net gamma-ray excess relative to an estimate of the uncertainty (our full analysis does not approximate). The off-Sun band shows the central 90$\%$ distribution obtained from the ``fake'' Suns. The observed excess from the solar disk is consistent with zero. The expected excess from two simulated models is shown for comparison (the 0.1 $\times$ CR upper bound is offset along the x-axis for clarity). We obtain upper limits roughly at the level of the 0.1 $\times$ CR upper bound model, shown below. 
}
\label{fig:counts}
\end{figure*}

For a given cosmic-ray map before the gamma-hadron cuts applied, we simplify and rewrite Eq. (\ref{eq:RI}) as
\begin{equation}
\label{RI}
\Delta \mathcal{N}_{\text{RoI}} =  N_{\gamma} + \Delta N_{\text{CR}},
\end{equation}
where $\Delta N_{\text{CR}} = N_{\text{CR}} - \langle N(\alpha,\delta)\rangle$. We write a similar expression for a post-cuts sky map with gamma-hadron separation,
\begin{equation}
\label{gh}
\Delta \mathcal{N}_{\text{cuts}} = \epsilon_{\gamma} N_{\gamma} + \epsilon_{\text{CR}}\Delta N_{\text{CR}}.
\end{equation}
Here, the number of gamma rays and cosmic rays are reduced by the efficiency factors $\epsilon_{\gamma}$ and $\epsilon_{\text{CR}}$, respectively. The efficiency factors are the fraction of photons and cosmic rays retained after the gamma-hadron cuts, shown in Table \ref{tab:bins}. The gamma-ray efficiencies for each bin are obtained from simulation as described in Ref. \cite{2017ApJ...843...39A}. The cosmic-ray efficiencies $\epsilon_{\mathrm{CR}}$ are calculated from off-RoI pixels using the ratio of background counts before and after the cuts. The efficiency of rejecting cosmic rays is a function of the measured shower size, which correlates with cosmic-ray energy \cite{2017ApJ...843...39A}. The ratio of efficiencies $\epsilon_{\gamma}/ \epsilon_{\mathrm{CR}}$, therefore improves with increasing energy-proxy bin number as illustrated in the left panel of Fig. \ref{fig:counts}. 

The relative counts $\Delta \mathcal{N}_{\text{RoI}}$ and $\Delta \mathcal{N}_{\text{cuts}}$ can be obtained from the respective maps. We then solve Eq. (\ref{RI}) and (\ref{gh}) for $N_{\gamma}$ to obtain an expression for the observed number of excess gamma counts in each bin near the Sun, 
\begin{equation}
\label{gam}
N_{\gamma} = \frac{\Delta \mathcal{N}_{\text{cuts}} - \epsilon_{\mathrm{CR}}\Delta \mathcal{N}_{\text{RoI}}}{\epsilon_{\gamma} - \epsilon_{\mathrm{CR}}}.
\end{equation}

The statistical uncertainty on $N_{\gamma}$ is obtained by propagating the Poisson errors on the observed data and background-quantities described above. Roughly, because the two terms in the numerator (Eq. \ref{gam}) are comparable when setting a limit, the uncertainty is $\sim \sqrt{2\langle N \rangle}$ (the full error analysis in this work does not use the approximation). The systematic uncertainty on $N_{\gamma}$ due to gamma rays passing as cosmic rays in $\epsilon_{\text{CR}}$ is constrained to $0.1\%$ of the total statistical uncertainty on $N_{\gamma}$ \cite{2017ApJ...842...85A}, which is negligible.


\subsubsection{Analysis Results}\label{subsubsec:results}

Figure \ref{fig:counts} (left panel) summarizes the results of the analysis, showing the number of events before and after applying gamma-hadron cuts for each bin. It also illustrates the increasing hadron suppression efficiency as a function of bin number, by comparing the counts in the CR map to those after gamma-hadron separation. 

We test the analysis method by recovering a simulated signal from a point source at the Sun's position. The simulation needs to account for the varying declination, and thus maximum zenith angle of the Sun throughout the year, as the sensitivity and energy threshold of HAWC depends on the zenith angle of a given source in the sky. This is because air showers coming from a large inclination with respect to the zenith traverse a greater slant depth through the atmosphere and are attenuated compared to vertical showers. The sensitivity is maximal for sources overhead at HAWC's latitude and falls by a factor of 8 for a change in zenith angle of $45^\circ$ \cite{Abeysekara:2017hyn}. To minimize the effect of the varying elevation of the Sun, we divide the data into groups with similar maximum solar zenith angle in a given time period. We achieve this by grouping events monthly, e.g., placing events from June {2015, 2016, 2017} in one group. We have twelve groups of data in total, one for each month. An expected signal is extracted from a simulated source  at the median position of the Sun in each group, before combining the results. 

The simulated Sun has the spectrum of the cosmic-ray upper bound discussed in Section \ref{sec:intro}, corresponding to a differential flux of $1.45 \times 10^{-11}$ TeV$^{-1}$ cm $^{-2}$ s$^{-1}$ at 1 TeV. We also test cases with $10\%$ and $1\%$ of the CR upper bound. We calculate the expected number of gamma-ray events from the simulated Suns following the method described in the previous section. Also shown in Figure \ref{fig:counts} (left panel), the combined expected signal can be compared to the number of gamma-ray events obtained from the data. 

To check whether the net calculated excess from the Sun is significantly higher than the expected background, we also search for gamma rays on 72 ``fake'' Suns.  They encompass background-only regions, half a degree apart, each at an angular distance of $d$ from the true position of the Sun. We choose $5^\circ \leq d \leq 40^\circ$ to obtain 72 off-Sun samples on either side of the Sun in $\alpha'$, effectively sampling the Sun's path at different times. With the off-Sun samples, we can estimate the expected fluctuations in the absence of signals.

\begin{figure*}
\centering
\makebox[0.4\width][c]{
\begin{tabular}{@{}cc@{}}
\includegraphics[width=0.55\textwidth]{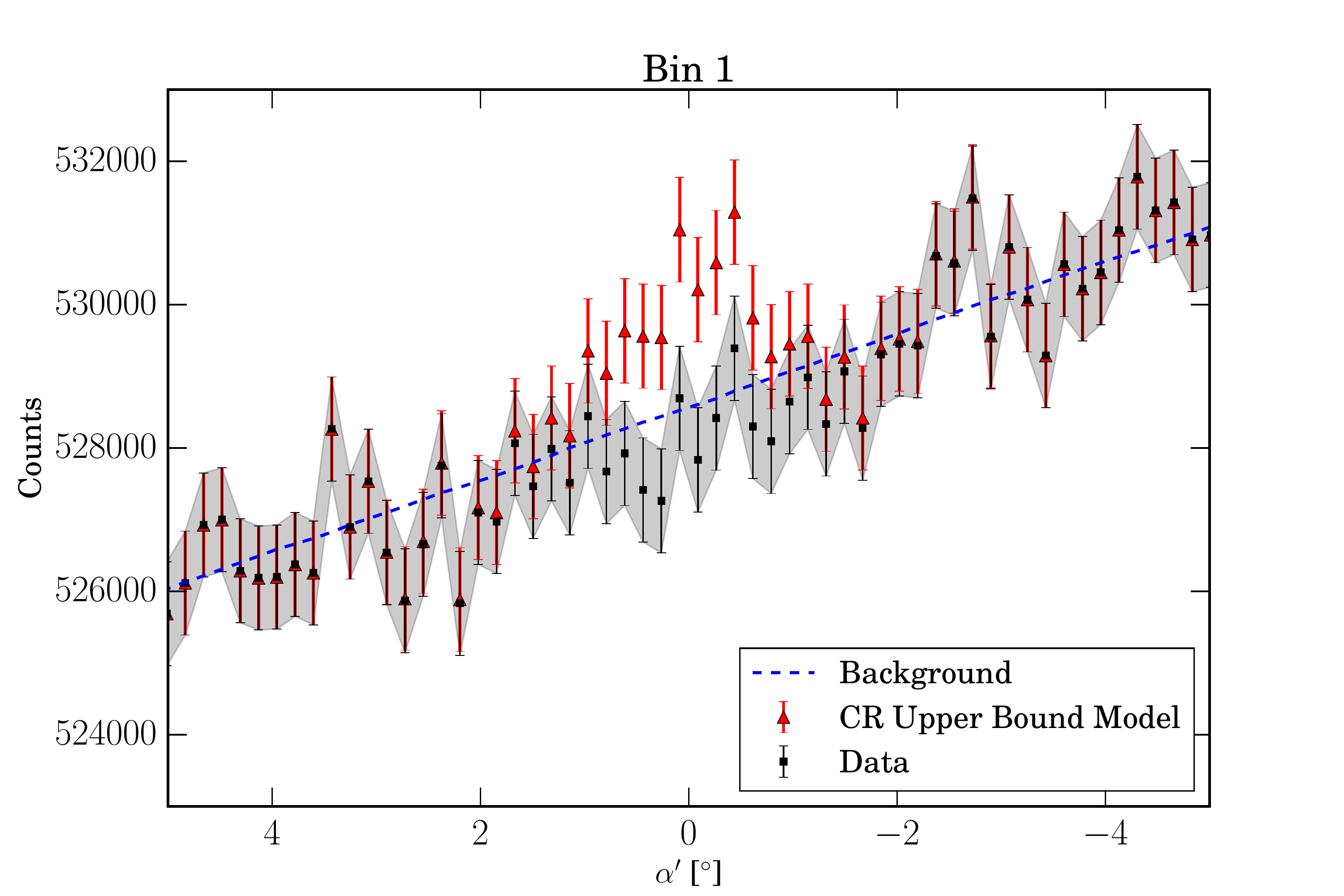} &
\includegraphics[width=0.55\textwidth]{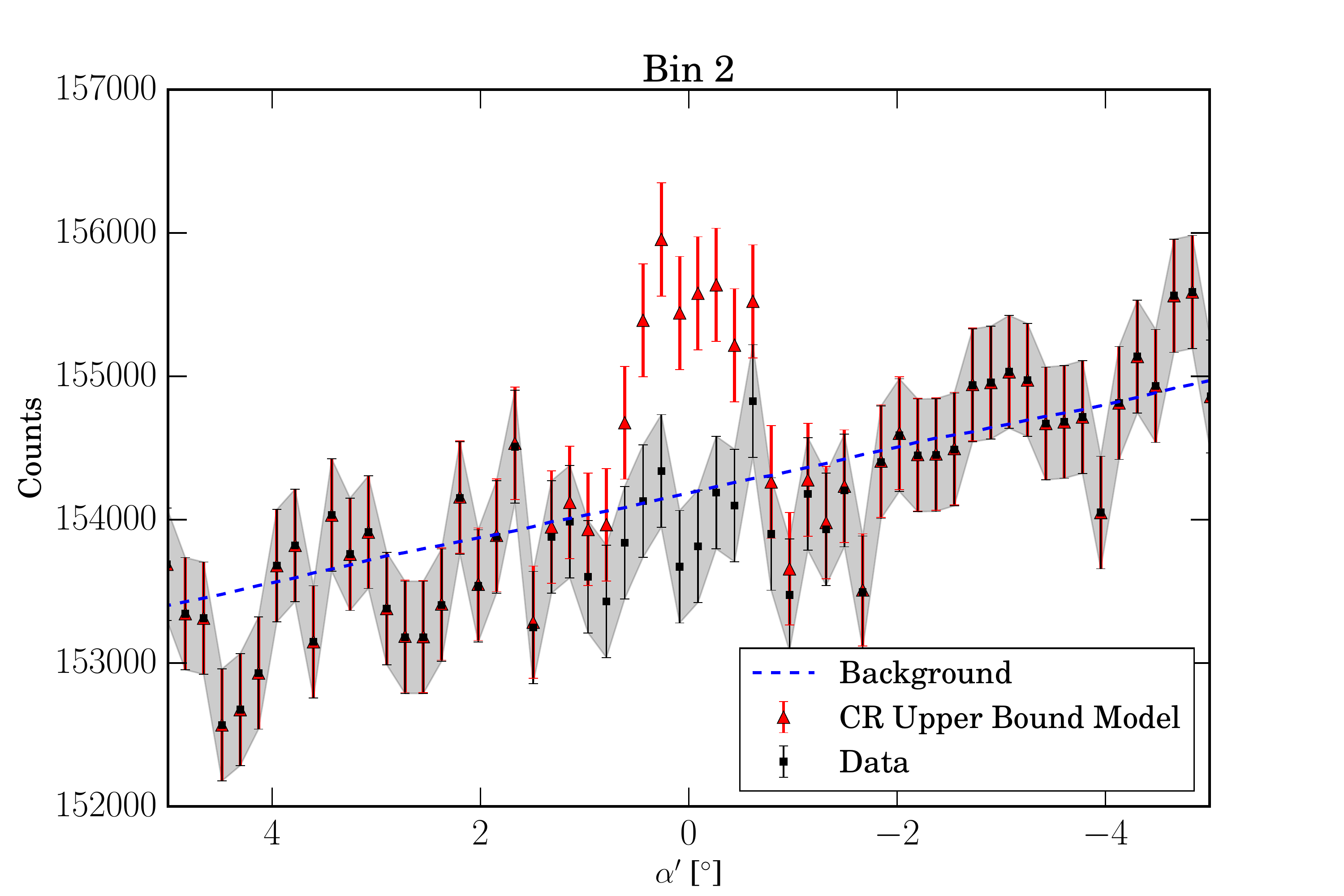} \\
\includegraphics[width=0.55\textwidth]{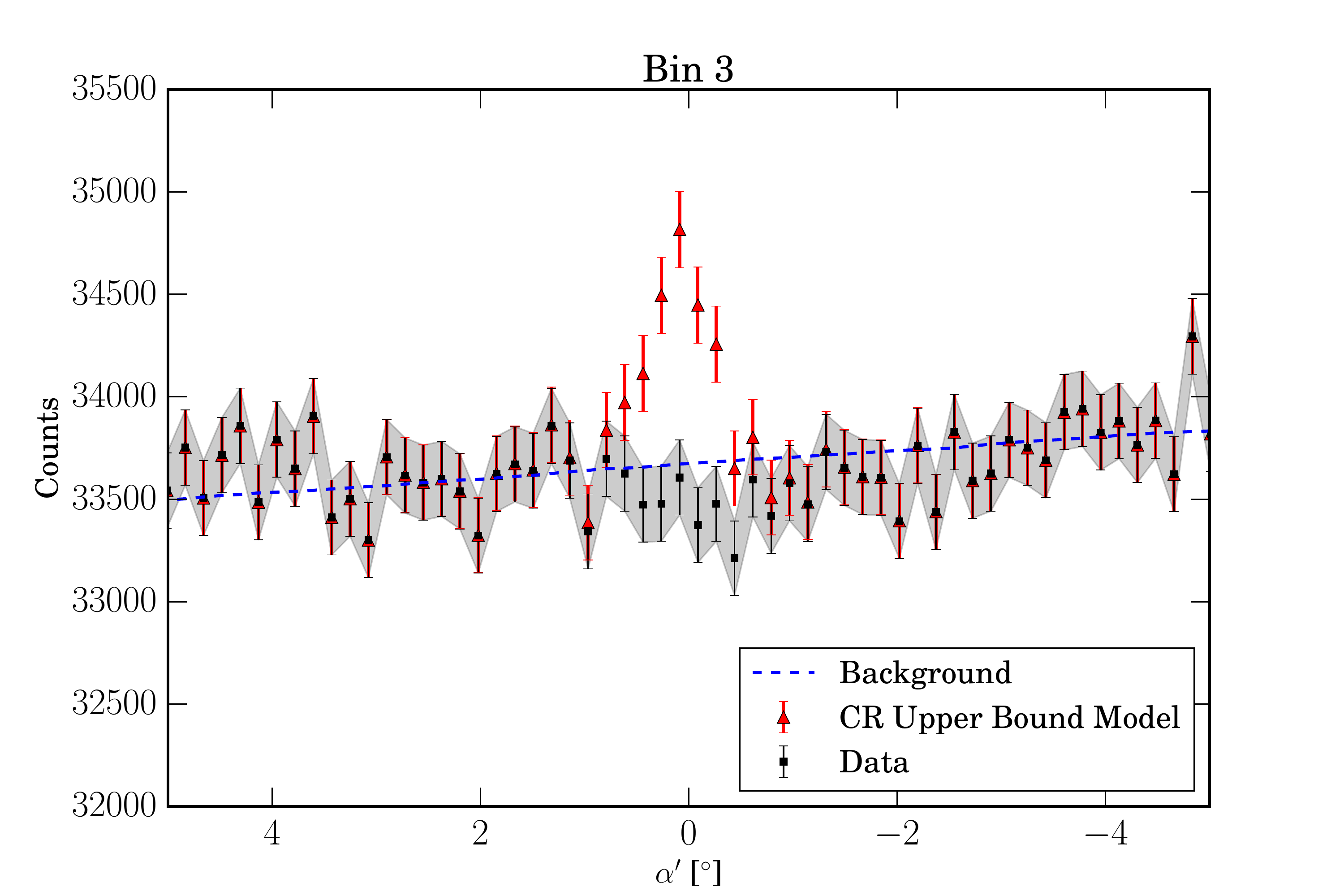} &
\includegraphics[width=0.55\textwidth]{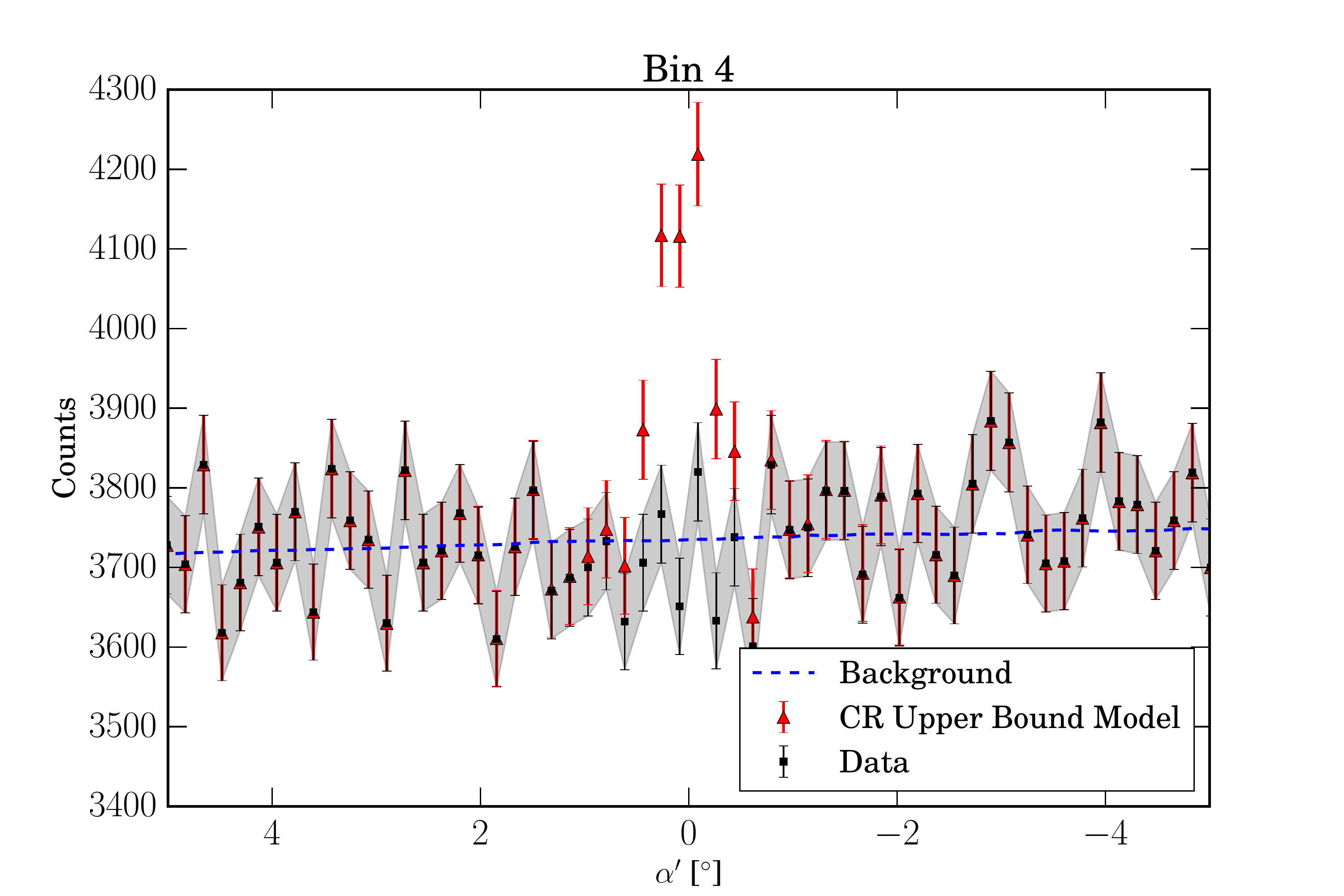} \\
\end{tabular}}
\caption{The data and background counts, after gamma-hadron separation, projected on the right ascension axis centered on the Sun at $\alpha ' = 0 $, for the examples of Bins 1--4. The total counts at every $\alpha '$ are summed in the declination range spanning the RoI: $-R_{\text{RoI}} \leq \delta' \leq R_{\text{RoI}}$ (see Table \ref{tab:bins}). For each bin (as labeled), the expected number of events from the CR upper bound model are also shown. The errors on the model and the data are given in each case by the square root of the number of events.}
\label{fig:databkg}
\end{figure*}

Figure \ref{fig:counts} (right panel) shows the excess points as a fraction of the approximate uncertainty (square root of expected background) for each bin for the total dataset. The full analysis does not use the aforementioned approximation. We observe that, despite fluctuations in individual bins, the total net excess is consistent with zero, and is within the fluctuations seen from the off-Sun regions.  Figure \ref{fig:databkg} shows the observed data, the expected background and the expected maximum signal for the first four bins. Given that no clear excess is seen, we proceed to convert the observed gamma-ray counts to upper limits on the differential flux in combined energy bins following the method in Ref. \cite{2017ApJ...842...85A}.


\subsection{\label{sub:limits}Upper Limits in Differential Energy Bins}
We compute differential upper limits in half-decade bins in $\log(E)$, where the width of each interval is comparable to the energy resolution of the detector. We choose an energy scale binned in intervals of $0.5$ in log space, centered at 0.88 TeV, 2.78 TeV, 8.81 TeV, 27.1 TeV and 88.1 TeV (see Table \ref{tab:res}). The flux injection in one bin largely remains there after considering detection effects \cite{2017ApJ...843...39A}. The observed data from the energy proxy bins is combined and rebinned into energy intervals following the weighting scheme discussed in Ref. \cite{2017ApJ...842...85A}. We then compare the observed counts from the solar disk with the expected counts from a Sun-like source of a known spectrum convolved with the HAWC detector response using the likelihood framework described in Ref. \cite{2015arXiv150807479Y}. The procedure defines a source model in a narrow energy range for each bin and calculates the expected counts after taking into account the point spread function and detector response at the location of the source. The point spread function of the detector is approximated by a double gaussian \cite{2017ApJ...843...39A} and the expected number of counts are used from a uniform disk of radius $0.25^\circ$. The observed number of counts, $N_{\gamma}$ from a source in a particular energy bin are proportional to the differential flux $F_{\gamma}$ at the bin energy,
\begin{equation}
\frac{F_{\gamma}}{F_0} = \frac{N_{\gamma}}{N_{\gamma 0}}.
\label{eq:exp}
\end{equation}
Here $F_{0}$ and $N_{\gamma 0}$ respectively denote the expected flux and counts from the nominal source model. For the expected flux $F_{0}$ in a differential energy bin, we assume a simple power law with a spectral index of $2.7$, following the cosmic-ray upper bound. This assumption does not appreciably affect the final result as the data is compared to a constant value of expected flux in each bin. We tested the effect of varying the assumed spectral index to 2.1 and 2.3 and notice no significant difference in the reported upper limits in each energy bin.
We compute the expected counts $N_{\gamma 0}$ from simulated Suns located at the median position in each month. The monthly counts are then summed to get the total expected counts for the duration of the data. The expected and observed counts, $N_{\gamma 0}$ and $N_{\gamma}$, are used to calculate a likelihood function, $\mathcal{L_{\odot}}$, for the source model, and $\mathcal{L_{\text{Bkg}}}$ for the null hypothesis (background only). The likelihood-ratio defines the Test Statistic (TS),
\begin{equation}
TS = 2 \ln{\frac{ \mathcal{L_{\odot}}}{\mathcal{L_{\text{Bkg}}}}} .
\label{eq:LL}
\end{equation}
Finally, a likelihood fit to the source model defined above gives the observed differential flux $F_{\gamma}$; its uncertainty --- corresponding to a change in the Test Statistic (TS) of 2.71 --- is used to construct the $95\%$ upper limits. Table \ref{tab:res} summarizes the resulting $95\%$ C.L. limits.   

\begin{table}[t!]
\centering
\begin{ruledtabular}
\begin{tabular}{cc}
\textbf{Energy [TeV]} &\textbf{$E^{2}$ dN/dE [TeV cm$^{-2}$s$^{-1}$] [95\%]}\\
\hline
\rule{0pt}{3ex}
0.5 -- 1.6 & $2.2\times10^{-12}$ \\
1.6 -- 5.0 & $8.8\times10^{-13}$\\
\phantom{k}5.0 -- 15.7 & $2.8\times 10^{-13}$\\
15.7 -- 50.0 &$8.1\times 10^{-14}$\\
\phantom{kkk} $>$ 50.0  &$6.3\times 10^{-14}$\\
\end{tabular}
\end{ruledtabular}
\caption{The energy range and the corresponding $95\%$ C.L. upper limits of solar gamma-ray flux obtained from 2014--2017 HAWC data in this work. }
\label{tab:res}
\end{table}

       
\section{\label{sec:sim}HAWC Sensitivity}    
\subsection{\label{sec:off}Sensitivity from Off-Sun Regions}
To check our limits, we compute the expected sensitivity. The sensitivity refers to the median upper limit that would be obtained from analyzing an ensemble of background-only datasets in the absence of an excess signal \cite{1998PhRvD..57.3873F}. 
Utilizing the off-source ``fake Sun'' regions~(Sec.~\ref{subsubsec:results}), we obtain a band of upper limits that we can compare with the limits obtained from the actual Sun position. The HAWC sensitivity is the central $90\%$ band of upper limits obtained from analyzing the ``fake Sun'' regions. If the observed flux from a ``fake Sun'' region is negative due to an under-fluctuation or shadow contamination (on-Sun RoI only), the maximum likelihood is scaled to match the pure background hypothesis. This ensures that the computed limits are physical. We notice no systematic discrepancy between the off-source band and the on-source limits, further ensuring that our results are consistent with a non-detection. The width of the sensitivity band is also an illustration of the Poisson fluctuations in the expected background, as well as, the systematic errors in the analysis, which are discussed in detail under systematic uncertainties in section \ref{sec:syst}.

As an additional cross-check, we also performed a maximum likelihood analysis in which several uniform disks of $0.25^\circ$ radius along the Sun's trajectory in celestial coordinates were fit to an extended source model with a simple power law spectrum. The spectral index was fixed to $2.11$, $2.3$, or $2.7$ and the normalization at 1 TeV was the free parameter of the likelihood fit. We again found the results consistent with the null hypothesis. 

\begin{figure*}[hbt!]
\centering
 \makebox[0.4\width][c]{
 \begin{tabular}{@{}ccc@{}}
 \includegraphics[width=0.37\textwidth]{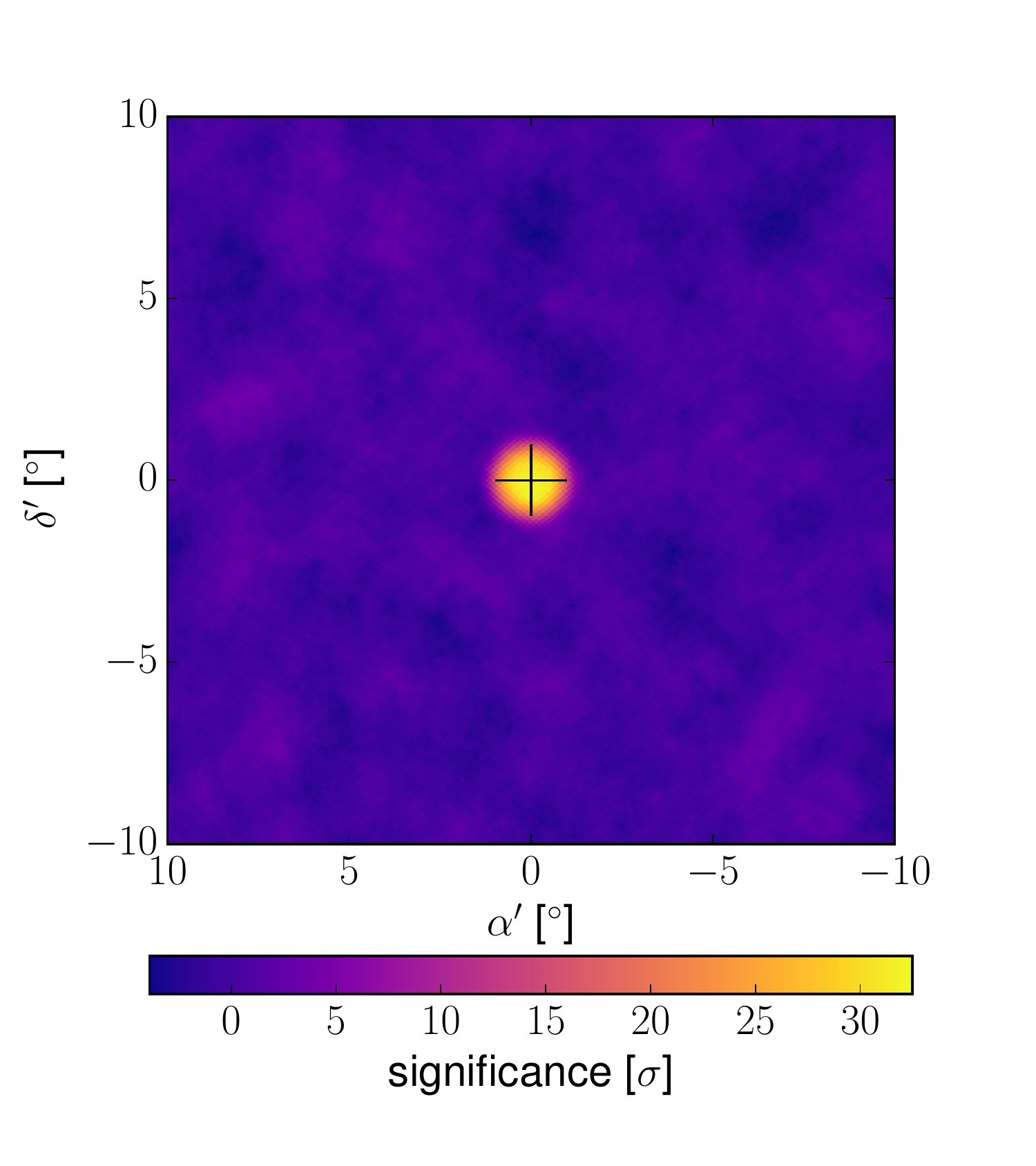} &
 \includegraphics[width=0.37\textwidth]{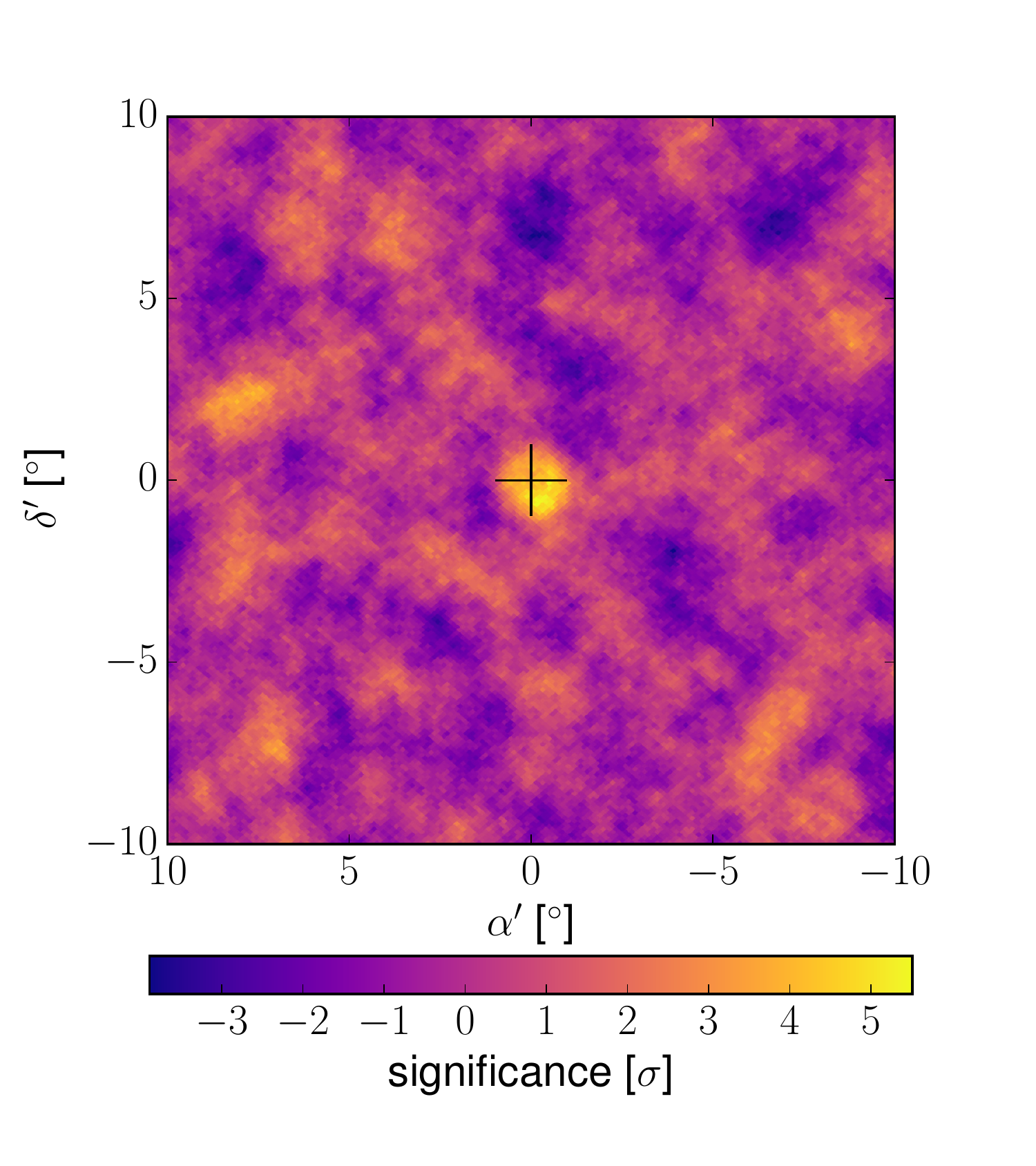} &
 \includegraphics[width=0.37\textwidth]{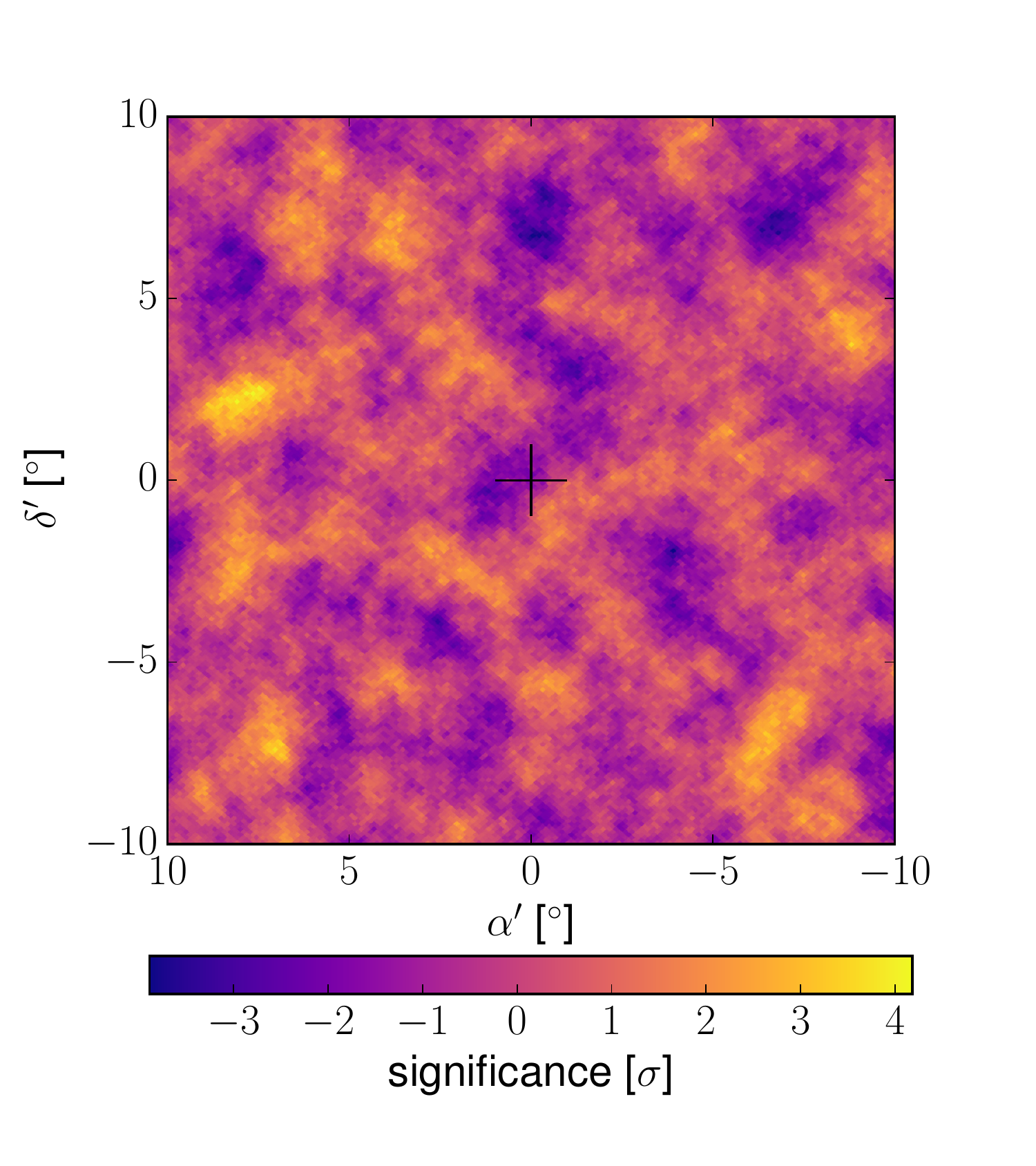}
 \end{tabular}
 }
\caption{Simulated post-cuts maps with injected Sun-like sources in Bin 4. The injected spectra are \textbf{Left:} $E^{-2.1}$ extrapolation of the Fermi-LAT measurements (see text), \textbf{Center:} $E^{-2.3}$, and \textbf{Right:} $E^{-2.7}$.  The Sun can be detected at $>20 \sigma$ for $E^{-2.1}$, and with lower significance ($5.5 \sigma$) at $E^{-2.3}$, but not for $E^{-2.7}$.}
 \label{fig:sim}
 \end{figure*}


\subsection{Validation and Data Challenge} 
Section \ref{sec:maps} describes the simulated signal from the CR upper bound as the fiducial flux for comparing HAWC results to a model. In contrast, the data challenge described in this section uses various extrapolations of the Fermi-LAT measurement. It serves as a way of validating our computed limits and testing the sensitivity to the projected solar minimum flux for a future analysis. We inject a source of a known spectrum at the position of the Sun into the HAWC data. We then perform the full analysis, accounting for the HAWC point spread function and reproduce the gamma-ray maps. We test three different hard spectra for the simulated source:\\
1. ($1.3\times10^{-11}$ TeV$^{-1}$ cm$^{-2}$ s$^{-1}$)(E/TeV)$^{-2.1}$ . This corresponds to the extrapolated spectrum at 1 TeV from the spectrum measured by Fermi-LAT during the solar minimum.\\
2. $10\%$ of the injected flux in Case 1.\\
3. $1\%$ of the injected flux in Case 1.\\
 
We observe that the Sun is visible as a bright source at over  $5\sigma$ in all bins for Case 1. For Case 2, the decreased flux is not detectable in the low energy bins but can be observed at $> 4 \sigma$ in Bin 6 and above. For Case 3, when the injected flux falls to $1\%$ of the maximum, there is no gamma-ray signal in any bin. 

We also test the effect of assuming a different spectral index for the simulated Sun with the same normalization. Figure \ref{fig:sim} shows simulated Suns with three different spectra: $E^{-2.1}$, $E^{-2.3}$ and $E^{-2.7}$ extrapolations of the Fermi-LAT data pivoted at 100 GeV during the solar minimum. The normalization at the pivot energy is $1.7\times10^{-9}$ TeV$^{-1}$ cm$^{-2}$ s$^{-1}$. The significance of the observation decreases as the spectrum becomes softer from $2.11$ to $2.3$. A source with an $E^{-2.6}$ or softer spectrum normalized to $1.7\times10^{-9}$ TeV$^{-1}$ cm$^{-2}$ s$^{-1}$ at 100 GeV is not observable with HAWC. The same simulation with a scaled down normalization to match the 2014--2017 Fermi-LAT data, yields no significant detection for a spectral index of $2.4$.  

These results are consistent with our upper limit calculations. The HAWC $95\%$ upper limit at $1$ TeV is $2.8 \times 10^{-12}$ TeV$^{-1}$ cm$^{-2}$ s $^{-1}$, which is about $10\%$ of the upper bound on the flux from cosmic-ray interactions. The tests with simulated sources show that HAWC would be able to detect an excess at high significance even if it is as low as $\sim 10^{-12}$ TeV$^{-1}$ cm$^{-2}$ s$^{-1}$ above 1 TeV. The results also constrain a naive 2.1--2.3 extrapolation of the Fermi-LAT spectrum observed during 2014--2017. 

\subsection{\label{sec:syst}Systematic Uncertainties}
 

\begin{figure*}[htb!]
\centering
\includegraphics[width=0.99\textwidth]{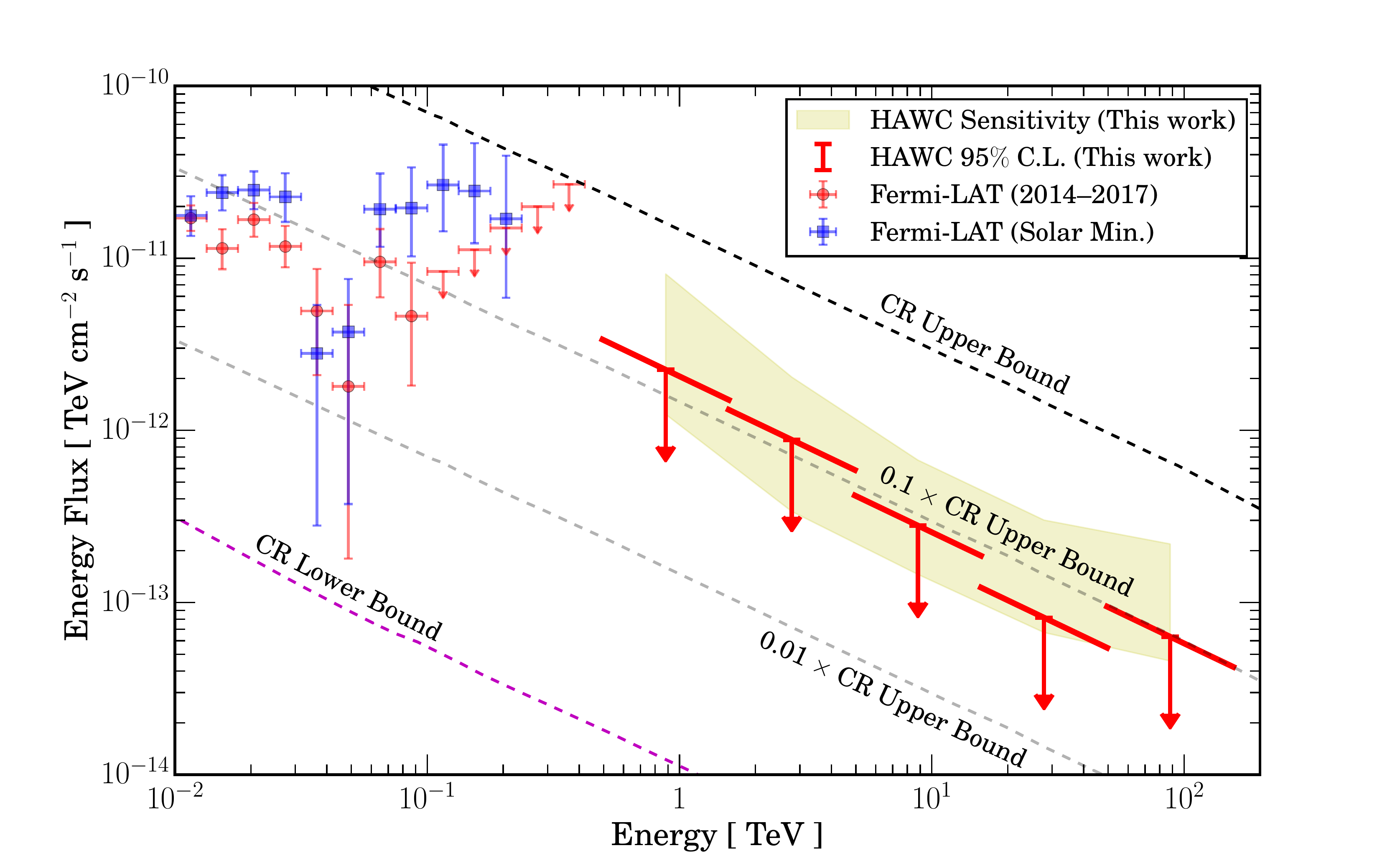}
\caption{ HAWC 95$\%$ upper limits from 3 years of data and the expected sensitivity, obtained in this work.  The Fermi-LAT observation during solar minimum~\cite{2018arXiv180305436L,2018arXiv180406846T} and during the same time period as this work~(2014--2017) are shown in blue squares and red circles. 
The dashed lines show the theoretical maximum and minimum fluxes produced by hadronic interactions \cite{2018arXiv180406846T,2016arXiv161202420Z}.} 
\label{fig:limits}
\end{figure*}


The ability to find a relative excess above the background is limited by the statistical uncertainties discussed above. Converting the measured excess to limits on the flux, involves a number of overall systematic effects that are presented below. 
\subsubsection{Source Declination and Analysis Tools}
This analysis is subject to the uncertainties of using techniques that are not fully tuned to the rapidly changing position of the Sun in the sky. In this work, we use the average values of the Sun's declination angle in the detector simulations and the computation of the flux, potentially smearing the signal flux by a factor of 2. Moreover, the binning of events by the fractional PMTs hit during an air shower is only a crude approximation of energy, which itself is correlated with a source's declination angle and is subject to large fluctuations (see Refs. \cite{2017ApJ...843...39A, 2017ApJ...843...39A, Marinelli:2017vzu}). The median energy of events in an analysis bin is a function of the source declination; the changing declination also contributes to the broadening of the energy resolution histogram. A $10\%$ change in the energy scale can affect the measured flux by $12\%$ in a differential energy bin. 
\subsubsection{Gamma-Hadron Separation and the Sun Shadow}
Another limitation of this analysis comes from an incomplete understanding of the shadow on short time scales. In low energy bins, the shadow is weak because the cosmic-ray flux is smeared out by the Sun's magnetic field \cite{PhysRevLett.120.031101}. A potential gamma-ray excess would be easier to detect over a weak shadow. However, the gamma-hadron separation efficiency is also the lowest in these bins, which limits the sensitivity. At higher energies, the gamma-hadron cuts are more efficient, but the increased strength of the cosmic-ray shadow makes the overall measurement more challenging. The gamma-hadron cuts also carry the effect of averaging the Sun's declination. The gamma-ray efficiencies $\epsilon_{\gamma}$ are obtained from simulations optimized by studying fixed sources in the sky \cite{2017ApJ...843...39A}. These issues can be addressed in a future analysis with an accurate modeling of the shadow with solar magnetic fields. The shadow, and the energy and zenith dependent limitations described above, combined with the nominal values used for the gamma-hadron cuts, make this a very conservative analysis. We will revisit these aspects with more data during the solar minimum.  
\subsubsection{Detector Performance}
All measurements are subject to the uncertainties inherent in the charge resolution and the quantum efficiency of the PMTs. These detector effects have been studied in detail and their impact on the measured photon flux and the energy resolution has been quantified \cite{2017ApJ...843...39A}. The pointing accuracy, angular resolution and the energy scale of the instrument have also been studied in observations of the Crab Nebula \cite{2017ApJ...843...39A} in gamma rays, and the Moon shadow in cosmic rays \cite{2014ApJ...796..108A,2017arXiv171000890H}. The overall effect of these uncertainties on the photon flux is $\pm 50\%$ and is contained within the sensitivity band and width of each energy bin in $\Delta$log$_{10}$(E/TeV).

\section{ \label{conc} Results and Discussion}
\subsection{HAWC 95\% Limits}

Figure \ref{fig:limits} summarizes the HAWC results in the context of past measurements and models of gamma-ray production from cosmic rays near the Sun. With a sensitivity less than $10\%$ of the Crab flux, the HAWC upper limits are already below the maximum expectation from cosmic-ray interactions and constrain fluxes to $\lesssim 10\%$ of the upper bound. The limits are also above the theoretical minimum flux from the solar limb as calculated in Ref. \cite{2016arXiv161202420Z} (see also Ref. \cite{Gao:2017bfv} for corroboration by an independent simulation). 

The current results are based on data collected by HAWC outside the solar minimum and the limits set here strongly constrain possible extensions of the spectrum measured with the Fermi Gamma-ray Space Telescope. During the three-year time period considered here, the spectrum measured with Fermi-LAT appears to be falling above 65 GeV. If that trend continued, the HAWC upper limits on the flux would not yet be sensitive enough to constrain the spectrum. However, the HAWC data do limit the appearance of any new, highest-energy component of the spectrum, such as if the drop near 80 GeV in the Fermi-LAT data were the beginning of a dip and rise like that seen near 40 GeV.  

Moreover, since the HAWC measurements were performed prior to the upcoming solar minimum, a null result also limits the cosmic-ray induced foreground for dark matter searches from the Sun. Gamma rays from cosmic-ray interactions in the Sun constitute the main foreground for solar dark matter searches. The low gamma-ray flux in the current period of observation leads to better constraints on dark matter than when the foreground flux is high during the solar minimum, as the dark matter flux should not change with the solar cycle. In a companion paper, we study the implications for dark matter searches from the Sun \cite{DMPaper}.

\subsection{Implications and Future Work}
We have presented HAWC's ability to perform a challenging measurement in close vicinity of the Sun. We rule out TeV gamma rays from the Sun up to a flux of a few $10^{-12}$ TeV$^{-1}$ cm$^{-2}$s$^{-1}$ at 1 TeV at the 95\% C.L. in periods of high solar activity. These are the strongest constraints on TeV gamma rays from the Sun to date. The results demonstrate that HAWC can probe a physically relevant parameter space that was not experimentally accessible until now. The HAWC sensitivity could be improved by additional exposure, improved analysis techniques, and ongoing detector upgrades \cite{2017arXiv170804032J}. The sensitivity laid out in this paper is crucial for an analysis in the next solar minimum, which began in early 2018. 

The prospects for a detection are significantly enhanced in the minimum of cycle 25. First, according to the Fermi-LAT data from the last solar minimum, the flux is expected to be much higher and the spectrum much harder, extending beyond 400 GeV before the exposure of Fermi-LAT becomes insufficient \cite{2018arXiv180305436L}. Second, because the flux is expected to be much brighter during a short period of 1--2 years, this gives a better signal-to-noise ratio. We expect to report new results within 2 years. Whether HAWC makes a measurement or sets an upper limit, this will powerfully constrain the mechanisms behind the solar-disk gamma-ray emission. At high enough energy, a cutoff in the gamma-ray spectrum is expected as solar magnetic fields will not be able to reflect the incoming cosmic rays.  Observation of such a cutoff will thus help identify the magnetic environment responsible for cosmic-ray mirroring, which is essential for understanding the underlying gamma-ray emission mechanism.

If the gamma-ray spectrum measured in the last solar minimum continues into the TeV range without a cutoff during the minimum of cycle 25, HAWC will be able to detect it with high significance. Long term monitoring from HAWC on the Sun shadow will also allow us to understand the effect of solar magnetic fields on TeV cosmic rays \cite{Amenomori:2013own, 2018PhRvL.120c1101A, Amenomori:2018xip}. Finally, hadronic cosmic rays interacting in the solar atmosphere will also produce solar atmospheric neutrinos \cite{2017PhRvD..96j3006N,2017JCAP...07..024A,Edsjo:2017kjk}, which are being searched for in IceCube~\cite{icecubesolar}. Gamma-ray constraints and observations from HAWC are crucial for understanding how cosmic rays interact with the solar atmosphere, and are also important for interpretations of solar atmospheric neutrino searches in neutrino telescopes~\cite{2017EPJC...77..146A,Adrian-Martinez:2016fdl}, and dark matter searches from the Sun \cite{1995NuPhS..43..265E,2004PhRvD..69l3505L,2009arXiv0908.0899F,2009PhRvD..79j3532P,2011JCAP...09..029R,Danninger:2014xza,2017PhRvD..95l3016L, Arina:2017sng, Smolinsky:2017fvb}.

\vspace{-0.5cm}
\section*{Acknowledgments}
\vspace{-0.5cm}
We acknowledge the support from: the US National Science Foundation (NSF) the US Department of Energy Office of High-Energy Physics;  the Laboratory Directed Research and Development (LDRD) program of Los Alamos National Laboratory; Consejo Nacional de Ciencia y Tecnolog\'{\i}a (CONACyT), M{\'e}xico (grants 271051, 232656, 260378, 179588, 239762, 254964, 271737, 258865, 243290, 132197, 281653 (C{\'a}tedras 873, 1563), Laboratorio Nacional HAWC de rayos gamma; L'OREAL Fellowship for Women in Science 2014; Red HAWC, M{\'e}xico; DGAPA-UNAM (grants IG100317, IN111315, IN111716-3, IA102715, 109916, IA102917, IN112218); VIEP-BUAP; PIFI 2012, 2013, PROFOCIE 2014, 2015; the University of Wisconsin Alumni Research Foundation; the Institute of Geophysics, Planetary Physics, and Signatures at Los Alamos National Laboratory; Polish Science Centre grant DEC-2014/13/B/ST9/945, DEC-2017/27/B/ST9/02272; Coordinaci{\'o}n de la Investigaci{\'o}n Cient\'{\i}fica de la Universidad Michoacana; Royal Society - Newton Advanced Fellowship 180385. Thanks to Scott Delay, Luciano D\'{\i}az and Eduardo Murrieta for technical support.

JFB is supported by (and BZ is partially supported by) NSF grant PHY-1714479. RKL is supported by the Office of High Energy Physics of the U.S. Department of Energy under Grant No.\;DE-SC00012567 and DE-SC0013999. KCYN is supported by Croucher Fellowship and Benoziyo Fellowship.
TL, AHGP and BZ are supported in part by NASA Grant No.\;80NSSC17K0754.

\bibliography{bib}{}

\begin{thebibliography}{84}%
\makeatletter
\providecommand \@ifxundefined [1]{%
 \@ifx{#1\undefined}
}%
\providecommand \@ifnum [1]{%
 \ifnum #1\expandafter \@firstoftwo
 \else \expandafter \@secondoftwo
 \fi
}%
\providecommand \@ifx [1]{%
 \ifx #1\expandafter \@firstoftwo
 \else \expandafter \@secondoftwo
 \fi
}%
\providecommand \natexlab [1]{#1}%
\providecommand \enquote  [1]{``#1''}%
\providecommand \bibnamefont  [1]{#1}%
\providecommand \bibfnamefont [1]{#1}%
\providecommand \citenamefont [1]{#1}%
\providecommand \href@noop [0]{\@secondoftwo}%
\providecommand \href [0]{\begingroup \@sanitize@url \@href}%
\providecommand \@href[1]{\@@startlink{#1}\@@href}%
\providecommand \@@href[1]{\endgroup#1\@@endlink}%
\providecommand \@sanitize@url [0]{\catcode `\\12\catcode `\$12\catcode
  `\&12\catcode `\#12\catcode `\^12\catcode `\_12\catcode `\%12\relax}%
\providecommand \@@startlink[1]{}%
\providecommand \@@endlink[0]{}%
\providecommand \url  [0]{\begingroup\@sanitize@url \@url }%
\providecommand \@url [1]{\endgroup\@href {#1}{\urlprefix }}%
\providecommand \urlprefix  [0]{URL }%
\providecommand \Eprint [0]{\href }%
\providecommand \doibase [0]{http://dx.doi.org/}%
\providecommand \selectlanguage [0]{\@gobble}%
\providecommand \bibinfo  [0]{\@secondoftwo}%
\providecommand \bibfield  [0]{\@secondoftwo}%
\providecommand \translation [1]{[#1]}%
\providecommand \BibitemOpen [0]{}%
\providecommand \bibitemStop [0]{}%
\providecommand \bibitemNoStop [0]{.\EOS\space}%
\providecommand \EOS [0]{\spacefactor3000\relax}%
\providecommand \BibitemShut  [1]{\csname bibitem#1\endcsname}%
\let\auto@bib@innerbib\@empty
\bibitem [{\citenamefont {{Ackermann}}\ \emph {et~al.}(2014)\citenamefont
  {{Ackermann}} \emph {et~al.}}]{2014ApJ...787...15A}%
  \BibitemOpen
  \bibfield  {author} {\bibinfo {author} {\bibfnamefont {M.}~\bibnamefont
  {{Ackermann}}} \emph {et~al.},\ }\href {\doibase 10.1088/0004-637X/787/1/15}
  {\bibfield  {journal} {\bibinfo  {journal} {\apj}\ }\textbf {\bibinfo
  {volume} {787}},\ \bibinfo {eid} {15} (\bibinfo {year} {2014})},\ \Eprint
  {http://arxiv.org/abs/1304.3749} {arXiv:1304.3749 [astro-ph.HE]} \BibitemShut
  {NoStop}%
\bibitem [{\citenamefont {Kafexhiu}\ \emph {et~al.}(2018)\citenamefont
  {Kafexhiu}, \citenamefont {Romoli}, \citenamefont {Taylor},\ and\
  \citenamefont {Aharonian}}]{Kafexhiu:2018wmh}%
  \BibitemOpen
  \bibfield  {author} {\bibinfo {author} {\bibfnamefont {E.}~\bibnamefont
  {Kafexhiu}}, \bibinfo {author} {\bibfnamefont {C.}~\bibnamefont {Romoli}},
  \bibinfo {author} {\bibfnamefont {A.~M.}\ \bibnamefont {Taylor}}, \ and\
  \bibinfo {author} {\bibfnamefont {F.}~\bibnamefont {Aharonian}},\ }\href@noop
  {} {\bibfield  {journal} {\bibinfo  {journal} {ArXiv e-prints}\ } (\bibinfo
  {year} {2018})},\ \Eprint {http://arxiv.org/abs/1803.02635} {arXiv:1803.02635
  [astro-ph.HE]} \BibitemShut {NoStop}%
\bibitem [{\citenamefont {Strong}(1991)}]{Strong327}%
  \BibitemOpen
  \bibfield  {author} {\bibinfo {author} {\bibfnamefont {K.}~\bibnamefont
  {Strong}},\ }\href {\doibase 10.1098/rsta.1991.0084} {\bibfield  {journal}
  {\bibinfo  {journal} {Philosophical Transactions of the Royal Society of
  London A: Mathematical, Physical and Engineering Sciences}\ }\textbf
  {\bibinfo {volume} {336}},\ \bibinfo {pages} {327} (\bibinfo {year}
  {1991})}\BibitemShut {NoStop}%
\bibitem [{\citenamefont {Lin}\ \emph {et~al.}(2002)\citenamefont {Lin} \emph
  {et~al.}}]{Lin2002}%
  \BibitemOpen
  \bibfield  {author} {\bibinfo {author} {\bibfnamefont {R.}~\bibnamefont
  {Lin}} \emph {et~al.},\ }\href {\doibase 10.1023/A:1022428818870} {\bibfield
  {journal} {\bibinfo  {journal} {Solar Physics}\ }\textbf {\bibinfo {volume}
  {210}},\ \bibinfo {pages} {3} (\bibinfo {year} {2002})}\BibitemShut {NoStop}%
\bibitem [{\citenamefont {Pesce-Rollins}\ \emph {et~al.}(2015)\citenamefont
  {Pesce-Rollins}, \citenamefont {Omodei}, \citenamefont {Petrosian},
  \citenamefont {Liu}, \citenamefont {da~Costa}, \citenamefont {Allafort},\
  and\ \citenamefont {Chen}}]{Pesce-Rollins:2015hpa}%
  \BibitemOpen
  \bibfield  {author} {\bibinfo {author} {\bibfnamefont {M.}~\bibnamefont
  {Pesce-Rollins}}, \bibinfo {author} {\bibfnamefont {N.}~\bibnamefont
  {Omodei}}, \bibinfo {author} {\bibfnamefont {V.}~\bibnamefont {Petrosian}},
  \bibinfo {author} {\bibfnamefont {W.}~\bibnamefont {Liu}}, \bibinfo {author}
  {\bibfnamefont {F.~R.}\ \bibnamefont {da~Costa}}, \bibinfo {author}
  {\bibfnamefont {A.}~\bibnamefont {Allafort}}, \ and\ \bibinfo {author}
  {\bibfnamefont {Q.}~\bibnamefont {Chen}},\ }\href {\doibase
  10.1088/2041-8205/805/2/L15} {\bibfield  {journal} {\bibinfo  {journal}
  {Astrophys. J.}\ }\textbf {\bibinfo {volume} {805}},\ \bibinfo {pages} {L15}
  (\bibinfo {year} {2015})},\ \Eprint {http://arxiv.org/abs/1505.03480}
  {arXiv:1505.03480 [astro-ph.SR]} \BibitemShut {NoStop}%
\bibitem [{\citenamefont {Share}\ \emph {et~al.}(2017)\citenamefont {Share},
  \citenamefont {Murphy}, \citenamefont {Tolbert}, \citenamefont {Dennis},
  \citenamefont {White}, \citenamefont {Schwartz},\ and\ \citenamefont
  {Tylka}}]{Share:2017tgw}%
  \BibitemOpen
  \bibfield  {author} {\bibinfo {author} {\bibfnamefont {G.~H.}\ \bibnamefont
  {Share}}, \bibinfo {author} {\bibfnamefont {R.~J.}\ \bibnamefont {Murphy}},
  \bibinfo {author} {\bibfnamefont {A.~K.}\ \bibnamefont {Tolbert}}, \bibinfo
  {author} {\bibfnamefont {B.~R.}\ \bibnamefont {Dennis}}, \bibinfo {author}
  {\bibfnamefont {S.~M.}\ \bibnamefont {White}}, \bibinfo {author}
  {\bibfnamefont {R.~A.}\ \bibnamefont {Schwartz}}, \ and\ \bibinfo {author}
  {\bibfnamefont {A.~J.}\ \bibnamefont {Tylka}},\ }\href@noop {} {\bibfield
  {journal} {\bibinfo  {journal} {ArXiv e-prints}\ } (\bibinfo {year}
  {2017})},\ \Eprint {http://arxiv.org/abs/1711.01511} {arXiv:1711.01511
  [astro-ph.SR]} \BibitemShut {NoStop}%
\bibitem [{\citenamefont {{Orlando}}\ and\ \citenamefont
  {{Strong}}(2008{\natexlab{a}})}]{2008A&A...480..847O}%
  \BibitemOpen
  \bibfield  {author} {\bibinfo {author} {\bibfnamefont {E.}~\bibnamefont
  {{Orlando}}}\ and\ \bibinfo {author} {\bibfnamefont {A.~W.}\ \bibnamefont
  {{Strong}}},\ }\href {\doibase 10.1051/0004-6361:20078817} {\bibfield
  {journal} {\bibinfo  {journal} {Astronomy and Astrophysics}\ }\textbf
  {\bibinfo {volume} {480}},\ \bibinfo {pages} {847} (\bibinfo {year}
  {2008}{\natexlab{a}})},\ \Eprint {http://arxiv.org/abs/0801.2178}
  {arXiv:0801.2178} \BibitemShut {NoStop}%
\bibitem [{\citenamefont {Abdo}\ \emph {et~al.}(2011)\citenamefont {Abdo} \emph
  {et~al.}}]{0004-637X-734-2-116}%
  \BibitemOpen
  \bibfield  {author} {\bibinfo {author} {\bibfnamefont {A.}~\bibnamefont
  {Abdo}} \emph {et~al.},\ }\href
  {http://stacks.iop.org/0004-637X/734/i=2/a=116} {\bibfield  {journal}
  {\bibinfo  {journal} {Astrophys. J.}\ }\textbf {\bibinfo {volume} {734}},\
  \bibinfo {pages} {116} (\bibinfo {year} {2011})}\BibitemShut {NoStop}%
\bibitem [{\citenamefont {{Orlando}}\ and\ \citenamefont
  {{Strong}}(2008{\natexlab{b}})}]{2008ICRC....2..505O}%
  \BibitemOpen
  \bibfield  {author} {\bibinfo {author} {\bibfnamefont {E.}~\bibnamefont
  {{Orlando}}}\ and\ \bibinfo {author} {\bibfnamefont {A.~W.}\ \bibnamefont
  {{Strong}}},\ }\href@noop {} {\bibfield  {journal} {\bibinfo  {journal}
  {International Cosmic Ray Conference}\ }\textbf {\bibinfo {volume} {2}},\
  \bibinfo {pages} {505} (\bibinfo {year} {2008}{\natexlab{b}})},\ \Eprint
  {http://arxiv.org/abs/0709.3841} {arXiv:0709.3841} \BibitemShut {NoStop}%
\bibitem [{\citenamefont {Orlando}\ and\ \citenamefont
  {Strong}(2007)}]{Orlando:2006zs}%
  \BibitemOpen
  \bibfield  {author} {\bibinfo {author} {\bibfnamefont {E.}~\bibnamefont
  {Orlando}}\ and\ \bibinfo {author} {\bibfnamefont {A.}~\bibnamefont
  {Strong}},\ }\bibfield  {booktitle} {\emph {\bibinfo {booktitle} {{The
  Multi-Messenger Approach to High-Energy Gamma-Ray Sources: 3rd Workshop on
  the Nature of Unidentified High-Energy Sources, Barcelona, Spain, 4-7 Jul,
  2006}}},\ }\href {\doibase 10.1007/s10509-007-9457-0} {\bibfield  {journal}
  {\bibinfo  {journal} {Astrophys. Space Sci.}\ }\textbf {\bibinfo {volume}
  {309}},\ \bibinfo {pages} {359} (\bibinfo {year} {2007})},\ \Eprint
  {http://arxiv.org/abs/astro-ph/0607563} {arXiv:astro-ph/0607563 [astro-ph]}
  \BibitemShut {NoStop}%
\bibitem [{\citenamefont {Moskalenko}\ \emph {et~al.}(2006)\citenamefont
  {Moskalenko}, \citenamefont {Porter},\ and\ \citenamefont
  {Digel}}]{Moskalenko:2006ta}%
  \BibitemOpen
  \bibfield  {author} {\bibinfo {author} {\bibfnamefont {I.~V.}\ \bibnamefont
  {Moskalenko}}, \bibinfo {author} {\bibfnamefont {T.~A.}\ \bibnamefont
  {Porter}}, \ and\ \bibinfo {author} {\bibfnamefont {S.~W.}\ \bibnamefont
  {Digel}},\ }\href {\doibase 10.1086/520882, 10.1086/509916} {\bibfield
  {journal} {\bibinfo  {journal} {Astrophys. J.}\ }\textbf {\bibinfo {volume}
  {652}},\ \bibinfo {pages} {L65} (\bibinfo {year} {2006})},\ \bibinfo {note}
  {[Erratum: Astrophys. J.664,L143(2007)]},\ \Eprint
  {http://arxiv.org/abs/astro-ph/0607521} {arXiv:astro-ph/0607521 [astro-ph]}
  \BibitemShut {NoStop}%
\bibitem [{\citenamefont {Orlando}\ and\ \citenamefont
  {Strong}(2013)}]{Orlando:2013pza}%
  \BibitemOpen
  \bibfield  {author} {\bibinfo {author} {\bibfnamefont {E.}~\bibnamefont
  {Orlando}}\ and\ \bibinfo {author} {\bibfnamefont {A.}~\bibnamefont
  {Strong}},\ }\href {\doibase
  https://doi.org/10.1016/j.nuclphysbps.2013.05.042} {\bibfield  {journal}
  {\bibinfo  {journal} {Nuclear Physics B - Proceedings Supplements}\ }\textbf
  {\bibinfo {volume} {239-240}},\ \bibinfo {pages} {266 } (\bibinfo {year}
  {2013})},\ \bibinfo {note} {proceedings of the 9th workshop on Science with
  the New Generation of High Energy Gamma-ray Experiments: From high energy
  gamma sources to cosmic rays, one century after their discovery}\BibitemShut
  {NoStop}%
\bibitem [{\citenamefont {Orlando}\ \emph {et~al.}(2017)\citenamefont
  {Orlando}, \citenamefont {Giglietto}, \citenamefont {Moskalenko},
  \citenamefont {Raino'},\ and\ \citenamefont {Strong}}]{Orlando:2017iyc}%
  \BibitemOpen
  \bibfield  {author} {\bibinfo {author} {\bibfnamefont {E.}~\bibnamefont
  {Orlando}}, \bibinfo {author} {\bibfnamefont {N.}~\bibnamefont {Giglietto}},
  \bibinfo {author} {\bibfnamefont {I.}~\bibnamefont {Moskalenko}}, \bibinfo
  {author} {\bibfnamefont {S.}~\bibnamefont {Raino'}}, \ and\ \bibinfo {author}
  {\bibfnamefont {A.}~\bibnamefont {Strong}},\ }\bibfield  {booktitle} {\emph
  {\bibinfo {booktitle} {{Proceedings, 35th International Cosmic Ray Conference
  (ICRC 2017): Bexco, Busan, Korea, July 12-20, 2017}}},\ }\href@noop {}
  {\bibfield  {journal} {\bibinfo  {journal} {PoS}\ }\textbf {\bibinfo {volume}
  {ICRC2017}},\ \bibinfo {pages} {693} (\bibinfo {year} {2017})},\ \Eprint
  {http://arxiv.org/abs/1712.09745} {arXiv:1712.09745 [astro-ph.HE]}
  \BibitemShut {NoStop}%
\bibitem [{\citenamefont {{Seckel}}\ \emph {et~al.}(1991)\citenamefont
  {{Seckel}}, \citenamefont {{Stanev}},\ and\ \citenamefont
  {{Gaisser}}}]{1991ApJ...382..652S}%
  \BibitemOpen
  \bibfield  {author} {\bibinfo {author} {\bibfnamefont {D.}~\bibnamefont
  {{Seckel}}}, \bibinfo {author} {\bibfnamefont {T.}~\bibnamefont {{Stanev}}},
  \ and\ \bibinfo {author} {\bibfnamefont {T.~K.}\ \bibnamefont {{Gaisser}}},\
  }\href {\doibase 10.1086/170753} {\bibfield  {journal} {\bibinfo  {journal}
  {Astrophys. J.}\ }\textbf {\bibinfo {volume} {382}},\ \bibinfo {pages} {652}
  (\bibinfo {year} {1991})}\BibitemShut {NoStop}%
\bibitem [{\citenamefont {{Johnson}}(1989)}]{1989gros.work.....J}%
  \BibitemOpen
  \bibinfo {editor} {\bibfnamefont {W.~N.}\ \bibnamefont {{Johnson}}},\ ed.,\
  \href@noop {} {\emph {\bibinfo {title} {Gamma Ray Observatory Science
  Workshop}}}\ (\bibinfo {year} {1989})\BibitemShut {NoStop}%
\bibitem [{\citenamefont {Laurence}\ \emph {et~al.}(1966)\citenamefont
  {Laurence} \emph {et~al.}}]{doi:10.1029/JZ071i023p05778}%
  \BibitemOpen
  \bibfield  {author} {\bibinfo {author} {\bibfnamefont {P.}~\bibnamefont
  {Laurence}} \emph {et~al.},\ }\href {\doibase 10.1029/JZ071i023p05778}
  {\bibfield  {journal} {\bibinfo  {journal} {Journal of Geophysical Research}\
  }\textbf {\bibinfo {volume} {71}},\ \bibinfo {pages} {5778} (\bibinfo {year}
  {1966})}\BibitemShut {NoStop}%
\bibitem [{\citenamefont {{Linden}}\ \emph {et~al.}(2018)\citenamefont
  {{Linden}}, \citenamefont {{Zhou}}, \citenamefont {{Beacom}}, \citenamefont
  {{Peter}}, \citenamefont {{Ng}},\ and\ \citenamefont
  {{Tang}}}]{2018arXiv180305436L}%
  \BibitemOpen
  \bibfield  {author} {\bibinfo {author} {\bibfnamefont {T.}~\bibnamefont
  {{Linden}}}, \bibinfo {author} {\bibfnamefont {B.}~\bibnamefont {{Zhou}}},
  \bibinfo {author} {\bibfnamefont {J.~F.}\ \bibnamefont {{Beacom}}}, \bibinfo
  {author} {\bibfnamefont {A.~H.~G.}\ \bibnamefont {{Peter}}}, \bibinfo
  {author} {\bibfnamefont {K.~C.~Y.}\ \bibnamefont {{Ng}}}, \ and\ \bibinfo
  {author} {\bibfnamefont {Q.-W.}\ \bibnamefont {{Tang}}},\ }\href@noop {}
  {\bibfield  {journal} {\bibinfo  {journal} {ArXiv e-prints}\ } (\bibinfo
  {year} {2018})},\ \Eprint {http://arxiv.org/abs/1803.05436} {arXiv:1803.05436
  [astro-ph.HE]} \BibitemShut {NoStop}%
\bibitem [{\citenamefont {Ng}\ \emph {et~al.}(2016)\citenamefont {Ng},
  \citenamefont {Beacom}, \citenamefont {Peter},\ and\ \citenamefont
  {Rott}}]{Ng:2015gya}%
  \BibitemOpen
  \bibfield  {author} {\bibinfo {author} {\bibfnamefont {K.~C.~Y.}\
  \bibnamefont {Ng}}, \bibinfo {author} {\bibfnamefont {J.~F.}\ \bibnamefont
  {Beacom}}, \bibinfo {author} {\bibfnamefont {A.~H.~G.}\ \bibnamefont
  {Peter}}, \ and\ \bibinfo {author} {\bibfnamefont {C.}~\bibnamefont {Rott}},\
  }\href {\doibase 10.1103/PhysRevD.94.023004} {\bibfield  {journal} {\bibinfo
  {journal} {Phys. Rev.}\ }\textbf {\bibinfo {volume} {D94}},\ \bibinfo {pages}
  {023004} (\bibinfo {year} {2016})},\ \Eprint
  {http://arxiv.org/abs/1508.06276} {arXiv:1508.06276 [astro-ph.HE]}
  \BibitemShut {NoStop}%
\bibitem [{\citenamefont {{Tang}}\ \emph {et~al.}(2018)\citenamefont {{Tang}},
  \citenamefont {{Ng}}, \citenamefont {{Linden}}, \citenamefont {{Zhou}},
  \citenamefont {{Beacom}},\ and\ \citenamefont
  {{Peter}}}]{2018arXiv180406846T}%
  \BibitemOpen
  \bibfield  {author} {\bibinfo {author} {\bibfnamefont {Q.-W.}\ \bibnamefont
  {{Tang}}}, \bibinfo {author} {\bibfnamefont {K.~C.~Y.}\ \bibnamefont {{Ng}}},
  \bibinfo {author} {\bibfnamefont {T.}~\bibnamefont {{Linden}}}, \bibinfo
  {author} {\bibfnamefont {B.}~\bibnamefont {{Zhou}}}, \bibinfo {author}
  {\bibfnamefont {J.~F.}\ \bibnamefont {{Beacom}}}, \ and\ \bibinfo {author}
  {\bibfnamefont {A.~H.~G.}\ \bibnamefont {{Peter}}},\ }\href@noop {}
  {\bibfield  {journal} {\bibinfo  {journal} {ArXiv e-prints}\ } (\bibinfo
  {year} {2018})},\ \Eprint {http://arxiv.org/abs/1804.06846} {arXiv:1804.06846
  [astro-ph.HE]} \BibitemShut {NoStop}%
\bibitem [{\citenamefont {{Abeysekara}}\ \emph {et~al.}(2017)\citenamefont
  {{Abeysekara}} \emph {et~al.}}]{2017ApJ...843...39A}%
  \BibitemOpen
  \bibfield  {author} {\bibinfo {author} {\bibfnamefont {A.~U.}\ \bibnamefont
  {{Abeysekara}}} \emph {et~al.} (\bibinfo {collaboration} {HAWC
  Collaboration}),\ }\href {\doibase 10.3847/1538-4357/aa7555} {\bibfield
  {journal} {\bibinfo  {journal} {Astrophys. J.}\ }\textbf {\bibinfo {volume}
  {843}},\ \bibinfo {eid} {39} (\bibinfo {year} {2017})},\ \Eprint
  {http://arxiv.org/abs/1701.01778} {arXiv:1701.01778 [astro-ph.HE]}
  \BibitemShut {NoStop}%
\bibitem [{\citenamefont {Zhe}\ \emph {et~al.}(2016)\citenamefont {Zhe} \emph
  {et~al.}}]{2016ARGOGAMMA}%
  \BibitemOpen
  \bibfield  {author} {\bibinfo {author} {\bibfnamefont {L.}~\bibnamefont
  {Zhe}} \emph {et~al.} (\bibinfo {collaboration} {ARGO-YBJ}),\ }\href@noop {}
  {\bibfield  {journal} {\bibinfo  {journal} {7th Workshop on Air Shower
  Detection at High Altitude}\ } (\bibinfo {year} {2016})}\BibitemShut
  {NoStop}%
\bibitem [{\citenamefont {{Zhou}}\ \emph {et~al.}(2017)\citenamefont {{Zhou}},
  \citenamefont {{Ng}}, \citenamefont {{Beacom}},\ and\ \citenamefont
  {{Peter}}}]{2016arXiv161202420Z}%
  \BibitemOpen
  \bibfield  {author} {\bibinfo {author} {\bibfnamefont {B.}~\bibnamefont
  {{Zhou}}}, \bibinfo {author} {\bibfnamefont {K.~C.~Y.}\ \bibnamefont {{Ng}}},
  \bibinfo {author} {\bibfnamefont {J.~F.}\ \bibnamefont {{Beacom}}}, \ and\
  \bibinfo {author} {\bibfnamefont {A.~H.~G.}\ \bibnamefont {{Peter}}},\ }\href
  {\doibase 10.1103/PhysRevD.96.023015} {\bibfield  {journal} {\bibinfo
  {journal} {Phys. Rev. D}\ }\textbf {\bibinfo {volume} {96}},\ \bibinfo {eid}
  {023015} (\bibinfo {year} {2017})},\ \Eprint
  {http://arxiv.org/abs/1612.02420} {arXiv:1612.02420 [astro-ph.HE]}
  \BibitemShut {NoStop}%
\bibitem [{\citenamefont {Gould}(1992)}]{Gould:1991hx}%
  \BibitemOpen
  \bibfield  {author} {\bibinfo {author} {\bibfnamefont {A.}~\bibnamefont
  {Gould}},\ }\href {\doibase 10.1086/171156} {\bibfield  {journal} {\bibinfo
  {journal} {Astrophys. J.}\ }\textbf {\bibinfo {volume} {388}},\ \bibinfo
  {pages} {338} (\bibinfo {year} {1992})}\BibitemShut {NoStop}%
\bibitem [{\citenamefont {Silk}\ \emph {et~al.}(1985)\citenamefont {Silk},
  \citenamefont {Olive},\ and\ \citenamefont {Srednicki}}]{PhysRevLett.55.257}%
  \BibitemOpen
  \bibfield  {author} {\bibinfo {author} {\bibfnamefont {J.}~\bibnamefont
  {Silk}}, \bibinfo {author} {\bibfnamefont {K.}~\bibnamefont {Olive}}, \ and\
  \bibinfo {author} {\bibfnamefont {M.}~\bibnamefont {Srednicki}},\ }\href
  {\doibase 10.1103/PhysRevLett.55.257} {\bibfield  {journal} {\bibinfo
  {journal} {Phys. Rev. Lett.}\ }\textbf {\bibinfo {volume} {55}},\ \bibinfo
  {pages} {257} (\bibinfo {year} {1985})}\BibitemShut {NoStop}%
\bibitem [{\citenamefont {Edsj{\"o}}(1995)}]{1995NuPhS..43..265E}%
  \BibitemOpen
  \bibfield  {author} {\bibinfo {author} {\bibfnamefont {J.}~\bibnamefont
  {Edsj{\"o}}},\ }\href {\doibase 10.1016/0920-5632(95)00487-T} {\bibfield
  {journal} {\bibinfo  {journal} {Nuclear Physics B Proceedings Supplements}\
  }\textbf {\bibinfo {volume} {43}},\ \bibinfo {pages} {265} (\bibinfo {year}
  {1995})},\ \Eprint {http://arxiv.org/abs/hep-ph/9504205} {hep-ph/9504205}
  \BibitemShut {NoStop}%
\bibitem [{\citenamefont {{Lundberg}}\ and\ \citenamefont
  {{Edsj{\"o}}}(2004)}]{2004PhRvD..69l3505L}%
  \BibitemOpen
  \bibfield  {author} {\bibinfo {author} {\bibfnamefont {J.}~\bibnamefont
  {{Lundberg}}}\ and\ \bibinfo {author} {\bibfnamefont {J.}~\bibnamefont
  {{Edsj{\"o}}}},\ }\href {\doibase 10.1103/PhysRevD.69.123505} {\bibfield
  {journal} {\bibinfo  {journal} {\prd}\ }\textbf {\bibinfo {volume} {69}},\
  \bibinfo {eid} {123505} (\bibinfo {year} {2004})},\ \Eprint
  {http://arxiv.org/abs/astro-ph/0401113} {astro-ph/0401113} \BibitemShut
  {NoStop}%
\bibitem [{\citenamefont {{Flacke}}\ \emph {et~al.}(2009)\citenamefont
  {{Flacke}}, \citenamefont {{Menon}}, \citenamefont {{Hooper}},\ and\
  \citenamefont {{Freese}}}]{2009arXiv0908.0899F}%
  \BibitemOpen
  \bibfield  {author} {\bibinfo {author} {\bibfnamefont {T.}~\bibnamefont
  {{Flacke}}}, \bibinfo {author} {\bibfnamefont {A.}~\bibnamefont {{Menon}}},
  \bibinfo {author} {\bibfnamefont {D.}~\bibnamefont {{Hooper}}}, \ and\
  \bibinfo {author} {\bibfnamefont {K.}~\bibnamefont {{Freese}}},\ }\href@noop
  {} {\bibfield  {journal} {\bibinfo  {journal} {ArXiv e-prints}\ } (\bibinfo
  {year} {2009})},\ \Eprint {http://arxiv.org/abs/0908.0899} {arXiv:0908.0899
  [hep-ph]} \BibitemShut {NoStop}%
\bibitem [{\citenamefont {{Peter}}(2009)}]{2009PhRvD..79j3532P}%
  \BibitemOpen
  \bibfield  {author} {\bibinfo {author} {\bibfnamefont {A.~H.~G.}\
  \bibnamefont {{Peter}}},\ }\href {\doibase 10.1103/PhysRevD.79.103532}
  {\bibfield  {journal} {\bibinfo  {journal} {\prd}\ }\textbf {\bibinfo
  {volume} {79}},\ \bibinfo {eid} {103532} (\bibinfo {year} {2009})},\ \Eprint
  {http://arxiv.org/abs/0902.1347} {arXiv:0902.1347 [astro-ph.HE]} \BibitemShut
  {NoStop}%
\bibitem [{\citenamefont {{Rott}}\ \emph {et~al.}(2011)\citenamefont {{Rott}},
  \citenamefont {{Tanaka}},\ and\ \citenamefont
  {{Itow}}}]{2011JCAP...09..029R}%
  \BibitemOpen
  \bibfield  {author} {\bibinfo {author} {\bibfnamefont {C.}~\bibnamefont
  {{Rott}}}, \bibinfo {author} {\bibfnamefont {T.}~\bibnamefont {{Tanaka}}}, \
  and\ \bibinfo {author} {\bibfnamefont {Y.}~\bibnamefont {{Itow}}},\ }\href
  {\doibase 10.1088/1475-7516/2011/09/029} {\bibfield  {journal} {\bibinfo
  {journal} {Journal of Cosmology and Astroparticle Physics}\ }\textbf
  {\bibinfo {volume} {9}},\ \bibinfo {eid} {029} (\bibinfo {year} {2011})},\
  \Eprint {http://arxiv.org/abs/1107.3182} {arXiv:1107.3182 [astro-ph.HE]}
  \BibitemShut {NoStop}%
\bibitem [{\citenamefont {Danninger}\ and\ \citenamefont
  {Rott}(2014)}]{Danninger:2014xza}%
  \BibitemOpen
  \bibfield  {author} {\bibinfo {author} {\bibfnamefont {M.}~\bibnamefont
  {Danninger}}\ and\ \bibinfo {author} {\bibfnamefont {C.}~\bibnamefont
  {Rott}},\ }\href {\doibase 10.1016/j.dark.2014.10.002} {\bibfield  {journal}
  {\bibinfo  {journal} {Phys. Dark Univ.}\ }\textbf {\bibinfo {volume} {5-6}},\
  \bibinfo {pages} {35} (\bibinfo {year} {2014})},\ \Eprint
  {http://arxiv.org/abs/1509.08230} {arXiv:1509.08230 [astro-ph.HE]}
  \BibitemShut {NoStop}%
\bibitem [{\citenamefont {Choi}\ \emph {et~al.}(2015)\citenamefont {Choi} \emph
  {et~al.}}]{Choi:2015ara}%
  \BibitemOpen
  \bibfield  {author} {\bibinfo {author} {\bibfnamefont {K.}~\bibnamefont
  {Choi}} \emph {et~al.} (\bibinfo {collaboration} {Super-Kamiokande}),\ }\href
  {\doibase 10.1103/PhysRevLett.114.141301} {\bibfield  {journal} {\bibinfo
  {journal} {Phys. Rev. Lett.}\ }\textbf {\bibinfo {volume} {114}},\ \bibinfo
  {pages} {141301} (\bibinfo {year} {2015})},\ \Eprint
  {http://arxiv.org/abs/1503.04858} {arXiv:1503.04858 [hep-ex]} \BibitemShut
  {NoStop}%
\bibitem [{\citenamefont {Aartsen}\ \emph {et~al.}(2017)\citenamefont {Aartsen}
  \emph {et~al.}}]{Aartsen:2016zhm}%
  \BibitemOpen
  \bibfield  {author} {\bibinfo {author} {\bibfnamefont {M.~G.}\ \bibnamefont
  {Aartsen}} \emph {et~al.} (\bibinfo {collaboration} {IceCube}),\ }\href
  {\doibase 10.1140/epjc/s10052-017-4689-9} {\bibfield  {journal} {\bibinfo
  {journal} {Eur. Phys. J.}\ }\textbf {\bibinfo {volume} {C77}},\ \bibinfo
  {pages} {146} (\bibinfo {year} {2017})},\ \Eprint
  {http://arxiv.org/abs/1612.05949} {arXiv:1612.05949 [astro-ph.HE]}
  \BibitemShut {NoStop}%
\bibitem [{\citenamefont {{Widmark}}(2017)}]{2017JCAP...05..046W}%
  \BibitemOpen
  \bibfield  {author} {\bibinfo {author} {\bibfnamefont {A.}~\bibnamefont
  {{Widmark}}},\ }\href {\doibase 10.1088/1475-7516/2017/05/046} {\bibfield
  {journal} {\bibinfo  {journal} {Journal of Cosmology and Astroparticle
  Physics}\ }\textbf {\bibinfo {volume} {5}},\ \bibinfo {eid} {046} (\bibinfo
  {year} {2017})},\ \Eprint {http://arxiv.org/abs/1703.06878} {arXiv:1703.06878
  [hep-ph]} \BibitemShut {NoStop}%
\bibitem [{\citenamefont {Garani}\ and\ \citenamefont
  {Palomares-Ruiz}(2017)}]{Garani:2017jcj}%
  \BibitemOpen
  \bibfield  {author} {\bibinfo {author} {\bibfnamefont {R.}~\bibnamefont
  {Garani}}\ and\ \bibinfo {author} {\bibfnamefont {S.}~\bibnamefont
  {Palomares-Ruiz}},\ }\href {\doibase 10.1088/1475-7516/2017/05/007}
  {\bibfield  {journal} {\bibinfo  {journal} {JCAP}\ }\textbf {\bibinfo
  {volume} {1705}},\ \bibinfo {pages} {007} (\bibinfo {year} {2017})},\ \Eprint
  {http://arxiv.org/abs/1702.02768} {arXiv:1702.02768 [hep-ph]} \BibitemShut
  {NoStop}%
\bibitem [{\citenamefont {Baum}\ \emph {et~al.}(2017)\citenamefont {Baum},
  \citenamefont {Visinelli}, \citenamefont {Freese},\ and\ \citenamefont
  {Stengel}}]{Baum:2016oow}%
  \BibitemOpen
  \bibfield  {author} {\bibinfo {author} {\bibfnamefont {S.}~\bibnamefont
  {Baum}}, \bibinfo {author} {\bibfnamefont {L.}~\bibnamefont {Visinelli}},
  \bibinfo {author} {\bibfnamefont {K.}~\bibnamefont {Freese}}, \ and\ \bibinfo
  {author} {\bibfnamefont {P.}~\bibnamefont {Stengel}},\ }\href {\doibase
  10.1103/PhysRevD.95.043007} {\bibfield  {journal} {\bibinfo  {journal} {Phys.
  Rev.}\ }\textbf {\bibinfo {volume} {D95}},\ \bibinfo {pages} {043007}
  (\bibinfo {year} {2017})},\ \Eprint {http://arxiv.org/abs/1611.09665}
  {arXiv:1611.09665 [astro-ph.CO]} \BibitemShut {NoStop}%
\bibitem [{\citenamefont {Meade}\ \emph {et~al.}(2010)\citenamefont {Meade},
  \citenamefont {Nussinov}, \citenamefont {Papucci},\ and\ \citenamefont
  {Volansky}}]{Meade:2009mu}%
  \BibitemOpen
  \bibfield  {author} {\bibinfo {author} {\bibfnamefont {P.}~\bibnamefont
  {Meade}}, \bibinfo {author} {\bibfnamefont {S.}~\bibnamefont {Nussinov}},
  \bibinfo {author} {\bibfnamefont {M.}~\bibnamefont {Papucci}}, \ and\
  \bibinfo {author} {\bibfnamefont {T.}~\bibnamefont {Volansky}},\ }\href
  {\doibase 10.1007/JHEP06(2010)029} {\bibfield  {journal} {\bibinfo  {journal}
  {JHEP}\ }\textbf {\bibinfo {volume} {06}},\ \bibinfo {pages} {029} (\bibinfo
  {year} {2010})},\ \Eprint {http://arxiv.org/abs/0910.4160} {arXiv:0910.4160
  [hep-ph]} \BibitemShut {NoStop}%
\bibitem [{\citenamefont {{Batell}}\ \emph {et~al.}(2010)\citenamefont
  {{Batell}}, \citenamefont {{Pospelov}}, \citenamefont {{Ritz}},\ and\
  \citenamefont {{Shang}}}]{2010PhRvD..81g5004B}%
  \BibitemOpen
  \bibfield  {author} {\bibinfo {author} {\bibfnamefont {B.}~\bibnamefont
  {{Batell}}}, \bibinfo {author} {\bibfnamefont {M.}~\bibnamefont
  {{Pospelov}}}, \bibinfo {author} {\bibfnamefont {A.}~\bibnamefont {{Ritz}}},
  \ and\ \bibinfo {author} {\bibfnamefont {Y.}~\bibnamefont {{Shang}}},\ }\href
  {\doibase 10.1103/PhysRevD.81.075004} {\bibfield  {journal} {\bibinfo
  {journal} {Phys. Rev. D}\ }\textbf {\bibinfo {volume} {81}},\ \bibinfo {eid}
  {075004} (\bibinfo {year} {2010})},\ \Eprint {http://arxiv.org/abs/0910.1567}
  {arXiv:0910.1567 [hep-ph]} \BibitemShut {NoStop}%
\bibitem [{\citenamefont {{Schuster}}\ \emph {et~al.}(2010)\citenamefont
  {{Schuster}}, \citenamefont {{Toro}},\ and\ \citenamefont
  {{Yavin}}}]{2010PhRvD..81a6002S}%
  \BibitemOpen
  \bibfield  {author} {\bibinfo {author} {\bibfnamefont {P.}~\bibnamefont
  {{Schuster}}}, \bibinfo {author} {\bibfnamefont {N.}~\bibnamefont {{Toro}}},
  \ and\ \bibinfo {author} {\bibfnamefont {I.}~\bibnamefont {{Yavin}}},\ }\href
  {\doibase 10.1103/PhysRevD.81.016002} {\bibfield  {journal} {\bibinfo
  {journal} {\prd}\ }\textbf {\bibinfo {volume} {81}},\ \bibinfo {eid} {016002}
  (\bibinfo {year} {2010})},\ \Eprint {http://arxiv.org/abs/0910.1602}
  {arXiv:0910.1602 [hep-ph]} \BibitemShut {NoStop}%
\bibitem [{\citenamefont {Bell}\ and\ \citenamefont
  {Petraki}(2011)}]{Bell:2011sn}%
  \BibitemOpen
  \bibfield  {author} {\bibinfo {author} {\bibfnamefont {N.~F.}\ \bibnamefont
  {Bell}}\ and\ \bibinfo {author} {\bibfnamefont {K.}~\bibnamefont {Petraki}},\
  }\href {\doibase 10.1088/1475-7516/2011/04/003} {\bibfield  {journal}
  {\bibinfo  {journal} {JCAP}\ }\textbf {\bibinfo {volume} {1104}},\ \bibinfo
  {pages} {003} (\bibinfo {year} {2011})},\ \Eprint
  {http://arxiv.org/abs/1102.2958} {arXiv:1102.2958 [hep-ph]} \BibitemShut
  {NoStop}%
\bibitem [{\citenamefont {Feng}\ \emph {et~al.}(2016)\citenamefont {Feng},
  \citenamefont {Smolinsky},\ and\ \citenamefont {Tanedo}}]{Feng:2016ijc}%
  \BibitemOpen
  \bibfield  {author} {\bibinfo {author} {\bibfnamefont {J.~L.}\ \bibnamefont
  {Feng}}, \bibinfo {author} {\bibfnamefont {J.}~\bibnamefont {Smolinsky}}, \
  and\ \bibinfo {author} {\bibfnamefont {P.}~\bibnamefont {Tanedo}},\ }\href
  {\doibase 10.1103/PhysRevD.93.115036, 10.1103/PhysRevD.96.099903} {\bibfield
  {journal} {\bibinfo  {journal} {Phys. Rev.}\ }\textbf {\bibinfo {volume}
  {D93}},\ \bibinfo {pages} {115036} (\bibinfo {year} {2016})},\ \bibinfo
  {note} {[Erratum: Phys. Rev.D96,(2017)]},\ \Eprint
  {http://arxiv.org/abs/1602.01465} {arXiv:1602.01465 [hep-ph]} \BibitemShut
  {NoStop}%
\bibitem [{\citenamefont {Adrian-Martinez}\ \emph
  {et~al.}(2016{\natexlab{a}})\citenamefont {Adrian-Martinez} \emph
  {et~al.}}]{Adrian-Martinez:2016gti}%
  \BibitemOpen
  \bibfield  {author} {\bibinfo {author} {\bibfnamefont {S.}~\bibnamefont
  {Adrian-Martinez}} \emph {et~al.} (\bibinfo {collaboration} {ANTARES}),\
  }\href {\doibase 10.1016/j.physletb.2016.05.019} {\bibfield  {journal}
  {\bibinfo  {journal} {Phys. Lett.}\ }\textbf {\bibinfo {volume} {B759}},\
  \bibinfo {pages} {69} (\bibinfo {year} {2016}{\natexlab{a}})},\ \Eprint
  {http://arxiv.org/abs/1603.02228} {arXiv:1603.02228 [astro-ph.HE]}
  \BibitemShut {NoStop}%
\bibitem [{\citenamefont {{Leane}}\ \emph {et~al.}(2017)\citenamefont
  {{Leane}}, \citenamefont {{Ng}},\ and\ \citenamefont
  {{Beacom}}}]{2017PhRvD..95l3016L}%
  \BibitemOpen
  \bibfield  {author} {\bibinfo {author} {\bibfnamefont {R.~K.}\ \bibnamefont
  {{Leane}}}, \bibinfo {author} {\bibfnamefont {K.~C.~Y.}\ \bibnamefont
  {{Ng}}}, \ and\ \bibinfo {author} {\bibfnamefont {J.~F.}\ \bibnamefont
  {{Beacom}}},\ }\href {\doibase 10.1103/PhysRevD.95.123016} {\bibfield
  {journal} {\bibinfo  {journal} {Phys. Rev. D}\ }\textbf {\bibinfo {volume}
  {95}},\ \bibinfo {eid} {123016} (\bibinfo {year} {2017})},\ \Eprint
  {http://arxiv.org/abs/1703.04629} {arXiv:1703.04629 [astro-ph.HE]}
  \BibitemShut {NoStop}%
\bibitem [{\citenamefont {Arina}\ \emph {et~al.}(2017)\citenamefont {Arina},
  \citenamefont {Backović}, \citenamefont {Heisig},\ and\ \citenamefont
  {Lucente}}]{Arina:2017sng}%
  \BibitemOpen
  \bibfield  {author} {\bibinfo {author} {\bibfnamefont {C.}~\bibnamefont
  {Arina}}, \bibinfo {author} {\bibfnamefont {M.}~\bibnamefont {Backović}},
  \bibinfo {author} {\bibfnamefont {J.}~\bibnamefont {Heisig}}, \ and\ \bibinfo
  {author} {\bibfnamefont {M.}~\bibnamefont {Lucente}},\ }\href {\doibase
  10.1103/PhysRevD.96.063010} {\bibfield  {journal} {\bibinfo  {journal} {Phys.
  Rev.}\ }\textbf {\bibinfo {volume} {D96}},\ \bibinfo {pages} {063010}
  (\bibinfo {year} {2017})},\ \Eprint {http://arxiv.org/abs/1703.08087}
  {arXiv:1703.08087 [astro-ph.HE]} \BibitemShut {NoStop}%
\bibitem [{\citenamefont {{Smolinsky}}\ and\ \citenamefont
  {{Tanedo}}(2017)}]{Smolinsky:2017fvb}%
  \BibitemOpen
  \bibfield  {author} {\bibinfo {author} {\bibfnamefont {J.}~\bibnamefont
  {{Smolinsky}}}\ and\ \bibinfo {author} {\bibfnamefont {P.}~\bibnamefont
  {{Tanedo}}},\ }\href {\doibase 10.1103/PhysRevD.95.075015} {\bibfield
  {journal} {\bibinfo  {journal} {\prd}\ }\textbf {\bibinfo {volume} {95}},\
  \bibinfo {eid} {075015} (\bibinfo {year} {2017})},\ \Eprint
  {http://arxiv.org/abs/1701.03168} {arXiv:1701.03168 [hep-ph]} \BibitemShut
  {NoStop}%
\bibitem [{\citenamefont {Moskalenko}\ and\ \citenamefont
  {Karakula}(1993)}]{0954-3899-19-9-019}%
  \BibitemOpen
  \bibfield  {author} {\bibinfo {author} {\bibfnamefont {I.~V.}\ \bibnamefont
  {Moskalenko}}\ and\ \bibinfo {author} {\bibfnamefont {S.}~\bibnamefont
  {Karakula}},\ }\href {http://stacks.iop.org/0954-3899/19/i=9/a=019}
  {\bibfield  {journal} {\bibinfo  {journal} {Journal of Physics G: Nuclear and
  Particle Physics}\ }\textbf {\bibinfo {volume} {19}},\ \bibinfo {pages}
  {1399} (\bibinfo {year} {1993})}\BibitemShut {NoStop}%
\bibitem [{\citenamefont {Ingelman}\ and\ \citenamefont
  {Thunman}(1996)}]{Ingelman:1996mj}%
  \BibitemOpen
  \bibfield  {author} {\bibinfo {author} {\bibfnamefont {G.}~\bibnamefont
  {Ingelman}}\ and\ \bibinfo {author} {\bibfnamefont {M.}~\bibnamefont
  {Thunman}},\ }\href {\doibase 10.1103/PhysRevD.54.4385} {\bibfield  {journal}
  {\bibinfo  {journal} {Phys. Rev.}\ }\textbf {\bibinfo {volume} {D54}},\
  \bibinfo {pages} {4385} (\bibinfo {year} {1996})},\ \Eprint
  {http://arxiv.org/abs/hep-ph/9604288} {arXiv:hep-ph/9604288 [hep-ph]}
  \BibitemShut {NoStop}%
\bibitem [{\citenamefont {Andersen}\ and\ \citenamefont
  {Klein}(2011)}]{Andersen:2011dz}%
  \BibitemOpen
  \bibfield  {author} {\bibinfo {author} {\bibfnamefont {K.~K.}\ \bibnamefont
  {Andersen}}\ and\ \bibinfo {author} {\bibfnamefont {S.~R.}\ \bibnamefont
  {Klein}},\ }\href {\doibase 10.1103/PhysRevD.83.103519} {\bibfield  {journal}
  {\bibinfo  {journal} {Phys. Rev.}\ }\textbf {\bibinfo {volume} {D83}},\
  \bibinfo {pages} {103519} (\bibinfo {year} {2011})},\ \Eprint
  {http://arxiv.org/abs/1103.5090} {arXiv:1103.5090 [astro-ph.HE]} \BibitemShut
  {NoStop}%
\bibitem [{\citenamefont {{Ng}}\ \emph {et~al.}(2017)\citenamefont {{Ng}},
  \citenamefont {{Beacom}}, \citenamefont {{Peter}},\ and\ \citenamefont
  {{Rott}}}]{2017PhRvD..96j3006N}%
  \BibitemOpen
  \bibfield  {author} {\bibinfo {author} {\bibfnamefont {K.~C.~Y.}\
  \bibnamefont {{Ng}}}, \bibinfo {author} {\bibfnamefont {J.~F.}\ \bibnamefont
  {{Beacom}}}, \bibinfo {author} {\bibfnamefont {A.~H.~G.}\ \bibnamefont
  {{Peter}}}, \ and\ \bibinfo {author} {\bibfnamefont {C.}~\bibnamefont
  {{Rott}}},\ }\href {\doibase 10.1103/PhysRevD.96.103006} {\bibfield
  {journal} {\bibinfo  {journal} {\prd}\ }\textbf {\bibinfo {volume} {96}},\
  \bibinfo {eid} {103006} (\bibinfo {year} {2017})},\ \Eprint
  {http://arxiv.org/abs/1703.10280} {arXiv:1703.10280 [astro-ph.HE]}
  \BibitemShut {NoStop}%
\bibitem [{\citenamefont {{Arg{\"u}elles}}\ \emph {et~al.}(2017)\citenamefont
  {{Arg{\"u}elles}}, \citenamefont {{de Wasseige}}, \citenamefont
  {{Fedynitch}},\ and\ \citenamefont {{Jones}}}]{2017JCAP...07..024A}%
  \BibitemOpen
  \bibfield  {author} {\bibinfo {author} {\bibfnamefont {C.~A.}\ \bibnamefont
  {{Arg{\"u}elles}}}, \bibinfo {author} {\bibfnamefont {G.}~\bibnamefont {{de
  Wasseige}}}, \bibinfo {author} {\bibfnamefont {A.}~\bibnamefont
  {{Fedynitch}}}, \ and\ \bibinfo {author} {\bibfnamefont {B.~J.~P.}\
  \bibnamefont {{Jones}}},\ }\href {\doibase 10.1088/1475-7516/2017/07/024}
  {\bibfield  {journal} {\bibinfo  {journal} {Journal of Cosmology and
  Astroparticle Physics}\ }\textbf {\bibinfo {volume} {7}},\ \bibinfo {eid}
  {024} (\bibinfo {year} {2017})},\ \Eprint {http://arxiv.org/abs/1703.07798}
  {arXiv:1703.07798 [astro-ph.HE]} \BibitemShut {NoStop}%
\bibitem [{\citenamefont {Edsjo}\ \emph {et~al.}(2017)\citenamefont {Edsjo},
  \citenamefont {Elevant}, \citenamefont {Enberg},\ and\ \citenamefont
  {Niblaeus}}]{Edsjo:2017kjk}%
  \BibitemOpen
  \bibfield  {author} {\bibinfo {author} {\bibfnamefont {J.}~\bibnamefont
  {Edsjo}}, \bibinfo {author} {\bibfnamefont {J.}~\bibnamefont {Elevant}},
  \bibinfo {author} {\bibfnamefont {R.}~\bibnamefont {Enberg}}, \ and\ \bibinfo
  {author} {\bibfnamefont {C.}~\bibnamefont {Niblaeus}},\ }\href {\doibase
  10.1088/1475-7516/2017/06/033} {\bibfield  {journal} {\bibinfo  {journal}
  {JCAP}\ }\textbf {\bibinfo {volume} {1706}},\ \bibinfo {pages} {033}
  (\bibinfo {year} {2017})},\ \Eprint {http://arxiv.org/abs/1704.02892}
  {arXiv:1704.02892 [astro-ph.HE]} \BibitemShut {NoStop}%
\bibitem [{\citenamefont {{Wiegelmann}}\ \emph {et~al.}(2014)\citenamefont
  {{Wiegelmann}}, \citenamefont {{Thalmann}},\ and\ \citenamefont
  {{Solanki}}}]{2014A&ARv..22...78W}%
  \BibitemOpen
  \bibfield  {author} {\bibinfo {author} {\bibfnamefont {T.}~\bibnamefont
  {{Wiegelmann}}}, \bibinfo {author} {\bibfnamefont {J.~K.}\ \bibnamefont
  {{Thalmann}}}, \ and\ \bibinfo {author} {\bibfnamefont {S.~K.}\ \bibnamefont
  {{Solanki}}},\ }\href {\doibase 10.1007/s00159-014-0078-7} {\bibfield
  {journal} {\bibinfo  {journal} {The Astronomy and Astrophysics Review}\
  }\textbf {\bibinfo {volume} {22}},\ \bibinfo {eid} {78} (\bibinfo {year}
  {2014})},\ \Eprint {http://arxiv.org/abs/1410.4214} {arXiv:1410.4214
  [astro-ph.SR]} \BibitemShut {NoStop}%
\bibitem [{\citenamefont {{Atwood}}\ \emph {et~al.}(2009)\citenamefont
  {{Atwood}} \emph {et~al.}}]{2009ApJ...697.1071A}%
  \BibitemOpen
  \bibfield  {author} {\bibinfo {author} {\bibfnamefont {W.~B.}\ \bibnamefont
  {{Atwood}}} \emph {et~al.},\ }\href {\doibase 10.1088/0004-637X/697/2/1071}
  {\bibfield  {journal} {\bibinfo  {journal} {\apj}\ }\textbf {\bibinfo
  {volume} {697}},\ \bibinfo {pages} {1071} (\bibinfo {year} {2009})},\ \Eprint
  {http://arxiv.org/abs/0902.1089} {arXiv:0902.1089 [astro-ph.IM]} \BibitemShut
  {NoStop}%
\bibitem [{\citenamefont {{Chang}}\ \emph {et~al.}(2017)\citenamefont {{Chang}}
  \emph {et~al.}}]{2017APh....95....6C}%
  \BibitemOpen
  \bibfield  {author} {\bibinfo {author} {\bibfnamefont {J.}~\bibnamefont
  {{Chang}}} \emph {et~al.},\ }\href {\doibase
  10.1016/j.astropartphys.2017.08.005} {\bibfield  {journal} {\bibinfo
  {journal} {Astroparticle Physics}\ }\textbf {\bibinfo {volume} {95}},\
  \bibinfo {pages} {6} (\bibinfo {year} {2017})},\ \Eprint
  {http://arxiv.org/abs/1706.08453} {arXiv:1706.08453 [astro-ph.IM]}
  \BibitemShut {NoStop}%
\bibitem [{\citenamefont {Adriani}\ \emph {et~al.}(2017)\citenamefont {Adriani}
  \emph {et~al.}}]{PhysRevLett.119.181101}%
  \BibitemOpen
  \bibfield  {author} {\bibinfo {author} {\bibfnamefont {O.}~\bibnamefont
  {Adriani}} \emph {et~al.} (\bibinfo {collaboration} {CALET Collaboration}),\
  }\href {\doibase 10.1103/PhysRevLett.119.181101} {\bibfield  {journal}
  {\bibinfo  {journal} {Phys. Rev. Lett.}\ }\textbf {\bibinfo {volume} {119}},\
  \bibinfo {pages} {181101} (\bibinfo {year} {2017})}\BibitemShut {NoStop}%
\bibitem [{\citenamefont {{Anderhub}}\ \emph {et~al.}(2013)\citenamefont
  {{Anderhub}} \emph {et~al.}}]{2013JInst...8P6008A}%
  \BibitemOpen
  \bibfield  {author} {\bibinfo {author} {\bibfnamefont {H.}~\bibnamefont
  {{Anderhub}}} \emph {et~al.},\ }\href {\doibase
  10.1088/1748-0221/8/06/P06008} {\bibfield  {journal} {\bibinfo  {journal}
  {Journal of Instrumentation}\ }\textbf {\bibinfo {volume} {8}},\ \bibinfo
  {eid} {P06008} (\bibinfo {year} {2013})},\ \Eprint
  {http://arxiv.org/abs/1304.1710} {arXiv:1304.1710 [astro-ph.IM]} \BibitemShut
  {NoStop}%
\bibitem [{\citenamefont {Lessard}(1999)}]{LESSARD1999243}%
  \BibitemOpen
  \bibfield  {author} {\bibinfo {author} {\bibfnamefont {R.}~\bibnamefont
  {Lessard}},\ }\href {\doibase https://doi.org/10.1016/S0927-6505(99)00057-2}
  {\bibfield  {journal} {\bibinfo  {journal} {Astroparticle Physics}\ }\textbf
  {\bibinfo {volume} {11}},\ \bibinfo {pages} {243 } (\bibinfo {year}
  {1999})},\ \bibinfo {note} {teV Astrophysics of Extragalactic
  Sources}\BibitemShut {NoStop}%
\bibitem [{\citenamefont {{Balzer}}\ \emph {et~al.}(2014)\citenamefont
  {{Balzer}} \emph {et~al.}}]{2014APh....54...67B}%
  \BibitemOpen
  \bibfield  {author} {\bibinfo {author} {\bibfnamefont {A.}~\bibnamefont
  {{Balzer}}} \emph {et~al.},\ }\href {\doibase
  10.1016/j.astropartphys.2013.11.007} {\bibfield  {journal} {\bibinfo
  {journal} {Astroparticle Physics}\ }\textbf {\bibinfo {volume} {54}},\
  \bibinfo {pages} {67} (\bibinfo {year} {2014})},\ \Eprint
  {http://arxiv.org/abs/1311.3486} {arXiv:1311.3486 [astro-ph.IM]} \BibitemShut
  {NoStop}%
\bibitem [{\citenamefont {{Ahnen}}\ \emph {et~al.}(2017)\citenamefont {{Ahnen}}
  \emph {et~al.}}]{2017APh....94...29A}%
  \BibitemOpen
  \bibfield  {author} {\bibinfo {author} {\bibfnamefont {M.~L.}\ \bibnamefont
  {{Ahnen}}} \emph {et~al.},\ }\href {\doibase
  10.1016/j.astropartphys.2017.08.001} {\bibfield  {journal} {\bibinfo
  {journal} {Astroparticle Physics}\ }\textbf {\bibinfo {volume} {94}},\
  \bibinfo {pages} {29} (\bibinfo {year} {2017})},\ \Eprint
  {http://arxiv.org/abs/1704.00906} {arXiv:1704.00906 [astro-ph.IM]}
  \BibitemShut {NoStop}%
\bibitem [{\citenamefont {Albert}\ \emph {et~al.}(2018)\citenamefont {Albert}
  \emph {et~al.}}]{DMPaper}%
  \BibitemOpen
  \bibfield  {author} {\bibinfo {author} {\bibfnamefont {A.}~\bibnamefont
  {Albert}} \emph {et~al.},\ }\href@noop {} {\bibfield  {journal} {\bibinfo
  {journal} {{ArXiv e-prints}}\ } (\bibinfo {year} {2018})},\ \Eprint
  {http://arxiv.org/abs/1808.05624} {arXiv:1808.05624 [hep-ph]} \BibitemShut
  {NoStop}%
\bibitem [{\citenamefont {Abeysekara}\ \emph {et~al.}(2014)\citenamefont
  {Abeysekara} \emph {et~al.}}]{Abeysekara:2013qka}%
  \BibitemOpen
  \bibfield  {author} {\bibinfo {author} {\bibfnamefont {A.~U.}\ \bibnamefont
  {Abeysekara}} \emph {et~al.} (\bibinfo {collaboration} {HAWC
  Collaboration}),\ }\href@noop {} {\bibfield  {journal} {\bibinfo  {journal}
  {Braz. J. Phys.}\ }\textbf {\bibinfo {volume} {44}} (\bibinfo {year}
  {2014})},\ \Eprint {http://arxiv.org/abs/1310.0071} {arXiv:1310.0071
  [astro-ph.HE]} \BibitemShut {NoStop}%
\bibitem [{\citenamefont {{Abeysekara}}\ \emph {et~al.}(2014)\citenamefont
  {{Abeysekara}} \emph {et~al.}}]{2014ApJ...796..108A}%
  \BibitemOpen
  \bibfield  {author} {\bibinfo {author} {\bibfnamefont {A.~U.}\ \bibnamefont
  {{Abeysekara}}} \emph {et~al.} (\bibinfo {collaboration} {HAWC
  Collaboration}),\ }\href {\doibase 10.1088/0004-637X/796/2/108} {\bibfield
  {journal} {\bibinfo  {journal} {Astrophys. J.}\ }\textbf {\bibinfo {volume}
  {796}},\ \bibinfo {eid} {108} (\bibinfo {year} {2014})},\ \Eprint
  {http://arxiv.org/abs/1408.4805} {arXiv:1408.4805 [astro-ph.HE]} \BibitemShut
  {NoStop}%
\bibitem [{\citenamefont {Abeysekara}\ \emph {et~al.}(2017)\citenamefont
  {Abeysekara} \emph {et~al.}}]{Abeysekara:2017hyn}%
  \BibitemOpen
  \bibfield  {author} {\bibinfo {author} {\bibfnamefont {A.~U.}\ \bibnamefont
  {Abeysekara}} \emph {et~al.},\ }\href {\doibase 10.3847/1538-4357/aa7556}
  {\bibfield  {journal} {\bibinfo  {journal} {Astrophys. J.}\ }\textbf
  {\bibinfo {volume} {843}},\ \bibinfo {pages} {40} (\bibinfo {year} {2017})},\
  \Eprint {http://arxiv.org/abs/1702.02992} {arXiv:1702.02992 [astro-ph.HE]}
  \BibitemShut {NoStop}%
\bibitem [{\citenamefont {Alfaro}\ \emph {et~al.}(2017)\citenamefont {Alfaro}
  \emph {et~al.}}]{2017arXiv171000890H}%
  \BibitemOpen
  \bibfield  {author} {\bibinfo {author} {\bibfnamefont {R.}~\bibnamefont
  {Alfaro}} \emph {et~al.} (\bibinfo {collaboration} {HAWC Collaboration}),\
  }\href {\doibase 10.1103/PhysRevD.96.122001} {\bibfield  {journal} {\bibinfo
  {journal} {Phys. Rev. D}\ }\textbf {\bibinfo {volume} {96}},\ \bibinfo
  {pages} {122001} (\bibinfo {year} {2017})}\BibitemShut {NoStop}%
\bibitem [{\citenamefont {Abeysekara}\ \emph {et~al.}(2018)\citenamefont
  {Abeysekara} \emph {et~al.}}]{Abeysekara:2018syp}%
  \BibitemOpen
  \bibfield  {author} {\bibinfo {author} {\bibfnamefont {A.~U.}\ \bibnamefont
  {Abeysekara}} \emph {et~al.} (\bibinfo {collaboration} {HAWC}),\ }\href
  {\doibase 10.1103/PhysRevD.97.102005} {\bibfield  {journal} {\bibinfo
  {journal} {Phys. Rev. D}\ } (\bibinfo {year} {2018}),\
  10.1103/PhysRevD.97.102005},\ \bibinfo {note} {[Phys.
  Rev.D97,102005(2018)]},\ \Eprint {http://arxiv.org/abs/1802.08913}
  {arXiv:1802.08913 [astro-ph.HE]} \BibitemShut {NoStop}%
\bibitem [{\citenamefont {Enriquez-Rivera}\ and\ \citenamefont
  {Lara}(2016)}]{Enriquez:2015nva}%
  \BibitemOpen
  \bibfield  {author} {\bibinfo {author} {\bibfnamefont {O.}~\bibnamefont
  {Enriquez-Rivera}}\ and\ \bibinfo {author} {\bibfnamefont {A.}~\bibnamefont
  {Lara}} (\bibinfo {collaboration} {HAWC}),\ }\bibfield  {booktitle} {\emph
  {\bibinfo {booktitle} {{Proceedings, 34th International Cosmic Ray Conference
  (ICRC 2015): The Hague, The Netherlands, July 30-August 6, 2015}}},\
  }\href@noop {} {\bibfield  {journal} {\bibinfo  {journal} {PoS}\ }\textbf
  {\bibinfo {volume} {ICRC2015}},\ \bibinfo {pages} {099} (\bibinfo {year}
  {2016})},\ \Eprint {http://arxiv.org/abs/1508.07351} {arXiv:1508.07351
  [astro-ph.SR]} \BibitemShut {NoStop}%
\bibitem [{\citenamefont {{G{\'o}rski}}\ \emph {et~al.}(2005)\citenamefont
  {{G{\'o}rski}}, \citenamefont {{Hivon}}, \citenamefont {{Banday}},
  \citenamefont {{Wandelt}}, \citenamefont {{Hansen}}, \citenamefont
  {{Reinecke}},\ and\ \citenamefont {{Bartelmann}}}]{2005ApJ...622..759G}%
  \BibitemOpen
  \bibfield  {author} {\bibinfo {author} {\bibfnamefont {K.~M.}\ \bibnamefont
  {{G{\'o}rski}}}, \bibinfo {author} {\bibfnamefont {E.}~\bibnamefont
  {{Hivon}}}, \bibinfo {author} {\bibfnamefont {A.~J.}\ \bibnamefont
  {{Banday}}}, \bibinfo {author} {\bibfnamefont {B.~D.}\ \bibnamefont
  {{Wandelt}}}, \bibinfo {author} {\bibfnamefont {F.~K.}\ \bibnamefont
  {{Hansen}}}, \bibinfo {author} {\bibfnamefont {M.}~\bibnamefont
  {{Reinecke}}}, \ and\ \bibinfo {author} {\bibfnamefont {M.}~\bibnamefont
  {{Bartelmann}}},\ }\href {\doibase 10.1086/427976} {\bibfield  {journal}
  {\bibinfo  {journal} {Astrophys. J.}\ }\textbf {\bibinfo {volume} {622}},\
  \bibinfo {pages} {759} (\bibinfo {year} {2005})},\ \Eprint
  {http://arxiv.org/abs/astro-ph/0409513} {astro-ph/0409513} \BibitemShut
  {NoStop}%
\bibitem [{\citenamefont {{Atkins}}\ \emph {et~al.}(2003)\citenamefont
  {{Atkins}} \emph {et~al.}}]{2003ApJ...595..803A}%
  \BibitemOpen
  \bibfield  {author} {\bibinfo {author} {\bibfnamefont {R.}~\bibnamefont
  {{Atkins}}} \emph {et~al.},\ }\href {\doibase 10.1086/377498} {\bibfield
  {journal} {\bibinfo  {journal} {Astrophys. J.}\ }\textbf {\bibinfo {volume}
  {595}},\ \bibinfo {pages} {803} (\bibinfo {year} {2003})},\ \Eprint
  {http://arxiv.org/abs/astro-ph/0305308} {astro-ph/0305308} \BibitemShut
  {NoStop}%
\bibitem [{\citenamefont {{Abdo}}\ \emph {et~al.}(2012)\citenamefont {{Abdo}}
  \emph {et~al.}}]{2012ApJ...750...63A}%
  \BibitemOpen
  \bibfield  {author} {\bibinfo {author} {\bibfnamefont {A.~A.}\ \bibnamefont
  {{Abdo}}} \emph {et~al.},\ }\href {\doibase 10.1088/0004-637X/750/1/63}
  {\bibfield  {journal} {\bibinfo  {journal} {\apj}\ }\textbf {\bibinfo
  {volume} {750}},\ \bibinfo {eid} {63} (\bibinfo {year} {2012})},\ \Eprint
  {http://arxiv.org/abs/1110.0409} {arXiv:1110.0409 [astro-ph.HE]} \BibitemShut
  {NoStop}%
\bibitem [{\citenamefont {Abdo}\ \emph {et~al.}(2009)\citenamefont {Abdo} \emph
  {et~al.}}]{0004-637X-698-2-2121}%
  \BibitemOpen
  \bibfield  {author} {\bibinfo {author} {\bibfnamefont {A.~A.}\ \bibnamefont
  {Abdo}} \emph {et~al.},\ }\href
  {http://stacks.iop.org/0004-637X/698/i=2/a=2121} {\bibfield  {journal}
  {\bibinfo  {journal} {The Astrophysical Journal}\ }\textbf {\bibinfo {volume}
  {698}},\ \bibinfo {pages} {2121} (\bibinfo {year} {2009})}\BibitemShut
  {NoStop}%
\bibitem [{\citenamefont {{Christopher}}(2011)}]{2011PhDT........70C}%
  \BibitemOpen
  \bibfield  {author} {\bibinfo {author} {\bibfnamefont {G.~E.}\ \bibnamefont
  {{Christopher}}},\ }\emph {\bibinfo {title} {{Physics from the Very-High
  Energy Cosmic-Ray Shadows of the Moon and Sun with Milagro}}},\ \href@noop {}
  {Ph.D. thesis},\ \bibinfo  {school} {New York University} (\bibinfo {year}
  {2011})\BibitemShut {NoStop}%
\bibitem [{\citenamefont {Amenomori}\ \emph {et~al.}(2013)\citenamefont
  {Amenomori} \emph {et~al.}}]{Amenomori:2013own}%
  \BibitemOpen
  \bibfield  {author} {\bibinfo {author} {\bibfnamefont {M.}~\bibnamefont
  {Amenomori}} \emph {et~al.} (\bibinfo {collaboration} {Tibet AS$\gamma$}),\
  }\href {\doibase 10.1103/PhysRevLett.111.011101} {\bibfield  {journal}
  {\bibinfo  {journal} {Phys. Rev. Lett.}\ }\textbf {\bibinfo {volume} {111}},\
  \bibinfo {pages} {011101} (\bibinfo {year} {2013})},\ \Eprint
  {http://arxiv.org/abs/1306.3009} {arXiv:1306.3009 [astro-ph.SR]} \BibitemShut
  {NoStop}%
\bibitem [{\citenamefont {{Amenomori}}\ and\ \citenamefont {{Tibet AS{$\gamma$}
  Collaboration}}(2018)}]{2018PhRvL.120c1101A}%
  \BibitemOpen
  \bibfield  {author} {\bibinfo {author} {\bibfnamefont {M.}~\bibnamefont
  {{Amenomori}}}\ and\ \bibinfo {author} {\bibnamefont {{Tibet AS{$\gamma$}
  Collaboration}}},\ }\href {\doibase 10.1103/PhysRevLett.120.031101}
  {\bibfield  {journal} {\bibinfo  {journal} {Physical Review Letters}\
  }\textbf {\bibinfo {volume} {120}},\ \bibinfo {eid} {031101} (\bibinfo {year}
  {2018})},\ \Eprint {http://arxiv.org/abs/1801.06942} {arXiv:1801.06942
  [astro-ph.SR]} \BibitemShut {NoStop}%
\bibitem [{\citenamefont {Amenomori}\ \emph
  {et~al.}(2018{\natexlab{a}})\citenamefont {Amenomori} \emph
  {et~al.}}]{Amenomori:2018xip}%
  \BibitemOpen
  \bibfield  {author} {\bibinfo {author} {\bibfnamefont {M.}~\bibnamefont
  {Amenomori}} \emph {et~al.} (\bibinfo {collaboration} {Tibet AS$\gamma$}),\
  }\href {\doibase 10.3847/1538-4357/aac2e6} {\bibfield  {journal} {\bibinfo
  {journal} {Astrophys. J.}\ }\textbf {\bibinfo {volume} {860}},\ \bibinfo
  {pages} {13} (\bibinfo {year} {2018}{\natexlab{a}})},\ \Eprint
  {http://arxiv.org/abs/1806.03387} {arXiv:1806.03387 [astro-ph.HE]}
  \BibitemShut {NoStop}%
\bibitem [{\citenamefont {{Abeysekara}}\ \emph {et~al.}(2017)\citenamefont
  {{Abeysekara}} \emph {et~al.}}]{2017ApJ...842...85A}%
  \BibitemOpen
  \bibfield  {author} {\bibinfo {author} {\bibfnamefont {A.~U.}\ \bibnamefont
  {{Abeysekara}}} \emph {et~al.},\ }\href {\doibase 10.3847/1538-4357/aa751a}
  {\bibfield  {journal} {\bibinfo  {journal} {Astrophysical Journal}\ }\textbf
  {\bibinfo {volume} {842}},\ \bibinfo {eid} {85} (\bibinfo {year} {2017})},\
  \Eprint {http://arxiv.org/abs/1703.01344} {arXiv:1703.01344 [astro-ph.HE]}
  \BibitemShut {NoStop}%
\bibitem [{\citenamefont {{Li}}\ and\ \citenamefont
  {{Ma}}(1983)}]{1983ApJ...272..317L}%
  \BibitemOpen
  \bibfield  {author} {\bibinfo {author} {\bibfnamefont {T.-P.}\ \bibnamefont
  {{Li}}}\ and\ \bibinfo {author} {\bibfnamefont {Y.-Q.}\ \bibnamefont
  {{Ma}}},\ }\href {\doibase 10.1086/161295} {\bibfield  {journal} {\bibinfo
  {journal} {Astrophys. J.}\ }\textbf {\bibinfo {volume} {272}},\ \bibinfo
  {pages} {317} (\bibinfo {year} {1983})}\BibitemShut {NoStop}%
\bibitem [{\citenamefont {{Younk}}\ \emph {et~al.}(2015)\citenamefont {{Younk}}
  \emph {et~al.}}]{2015arXiv150807479Y}%
  \BibitemOpen
  \bibfield  {author} {\bibinfo {author} {\bibfnamefont {P.~W.}\ \bibnamefont
  {{Younk}}} \emph {et~al.},\ }\href@noop {} {\bibfield  {journal} {\bibinfo
  {journal} {ArXiv e-prints}\ } (\bibinfo {year} {2015})},\ \Eprint
  {http://arxiv.org/abs/1508.07479} {arXiv:1508.07479 [astro-ph.IM]}
  \BibitemShut {NoStop}%
\bibitem [{\citenamefont {{Feldman}}\ and\ \citenamefont
  {{Cousins}}(1998)}]{1998PhRvD..57.3873F}%
  \BibitemOpen
  \bibfield  {author} {\bibinfo {author} {\bibfnamefont {G.~J.}\ \bibnamefont
  {{Feldman}}}\ and\ \bibinfo {author} {\bibfnamefont {R.~D.}\ \bibnamefont
  {{Cousins}}},\ }\href {\doibase 10.1103/PhysRevD.57.3873} {\bibfield
  {journal} {\bibinfo  {journal} {Phys. Rev. D}\ }\textbf {\bibinfo {volume}
  {57}},\ \bibinfo {pages} {3873} (\bibinfo {year} {1998})},\ \Eprint
  {http://arxiv.org/abs/physics/9711021} {physics/9711021} \BibitemShut
  {NoStop}%
\bibitem [{\citenamefont {Marinelli}\ and\ \citenamefont
  {Goodman}(2017)}]{Marinelli:2017vzu}%
  \BibitemOpen
  \bibfield  {author} {\bibinfo {author} {\bibfnamefont {S.~S.}\ \bibnamefont
  {Marinelli}}\ and\ \bibinfo {author} {\bibfnamefont {J.}~\bibnamefont
  {Goodman}} (\bibinfo {collaboration} {HAWC}),\ }in\ \href
  {https://inspirehep.net/record/1615737/files/arXiv:1708.03502.pdf} {\emph
  {\bibinfo {booktitle} {{Proceedings, 35th International Cosmic Ray Conference
  (ICRC 2017): Bexco, Busan, Korea, July 12-20, 2017}}}}\ (\bibinfo {year}
  {2017})\ \Eprint {http://arxiv.org/abs/1708.03502} {arXiv:1708.03502
  [astro-ph.IM]} \BibitemShut {NoStop}%
\bibitem [{\citenamefont {Amenomori}\ \emph
  {et~al.}(2018{\natexlab{b}})\citenamefont {Amenomori} \emph
  {et~al.}}]{PhysRevLett.120.031101}%
  \BibitemOpen
  \bibfield  {author} {\bibinfo {author} {\bibfnamefont {M.}~\bibnamefont
  {Amenomori}} \emph {et~al.} (\bibinfo {collaboration} {The Tibet AS$\gamma$
  Collaboration}),\ }\href {\doibase 10.1103/PhysRevLett.120.031101} {\bibfield
   {journal} {\bibinfo  {journal} {Phys. Rev. Lett.}\ }\textbf {\bibinfo
  {volume} {120}},\ \bibinfo {pages} {031101} (\bibinfo {year}
  {2018}{\natexlab{b}})}\BibitemShut {NoStop}%
\bibitem [{\citenamefont {Gao}\ \emph {et~al.}(2018)\citenamefont {Gao},
  \citenamefont {Chen}, \citenamefont {Li}, \citenamefont {Yu}, \citenamefont
  {Liu},\ and\ \citenamefont {He}}]{Gao:2017bfv}%
  \BibitemOpen
  \bibfield  {author} {\bibinfo {author} {\bibfnamefont {B.}~\bibnamefont
  {Gao}}, \bibinfo {author} {\bibfnamefont {S.}~\bibnamefont {Chen}}, \bibinfo
  {author} {\bibfnamefont {Z.}~\bibnamefont {Li}}, \bibinfo {author}
  {\bibfnamefont {C.}~\bibnamefont {Yu}}, \bibinfo {author} {\bibfnamefont
  {K.}~\bibnamefont {Liu}}, \ and\ \bibinfo {author} {\bibfnamefont
  {H.}~\bibnamefont {He}},\ }\href {\doibase 10.22323/1.301.0878} {\bibfield
  {journal} {\bibinfo  {journal} {PoS}\ }\textbf {\bibinfo {volume}
  {ICRC2017}},\ \bibinfo {pages} {878} (\bibinfo {year} {2018})}\BibitemShut
  {NoStop}%
\bibitem [{\citenamefont {{Joshi}}\ and\ \citenamefont
  {{Jardin-Blicq}}(2017)}]{2017arXiv170804032J}%
  \BibitemOpen
  \bibfield  {author} {\bibinfo {author} {\bibfnamefont {V.}~\bibnamefont
  {{Joshi}}}\ and\ \bibinfo {author} {\bibfnamefont {A.}~\bibnamefont
  {{Jardin-Blicq}}},\ }\href@noop {} {\bibfield  {journal} {\bibinfo  {journal}
  {ArXiv e-prints}\ } (\bibinfo {year} {2017})},\ \Eprint
  {http://arxiv.org/abs/1708.04032} {arXiv:1708.04032 [astro-ph.IM]}
  \BibitemShut {NoStop}%
\bibitem [{\citenamefont {In}\ and\ \citenamefont {Rott}(2017)}]{icecubesolar}%
  \BibitemOpen
  \bibfield  {author} {\bibinfo {author} {\bibfnamefont {S.}~\bibnamefont
  {In}}\ and\ \bibinfo {author} {\bibfnamefont {C.}~\bibnamefont {Rott}}
  (\bibinfo {collaboration} {IceCube}),\ }in\ \href
  {https://pos.sissa.it/301/965/pdf} {\emph {\bibinfo {booktitle}
  {{Proceedings, 35th International Cosmic Ray Conference (ICRC 2017): Bexco,
  Busan, Korea, July 12-20, 2017}}}}\ (\bibinfo {year} {2017})\ \Eprint
  {http://arxiv.org/abs/1710.01194} {arXiv:1710.01194} \BibitemShut {NoStop}%
\bibitem [{\citenamefont {{Aartsen}}\ \emph {et~al.}(2017)\citenamefont
  {{Aartsen}} \emph {et~al.}}]{2017EPJC...77..146A}%
  \BibitemOpen
  \bibfield  {author} {\bibinfo {author} {\bibfnamefont {M.~G.}\ \bibnamefont
  {{Aartsen}}} \emph {et~al.},\ }\href {\doibase
  10.1140/epjc/s10052-017-4689-9} {\bibfield  {journal} {\bibinfo  {journal}
  {European Physical Journal C}\ }\textbf {\bibinfo {volume} {77}},\ \bibinfo
  {eid} {146} (\bibinfo {year} {2017})},\ \Eprint
  {http://arxiv.org/abs/1612.05949} {arXiv:1612.05949 [astro-ph.HE]}
  \BibitemShut {NoStop}%
\bibitem [{\citenamefont {Adrian-Martinez}\ \emph
  {et~al.}(2016{\natexlab{b}})\citenamefont {Adrian-Martinez} \emph
  {et~al.}}]{Adrian-Martinez:2016fdl}%
  \BibitemOpen
  \bibfield  {author} {\bibinfo {author} {\bibfnamefont {S.}~\bibnamefont
  {Adrian-Martinez}} \emph {et~al.} (\bibinfo {collaboration} {KM3Net}),\
  }\href {\doibase 10.1088/0954-3899/43/8/084001} {\bibfield  {journal}
  {\bibinfo  {journal} {J. Phys.}\ }\textbf {\bibinfo {volume} {G43}},\
  \bibinfo {pages} {084001} (\bibinfo {year} {2016}{\natexlab{b}})},\ \Eprint
  {http://arxiv.org/abs/1601.07459} {arXiv:1601.07459 [astro-ph.IM]}
  \BibitemShut {NoStop}%
\end{thebibliography}%
\end{document}